\documentclass{aa}  
\usepackage{hyperref}
\hypersetup{colorlinks, allcolors=blue}
\usepackage{graphicx}    
\usepackage{color}  
\usepackage{txfonts}
\usepackage{inputenc}
\usepackage{multicol}
\usepackage{multirow}
\usepackage{longtable}
\usepackage{subcaption}  % use for side-by-side figures
\usepackage{placeins}
\usepackage{siunitx}
\usepackage{amsmath}
\usepackage{xcolor}
\usepackage{booktabs}
\usepackage{color} %DIF > 

\usepackage{algorithm} 
\usepackage{algpseudocode} 
\begin{document}

   \title{Survey of Profile Parameters of the {6196 \AA} Diffuse Interstellar Band}
   \subtitle{From Uniform Profiles to Doppler Splitting and Blueshifts}

   \author{M. Piecka\inst{1}
          \and
          S. Hutschenreuter\inst{1}
          \and
          J. Alves\inst{1,2}
          }

   \institute{University of Vienna, Department of Astrophysics,
              T\"urkenschanzstrasse 17, 1180 Vienna, Austria\\
              \email{martin.piecka@univie.ac.at}
         \and
             University of Vienna, Research Network Data Science at Uni Vienna, Kolingasse 14-16, 1090 Vienna, Austria}

   \date{Received X XX, XXXX; accepted X XX, XXXX}

%\abstract{}{}{}{}{} 
% 5 {} token are mandatory

  \abstract{

The diffuse interstellar band (DIB) at {6196~\AA} exhibits notable profile variations across the Milky Way. This study addresses three open issues: the unusual broadening of the DIB profile towards Upper~Sco, the lack of profile variations towards stars near $\eta$~Car, and the origin of the blueshift observed in Sco~OB1.
Using archival spectra of 453 early-type stars across the Galactic disk and in its proximity, we created a catalogue of the DIB's profile parameters.
Our analysis identified Doppler-split components within the DIB profiles across most regions with no evidence for these splits being able to account for the observed broadening ($\sim 23$~km\,s$^{-1}$) in Upper~Sco or other regions such as Orion, Vela~OB2, and Melotte~20 ($\alpha$~Per cluster).
We propose that neither the ages of the studied stellar populations nor the distances between clusters and nearby clouds significantly contribute to the broadening.
However, we detect a gradient in the full width at half maximum within the Sco-Cen and Orion regions, where broadening decreases with distance from the star-forming centres. This result points to a possible connection between the DIB broadening and star formation (likely via the impact of recent supernovae).
Regarding the Carina~Nebula, we confirm the lack of DIB profile variations in a small region near $\eta$~Car, although an adjacent southern area exhibits significant variations, comparable to those in Upper~Sco. In addition to the Carina~Nebula, we find that the Rosette~Nebula and NGC~6405 also show consistently narrow profiles ($<20$~km\,s$^{-1}$) with minimal deviations from the median over spatial scales of a few parsecs.
Finally, regarding the origin of the blueshift observed in Sco~OB1, we used a comparison with the Lagoon~Nebula and argue that the most natural explanation is the presence of an unresolved kinematic component in the profile of the DIB, shifting the measured centre of the band.

  }


   \maketitle
%
%-------------------------------------------------------------------
%-------------------------------------------------------------------
%-------------------------------------------------------------------
%-------------------------------------------------------------------
%-------------------------------------------------------------------
\section{Introduction}\label{section:1}

Diffuse interstellar bands (DIBs) are spectral absorption features that can be easily identified in the lines of sight towards OB stars \citep{1995ARA&A..33...19H}. They were originally discovered by \citet{1922LicOB..10..141H}, although further research started a decade later when the carriers were identified as interstellar medium (ISM) components \citep{1934PASP...46..206M,1938ApJ....87....9M}. The equivalent widths (EWs) of DIBs were found to be correlated with the interstellar reddening, suggesting a connection to the interstellar dust. It is predominantly believed that the carriers of DIBs are organic or carbonaceous molecules \citep{1985A&A...146...81L,1993ApJ...407..142H,1999ApJ...526..265S}. Up to this date, hundreds of DIBs have been catalogued in the optical and the near-IR parts of the spectrum \citep{2008ApJ...680.1256H,2009ApJ...705...32H,2014A&A...569A.117C,2017MNRAS.467.3099G,2019ApJ...878..151F,2023A&A...672A.181S}, with only a handful of them being unambiguously assigned to a specific molecule \citep{1997A&A...317L..59F, 2015Natur.523..322C,2019ApJ...875L..28C,2022MNRAS.509.4908N}.

Our understanding of the carriers of the DIBs and the interstellar properties the DIBs trace remains minimal. Some of the most important puzzle pieces added since the late 1990s include the ongoing discussion regarding the profile of the {6614~\AA} DIB \citep{1996ApJS..106..533W,2002A&A...384..215G,2004ApJ...611L.113C,2015ApJ...801....6B,2015MNRAS.453.3912M,2022A&A...662A..24M,2024A&A...686A..50E}, the discovery of a strong correlation between the {6196~\AA} and the {6614~\AA} DIBs \citep{1999A&A...351..680M,2010ApJ...708.1628M,2011ApJ...727...33F,2016A&A...585A..12B,2016MNRAS.460.2706K,2020MNRAS.496.2231B}, the identification of a specific DIB carrier \citep[C$_{60}^+$,][]{2015Natur.523..322C}, the studies of DIB profile variability \citep{1999AstL...25..656G,2017MNRAS.470.2835L,2023A&A...678A.148F}, and the production of 3D maps of DIB EWs \citep{2014Sci...345..791K,2015ApJ...798...35Z,2019NatAs...3..922F,2024A&A...689A..38C}. The review by \citet{2018PASP..130g1001K} includes additional information about the observational properties of DIBs. The complexity of the carriers of DIBs raises several questions that are important for our understanding of the ISM, including the carrier formation processes, the role of carriers in the ISM, and the properties of the ISM reflected in the DIB profiles.  

Several unexplained and unexplored observational facts have been reported over the last two decades. For example, some lines of sight appear to show differences in the structure of the {6196~\AA} DIB. \citet{2021MNRAS.508.4241K} find significant variations in the profile structure of several strong DIBs, including a shift in the central wavelength. These difference are mostly visible in the red wings of the DIBs. \citet{2023MNRAS.523.4158G} compare six stars from the Upper~Scorpius (USco) region in the Sco-Cen OB association with three different regions: a set of four stars in the direction of the Perseus complex of clouds, a set of three stars located within the Galactic disk below Sco-Cen, and a single target in the Galactic 1st quadrant. USco seems to displays a double-peaked profile of the {6196~\AA} DIB and a significant broadening compared with the other regions (especially Perseus). Based on the lack of splitting in the CH line at {4300~\AA}, \citet{2023MNRAS.523.4158G} argue against the presence of multiple clouds as the cause of the DIB broadening. The effect of rotational temperature on the profile of the {6196~\AA} DIB in USco is also found to be an unlikely source of the broadening. Instead, the authors suggest that exploring the properties of the cloud between us and USco might provide an additional insight into this mystery.

% ------------------- START TABLE
\begin{table*}
\caption{Regions investigated using the profile of the {6196~\AA} DIB}\label{table:1}
\small
\centering
\begin{tabular}{llll}
\hline\hline
Region ID & Region label & Targets & Included objects  \\
  \hline
  
1 & Eagle & 56 & NGC (6589, 6604); Omega~Nebula; Eagle~Nebula  \\
2 & Lagoon & 72 & Gum~(64, 76); NGC~(6357, 6405, 6475); Lagoon~Nebula; Sco~OB1 \\
3 & Sco-Cen & 89 & USco; UCL ($\eta$~Lup, Norma); NGC~(6067, 6188); LCC (disk, $\alpha$~Cen group); Chamaeleon \\
4 & Carina & 99 & IC~2944; NGC~(3572, 3576); Carina~Nebula  \\
5 & Vela & 27 & NGC~2516; Vela~OB2; Gum~(10, 14, 15, 17) \\
6 & CMa-Mon & 33 & Sh2-310; NGC~(2384, 2422); CMa~OB1; Rosette~Nebula \\
7 & Orion & 33 & Orion~(A, B); Briceno~1; Collinder~69; $\mu$~Tau \\
8 & Per-Tau & 10 & Pleiades; Perseus~cloud; Melotte~20 \\
9 & Gem-Aur & 8 & Sh2-247; IC~(405, 410, 417) \\
10 & Heart & 5 & IC~1848; Heart Nebula \\
11 & Sadr & 7 & Sh2-119; North~America; IC~1318 \\

\hline
\end{tabular}
\tablefoot{
The profile of the DIB was extracted for more stars than included in this table. However, these are either unrelated to a region of interest or do not provide sufficient information for studying a region (e.g. HD~209744). Regions 9--11 are poorly sampled.
}
\end{table*}
% ------------------- END TABLE

In addition to profile broadening, the central wavelengths of DIBs are known to display an offset from the central wavelengths of other interstellar species. \citet{2008PASP..120..178G} noticed a relative blueshift (up to 10 km\,s$^{-1}$) of the {6196~\AA} and a few other examined DIBs in Sco~OB1 association (located over 1~kpc beyond Sco-Cen) compared with the CH line at {4300~\AA} and the K~\textsc{i} line at {7699~\AA}. The broadening of all the DIBs was also noted. More recently, \citet{2015MNRAS.451.3210K} identified a redshift ($\sim 25$ km\,s$^{-1}$) in the positions of many DIBs seen towards Orion~OB1. In this case, the shift cannot be identified in the position of the {6196~\AA} DIB, which appears to be "perfectly aligned with the atomic lines".

In contrast with the known correlation between the extinction and the absorption strength of DIBs, \citet{2024AcA....74...69K} suggest that the carriers of DIBs might not always occupy the same space in the ISM as the dust. Their analysis is based on three targets observed in the Carina Nebula, in the near vicinity of $\eta$~Car itself. The authors show that the properties of five strong DIBs ({5780~\AA}, {5797~\AA}, {6196~\AA}, {6379~\AA}, and {6614~\AA}) do not vary between the lines of sight. On the other hand, atomic lines (Ca~\textsc{ii}, Ca~\textsc{i}, K~\textsc{i}) and the interstellar extinction are found to vary considerably. It is suggested that the best explanation for this discrepancy is the presence of most line-of-sight DIB carriers in a cloud in front of (but not too far away from) the Carina Nebula. This is possibly related to the existence of small-scale interstellar structures mentioned in \citet{2016A&A...585A..12B}.

In this paper, we aim to explore the profile of the {6196 \AA} DIB in various regions in the sky, focusing especially on young star-forming regions and star clusters. Our first objective is to identify lines of sight displaying similar abnormal properties of the DIB as seen in USco. 
Secondly, we aim to examine the Carina~Nebula and see whether we can find evidence to support the recently suggested idea of the existence of separate DIB and dust clouds. Finally, we use the analyses of the individual sky regions and provide a natural explanation for the blueshift of the {6196~\AA} DIB observed towards Sco~OB1. In Section~\ref{section:2}, we describe the data we use to extract information of the profile of the DIB. Section~\ref{section:3} focuses on methodology. The individual regions are studied in Section~\ref{section:4} using available spectra, dust maps, and literature information. We extract information from our analysis and attempt to answer the questions raised by our objectives in Section~\ref{section:5}.

%-------------------------------------------------------------------
\section{Archival data}\label{section:2}

To achieve the goals of our project, we took advantage of the spectra obtained from the ESO archives. Based on the available literature \citep[for example,][]{2021MNRAS.508.4241K}, three instruments were found to be most suited for studying the rough structure of the profiles of DIBs: the Fiber-fed Extended Range Optical Spectrograph \citep[FEROS;][]{1999Msngr..95....8K}, the High Accuracy Radial velocity Planet Searcher \citep[HARPS;][]{2003Msngr.114...20M}, and the Ultraviolet and Visual Echelle Spectrograph \citep[UVES;][]{2003Msngr.114...10B}. We made use of all three instruments since the width of the investigated DIB is much larger than the instrumental resolution limits. However, for each target star we used only a single instrument, mostly to avoid the need of downsampling spectra (with the exception of a few UVES observations). Specifically, the instrument was chosen by making use of the highest available resolution, and in the case of a similar resolution the preference was shifted to higher signal-to-noise ratio (SNR) spectra.

To gain additional insight into the properties of the intervening ISM, we investigated the profiles of other DIBs, atomic, and simple molecular lines when possible. FEROS and HARPS cover the following useful features: Ca~\textsc{ii} ({3933~\AA}), Ca~\textsc{i} ({4227~\AA}), CH$^+$ ({4232~\AA}), CH$^*$ ({4300~\AA}), and the DIBs ({5780~\AA}, {5797~\AA}, {6379~\AA}, {6614~\AA}). However, only FEROS can be used to study the K~\textsc{i} line at {7699~\AA}. The available UVES spectra cover various ranges of wavelengths, most often with an {$\approx 80$~\AA} gap covering the {5780~\AA} and {5797~\AA} DIBs, and the presence of most above-mentioned features depends on this range.

In order to cover targets at declinations $\delta > 30^{\circ}$, we decided to also include several observations made with the ELODIE echelle spectrograph \citep{1996A&AS..119..373B}. This enabled us to add more measurements from the Perseus region and include the Heart~Nebula and IC~1318 in our analysis. Since ELODIE covers the wavelengths between {4000~\AA} and {6800~\AA}, we could not investigate the Ca~\textsc{ii} and the K~\textsc{i} lines.

Distance information for the observed targets was extracted from the astrometric data provided by {\it Gaia} Data Release~3 \citep{2021A&A...649A...1G,2023A&A...674A...1G} and Hipparcos \citep[only bright nearby stars,][]{1997A&A...323L..49P,2007A&A...474..653V}. These distances serve as upper limits for the distance towards the DIB cloud(s) along the lines of sight. In Section~\ref{section:4}, we show that this information is invaluable for investigating the structure of the DIB profile.

Making a comprehensive comparison between profile parameters in different regions is impossible without a good understanding of the ISM along the lines of sight. Since many of the spectra we used miss at least one of the most important ISM tracers (Ca~\textsc{ii}, K~\textsc{i}), we need to find a different source of information. Recently, \citet{2024A&A...685A..82E} published detailed 3D~maps of the interstellar dust extracted from the Gaia data. Two versions of these maps are available -- one reaching to the distance of 1.25~kpc (high resolution) and the other reaching 2~kpc (lower resolution). The 1.25-kpc map probes the nearby ISM in a better detail, resolving the structure of the nearby clouds -- we used this version of the map in our analysis of the DIB profiles.

%-------------------------------------------------------------------
\section{Extracting the profile of the {6196 \AA} DIB}\label{section:3}

To investigate the broadening of the DIB, we decided to first visually inspect all available spectra in various star-forming regions covered by the instruments mentioned in Section~\ref{section:2}. Since our aim is to investigate hundreds of targets, the spectra needed to be automatically processed in the second step. Three criteria were used to pick the ideal targets:
\begin{itemize}
    \item The influence of the stellar features on the results must be minimal (resulting in most of the target stars being of a spectral type O, B, or~A)
    \item To avoid a problematic detection of the DIB using an automatised approach, only targets with a clear presence of the DIB are included
    \item Any spectrum with an instrumental artifact present within the profile of the DIB must be excluded
\end{itemize}
This resulted in a list of 453 targets covering different regions in the sky.

\subsection{Investigated regions}\label{section:3.1}

Based on the available archival spectra, we identified multiple groups of observations aimed at studying specific regions in our Galaxy. These include star-forming regions such as the Eagle~Nebula and the Orion~Nebula -- see Table~\ref{table:1} for a full list of the studied regions. Each region was labelled either according to the presence of a prominent object within (for example, the Orion~Nebula in Orion) or based on the constellation(s) the region spans. For each object in a region, we attempted to find at least one external target at a similar Gaia distance (for example, HD~26912 for Orion, or HD~87643 for Carina).

Unfortunately, the number of targets in the 4th quadrant of our Galaxy is severely lacking. Despite the low number statistics, we attempted to provide a general description of the behaviour of the studied DIB. This was done by considering the information we extracted from the more richly studied objects, such as USco, the Eagle~Nebula, or the Lagoon~Nebula.

Overall, the list of included objects in Table~\ref{table:1} covers distances from 130~pc (USco) up to 2.5~kpc (Carina~Nebula), with a few lines of sight reaching to 4~kpc (HD~91597). Since many of these objects cover a similar region in the sky, we have the opportunity to study the variations in the profile of the {6196~\AA} DIB as a function of the distance (for example: NGC~6405 at 500~pc, the Lagoon~Nebula at 1.3~kpc, the Eagle~Nebula at 1.7~kpc). We would like to point out that our approach cannot compete with the maps produced for the Gaia DIBs or the infrared DIBs. However, there is a general lack of works focused on individual regions of our Galaxy using optical DIBs ({$4000-7000$~\AA}) that might provide important information about the DIB carriers \citep[for example][]{2011A&A...533A.129V}.

\subsection{Automatic extraction of profile parameters}\label{section:3.2}

The most important aspect of extracting profile parameters from interstellar lines is the subtraction of blended lines and the continuum normalisation \citep[see][]{2013A&A...555A..25P}. Since the DIB at {6196~\AA} is mostly free from stellar and telluric contamination when observing OBA stars, our task was significantly simplified. The only possible issue can arise from the presence of a weaker DIB located {$\approx 1.3$~\AA} blue-wards from the DIB of interest. If the presence of the weaker DIB is an issue, the problem can be solved by simultaneously fitting both DIBs in the final step. However, a (local) continuum normalisation procedure can still be problematic to properly automatise.

The normalisation procedure applied by \citet{2023MNRAS.523.4158G} is based on fitting a cubic spline to the manually selected regions of the spectrum that are believed to represent the continuum. An automatisation of this approach would prove to be extremely difficult when considering the use of data products provided by different instruments. Instead, we decided to treat the whole wavelength range around the DIB ({$6190.8 - 6200.8$~\AA}) as continuum. This can be achieved by smoothing the observed spectrum, for example, by calculating the median flux within $N$ point from any given point. Such a procedure should behave well in the absence of strong absorption or emission features, which means that the close area around the DIB needs to be masked.

Masking of the DIB requires the identification of the following three important wavelengths: the blue end, the centre, and the red end of the feature. A rough estimate of the centre of the DIB can be obtained by fitting a simple line profile (for example, a Gaussian) to the non-normalised spectrum. This position of the line can then be used to identify the end-points of the DIB. We provide a detailed description of this procedure in Appendix~\ref{section:A}. The continuum within the profile of the DIB was evaluated by linear interpolation between the end-points.

The normalised spectra obtained by the same instrument for a given target star were stacked together. For FEROS and HARPS this procedure is trivial. For ELODIE and UVES, the spectra first needed to be shifted to the barycentric frame of reference. The wavelength-corrected spectra were then used to calculate the flux at predetermined wavelengths (based on the average resolution around the DIB) using linear interpolation. We note that the archival UVES spectra have varying resolution. For this reason, we tried to avoid the stacking process by only picking the highest mean SNR data product.

In the final step, the stacked normalised spectrum contains only the profile of the {6196~\AA} DIB. Since the profile can be Doppler split in some lines of sight and we lack a prior knowledge of its general structure, we decided to model the profile with a combination of two generalised Gaussian functions
\begin{equation}
    G(\lambda) = \sum_{i=1}^{2} A_i \exp{\left[-\frac{1}{2}\left(\frac{|\lambda - \lambda_i|}{\sigma_i}\right)^{\beta_i}\right]} \,\,.
\end{equation}
We used the following definitions for the basic profile parameters. Let the centre of the DIB (barycentric) be the median of the fitted profile, and let the line depth correspond to the maximum intensity. The FWHM of the profile can then be calculated as the difference between the left-most and the right-most points at half of the line depth. The EW was determined by integrating the area of the fitted profile.

\begin{figure}
 \includegraphics[width=\columnwidth]{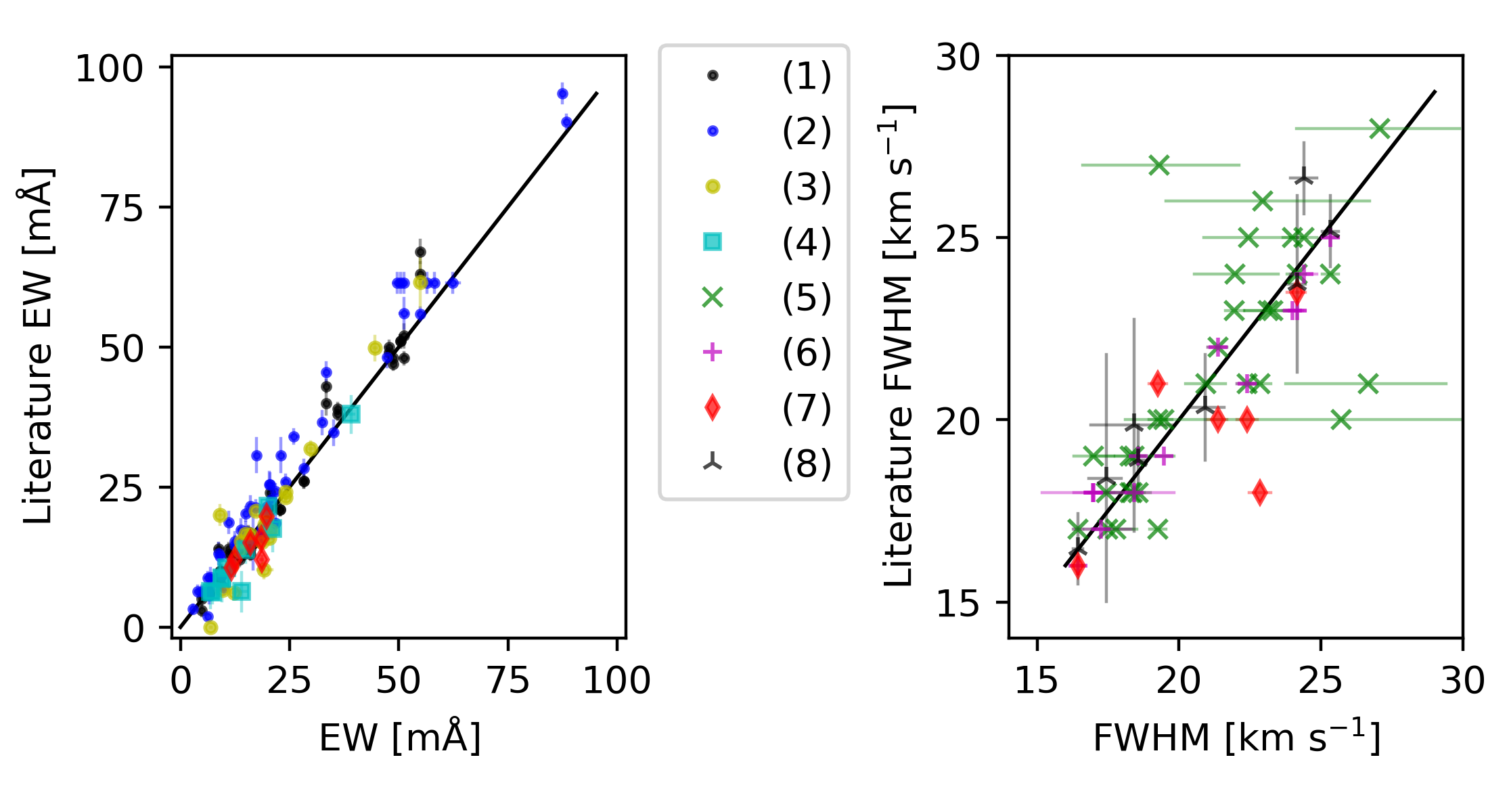}
 \caption{Validation of the procedure described in Section~\ref{section:3.2} and Appendix~\ref{section:A}. We used eight literature sources to compare with the results of our procedure: (1) \citet{2023A&A...678A.148F}, (2) \citet{2010ApJ...708.1628M}, (3) \citet{2005MNRAS.358..563M}, (4) \citet{2013A&A...555A..25P}, (5) \citet{2021MNRAS.508.4241K}, (6) \citet{2023MNRAS.523.4158G}, (7) \citet{2002A&A...384..215G}, and (8) \citet{2009A&A...498..785K}.}
 \label{fig:1}
\end{figure}

A bootstrapping procedure with $n=1000$ iterations was used to estimate the uncertainties of the four determined profile parameters. As described in \citet{2024A&A...689A..84P}, this approach works well for reasonably strong features. If the noise is high relative to the extracted line depth ({mostly for lines of sight with $\textrm{EW}_{6196} \lessapprox 5$~m\AA}), the uncertainties might be underestimated.

\subsection{Validation of profile parameters}\label{section:3.3}

For all measurements, we visually checked the quality of the result provided by our procedure, including all the steps of the continuum normalisation. Significant deviations from an acceptable fit were seen in 21 cases, all of them affected by an incorrect choice of the end-points. For these targets, we picked the end-points manually and ran the parameter extraction again. All of the results are available at the CDS.

The extracted profile parameters were compared with available literature data. Fig.~\ref{fig:1} clearly shows that the data points are distributed very close to the one-to-one line for the EW and the FWHM parameters. We notice a trend in the EW data, where our values seem to be systematically lower than what we find in the literature. In most cases, this is the result of a bad identification of the red wing end-point. Although this can noticeably affect the EW measurement, the FWHM should remain unaffected. We keep this short-coming in mind in the analysis below. The line centre parameter is often omitted in the literature and its definition may vary -- for example, the definition from \citet{2021MNRAS.508.4241K} differs significantly from our definition (Section~\ref{section:3.2}). Since no accepted model of the {6196~\AA} DIB profile is available, the ambiguous choice of the line centre prevents us from making an informative comparison with the literature.

%-------------------------------------------------------------------
\section{Probing our Galaxy with the {6196~\AA} DIB}\label{section:4}

Looking at the regions described in Table~\ref{table:1} (roughly sorted along the Galactic longitude, starting from $l=17^{\circ}$ and continuing westwards), we were able to identify multiple sets of targets sharing similar FWHM, EW, or the distance from the stars. For each of these sets, we calculated the median for the FWHM and the EW. We also computed the standard deviation for the FWHM to see how large are the variations in the width of the DIB depending on the line of sight within a sub-region. In the case of the EW, we found that the range of computed values is most informative. When necessary, we discuss the actual set of values and provide comments on the individual distances. All results can be found in Tables~\ref{table:2} and \ref{table:3}.

When comparing the profile of an interstellar feature in two different lines of sight, the key step in analysing the profiles is the identification of the rest wavelength. Our initial assumption is that most of the DIB carriers responsible for the observed absorption are located at the same location as the dense atomic gas probed by the K~\textsc{i} line. When possible, we compare this rest wavelength with the one extracted from the CH$^*$ line. The lack of information about the K~\textsc{i} line (Section~\ref{section:2}) resulted in a limited sample of stars usable for a direct comparison.

The dust maps from \citet{2024A&A...685A..82E} were used to analyse the regions in the sky mentioned in Tables~\ref{table:2} and \ref{table:3}. For each target of interest, we extracted the average extinctions in $A_V$ and plotted the (cumulative) extinction along the line of sight as a function of distance. The distance limit of the used dust map (Section~\ref{section:2}) implies a distance limit for the structures probed in front of the more distant objects, such as the Eagle~Nebula. However, identification of ISM structures within 1.25~kpc is still possible.

In the subsections below, we summerise our findings for the individual objects from Tables~\ref{table:2} and \ref{table:3}. We focused on investigating each region separately. Any behaviour of the profile of the DIB that can be identified towards most of the objects is pointed out in Section~\ref{section:5}.

\subsection{Eagle}\label{section:4.1}

\begin{figure}
 \includegraphics[width=\columnwidth]{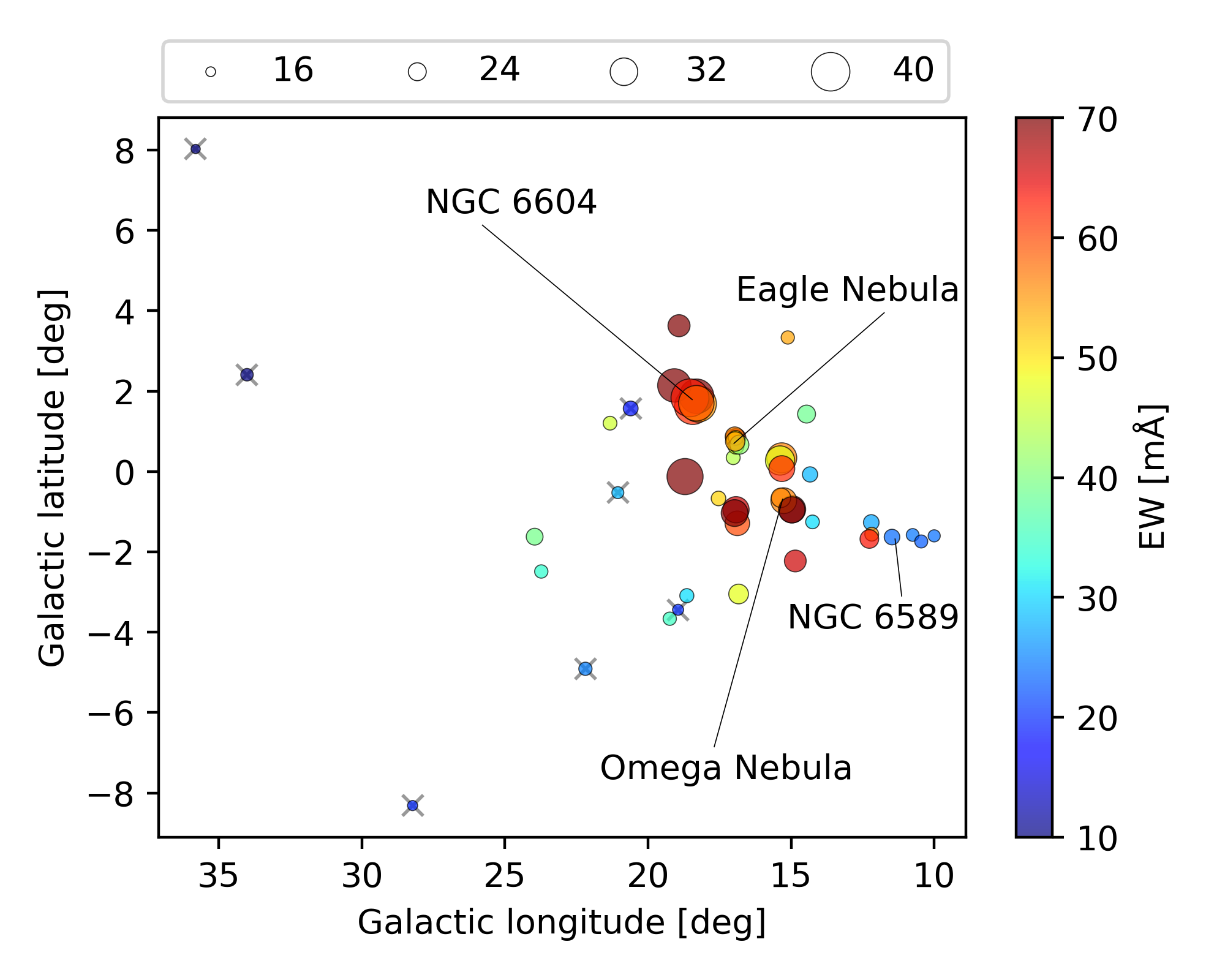}
 \caption{Sky map of the targets within the Eagle region. The size of a data point indicates the measured FWHM (values in~km\,s$^{-1}$), while the colour indicates the strength of the DIB. The most prominent groups of stars are highlighted. Grey crosses indicate stars located within 300~pc from the Sun.}
 \label{fig:2}
\end{figure}

We identify strong variations in the FWHM within this region (Fig.~\ref{fig:2}). Starting with the stars located in the vicinity of NGC~6589, we find a statistically significant jump from $\sim~20$~km\,s$^{-1}$ to $\sim~30$~km\,s$^{-1}$ near the Omega~Nebula (NGC~6618). The three stars located between the Omega~Nebula and the Eagle~Nebula also show the same value of the median FWHM, highlighting the increased broadening of the DIB in the region. Furthermore, NGC~6604 displays the broadest profile of the {6196~\AA} DIB in our data set (38~km\,s$^{-1}$). On the other hand, the DIB becomes significantly narrower as we look closer to the Eagle~Nebula located at the centre of the region.

In contrast with the other three prominent star clusters, the cluster NGC~6611 embedded within the Eagle~Nebula probes the foreground ISM with a good sky resolution. We are unable to identify any internal variations in the profile of the DIB. On the other hand, the peripheries of the Eagle region show a decrease in the FWHM as a function of the angular distance from the region, similar to what we see in the case of the Omega~Nebula and NGC~6589.

\begin{figure}
 \includegraphics[width=\columnwidth]{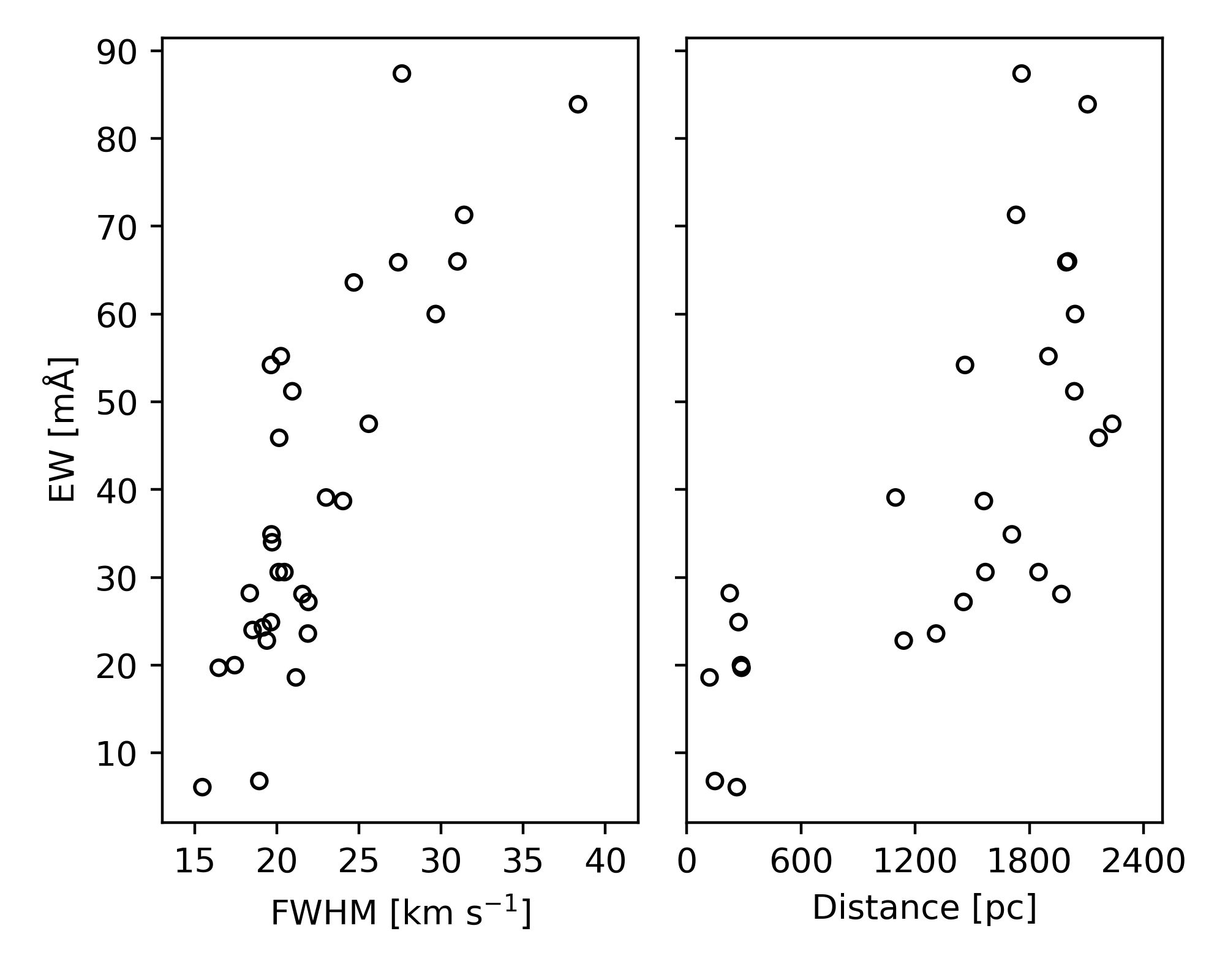}
 \caption{Correlations found in the peripheries of the Eagle region, including NGC~6589.}
 \label{fig:3}
\end{figure}

The line-of-sight distances from the individual objects and stars provide valuable information. The clusters NGC~6604, NGC~6611, and NGC~6618 are known to be located 1.7--2.0~kpc away from the Sun \citep[see][and references therein]{2000A&AS..144..451B,2023A&A...670A.108S,2024A&A...681A..21S}. On the other hand, most of the NGC~6589 targets are located at distances between 1.1 and 1.5~kpc, suggesting a position at least 100--200~pc closer than the Omega~Nebula, the Eagle~Nebula, or NGC~6604. Furthermore, our sample of targets located outside the star-forming regions includes seven stars located within 300~pc from the Sun. We note the existence of a linear correlation between the EW and the FWHM parameters in these outer parts of the region (Fig.~\ref{fig:3}). The EW does not vary significantly at smaller distances, but a notable jump from {20~m\AA} to {40--50~m\AA} can be observed at $d \gtrsim 1.5$~kpc.

\begin{figure}
 \includegraphics[width=\columnwidth]{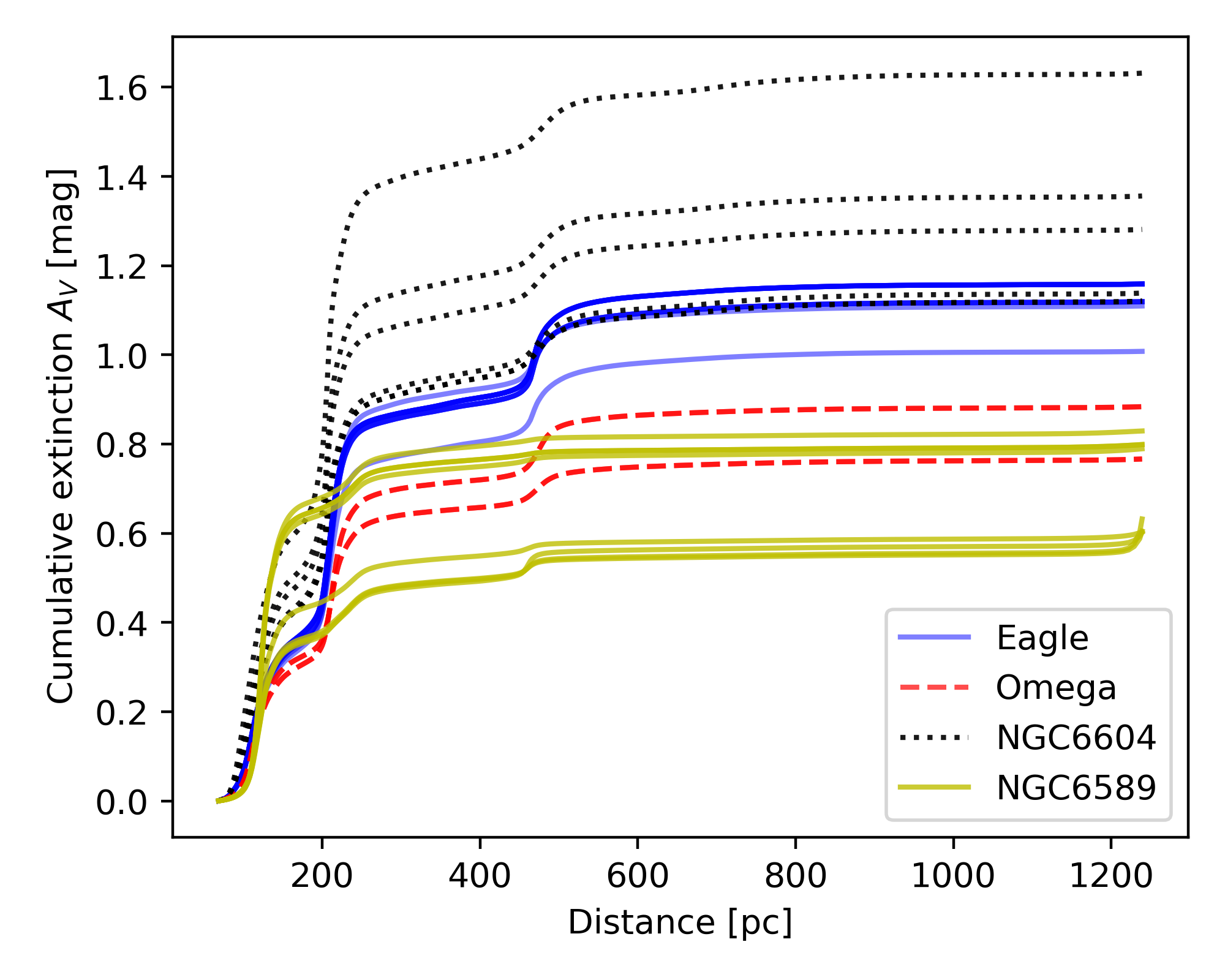}
 \caption{Line-of-sight extinction towards different targets in the Eagle region. The presence of two moderate-extinction clouds within 250~pc is noted, together with the diffuse cloud at 470~pc.}
 \label{fig:4}
\end{figure}

All of the information extracted from the Eagle region suggests that the observed DIB profile is composed of Doppler split components. The first component corresponds to a nearby cloud, located 100--250~pc away from the Sun. The presence of two distinct clouds at these distances is seen in dust maps (see Fig.~\ref{fig:4}), although the precise distances and the amount of extinction ($A_V$ between 0.4~mag and 0.8~mag) depend on the specific line of sight. Except for four targets located in the outskirts of the region, all directions show the lack of additional moderate-extinction clouds at distances 0.4--1.3~kpc. A more diffuse cloud is identified at 470~pc and is most distinguished towards the Eagle~Nebula ($A_V \lesssim 0.2$~mag). This cloud might be responsible for the presence of a weaker DIB component that almost overlaps with the one produced in the nearby cloud (similar velocities). Based on the knowledge we extracted from the {6196~\AA} DIB, a third distinct kinematic component must be located at a distance between 1.3~kpc and 1.7~kpc, otherwise we cannot explain the increase in the EW.

\begin{figure}
 \includegraphics[width=\columnwidth]{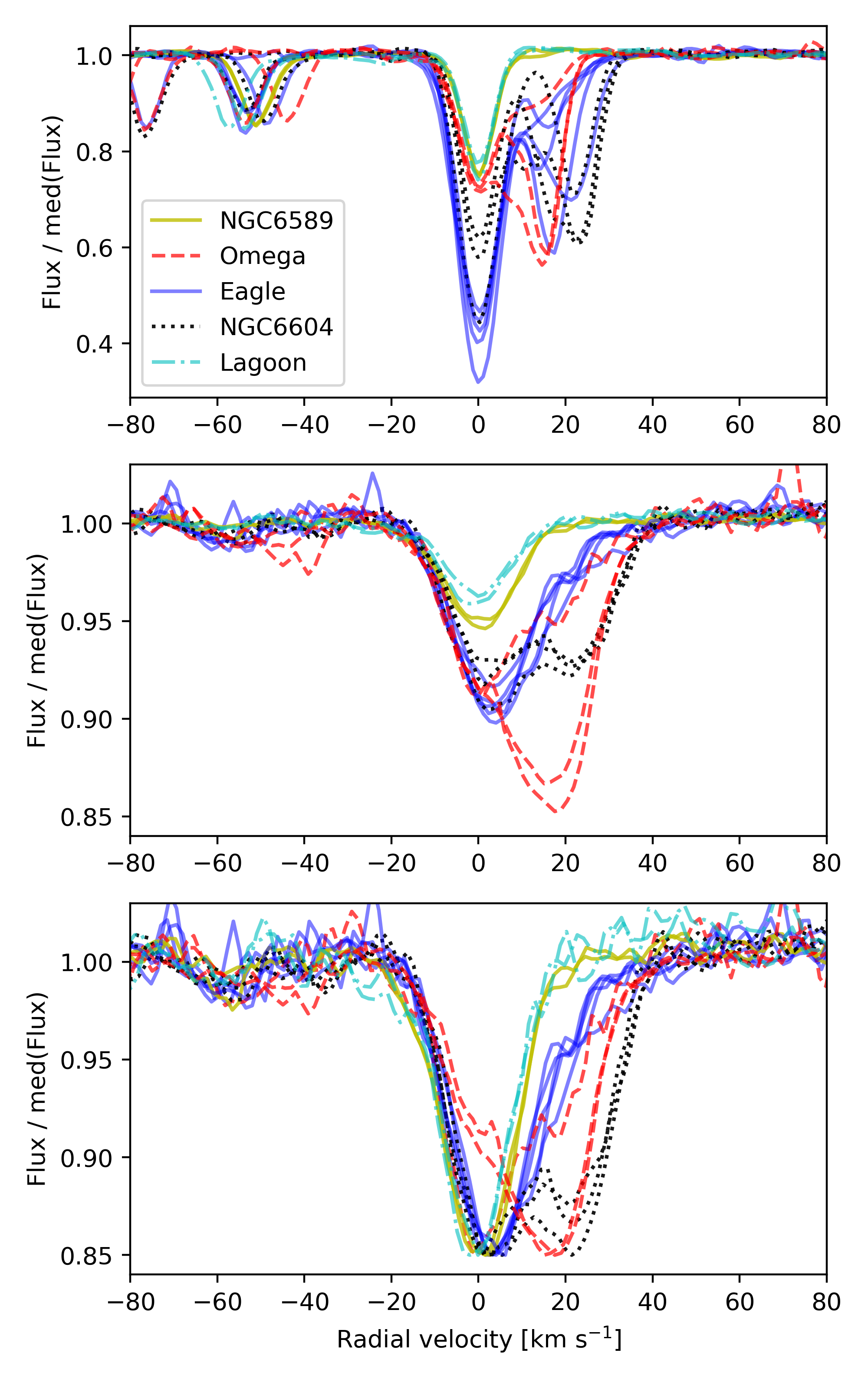}
 \caption{FEROS spectra of targets in the Eagle region and the Lagoon~Nebula. Wavelengths were shifted to the rest position of the K~\textsc{i} line (top) component corresponding to the motion of the local ISM. The middle panel shows the profile of the {6196~\AA} DIB. The bottom panel shows the same feature with intensities scaled to the same line depth. Continuum normalisation was not applied when producing this figure -- the fluxes were only re-scaled using the median of the flux within {5~\AA} from the shown features.}
 \label{fig:5}
\end{figure}

FEROS spectra obtained for some of the targets in the Eagle~region are displayed in Fig.~\ref{fig:5}. The FEROS instrument was chosen based on our aim to use the K~\textsc{i} line to identify the motion of the local ISM and to use it as the rest wavelength. We assume that other kinematic components seen in the interstellar features are produced by more distant clouds. Continuum normalisation is not performed in the production of Fig.~\ref{fig:5} --  our goal is to show the profile of the DIB without any complex flux reduction process. To be able to compare different spectra, we only use the median flux within {5~\AA} from the DIB (or K~\textsc{i}) for each individual star and divide the spectrum by this scalar value. We make use of a scaling procedure that returns the same line depth for each spectrum (Fig.~\ref{fig:5}, bottom panel) in order to search for variations in the DIB broadening.

The DIB profile appears to split (15--20~km\,s$^{-1}$) in the red wing, an effect clearly visible also in the profile of the K~\textsc{i} line. The split is most apparent in NGC~6604, leaving no doubt for this being the primary process responsible for the broadening of the DIB in this region. By comparing the intensities of the redshifted components of the DIB, we can highlight that the Omega~Nebula is the strongest, followed by NGC~6604, and the Eagle~Nebula being the weakest. Looking at the middle and the bottom panels of Fig~\ref{fig:5}, we note that the DIB in the Eagle~Nebula is slightly stronger (higher EW) when compared to NGC~6589. This discrepancy can be explained by the presence of another velocity component connected to the dust cloud at 470~pc. The cloud is significantly weaker towards NGC~6589 targets, which fits with the relatively increased intensity of the DIB at {6196.0~\AA} when looking towards the other objects in this region.

The profile of the DIB towards the young stellar object BD$-$16$^{\circ}$4826 is quite different from the other analysed targets in the Omega~Nebula, suggesting a difference between the north-eastern and the south-western parts of the nebula. This is the only piece of evidence hinting at a small scale variation in the profile of the DIB within the Eagle region. Unfortunately, we cannot base this finding on a statistically significant number of measurements. We note the apparent absence of a CH$^*$ component in the Omega~Nebula at the rest wavelength of the K~\textsc{i} line.

\subsection{Lagoon}\label{section:4.2}

\begin{figure}
 \includegraphics[width=\columnwidth]{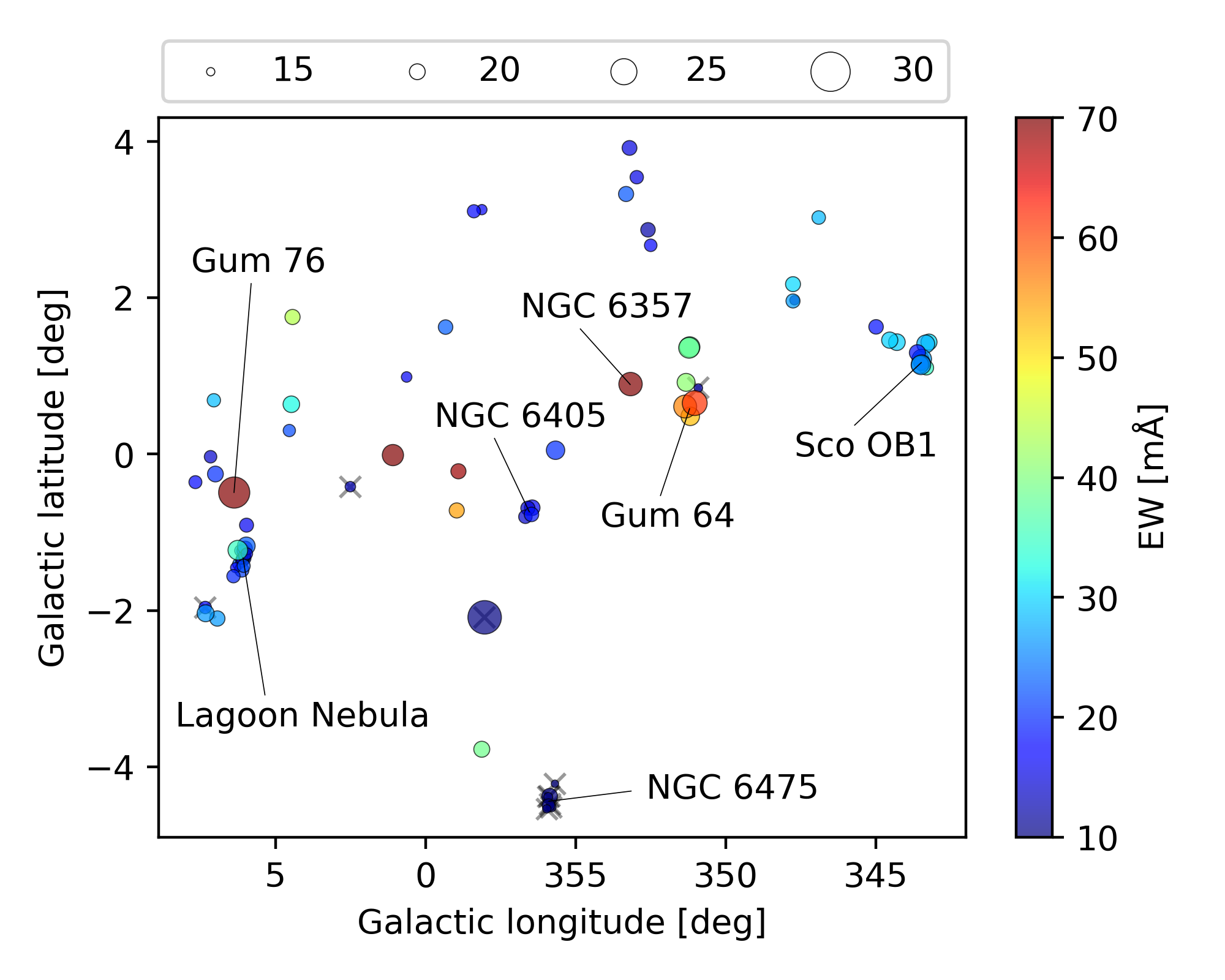}
 \caption{Same as Fig.~\ref{fig:2}, but focusing on the Lagoon region.}
 \label{fig:6}
\end{figure}

As can be seen in Fig.~\ref{fig:6}, we included a large number of objects in the Lagoon region. Table~\ref{table:2} shows that the DIB is fairly narrow (19~km\,s$^{-1}$) towards most of these objects. Although some of the targets are located close to the Sun (NGC~6405, NGC~6475), most of them can be found at distances larger than 1~kpc (the Lagoon~Nebula, Gum~64, Sco~OB1).

\begin{figure}
 \includegraphics[width=\columnwidth]{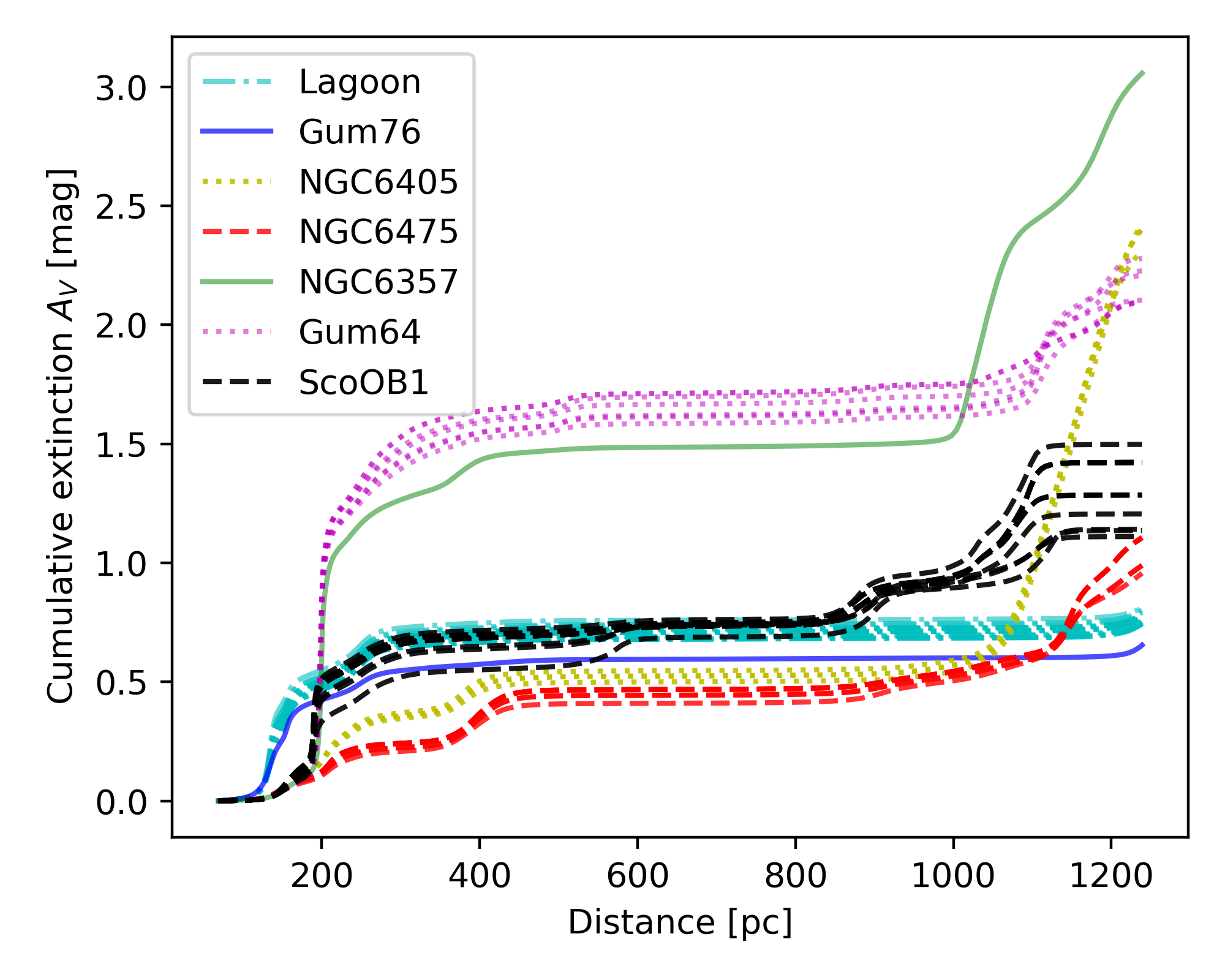}
 \caption{Line-of-sight extinction towards different targets in the Lagoon region. Most of the dust is located within 300~pc, although several walls can be seen at distances beyond 1~kpc, depending on the line of sight.}
 \label{fig:7}
\end{figure}

The lines of sight towards the Lagoon~Nebula show a DIB that is narrow and relatively weak, despite the nebula being located 1.3~kpc away \citep{2019MNRAS.486.2477W}. The spectra of most stars show FWHM values $\lesssim 19$~km\,s$^{-1}$, with the exception of Herschel~36, an O-type star located in a very dusty part of the nebula with an anomalous absorption in the red wings of {5780~\AA}, {5797~\AA}, and {6614~\AA} DIBs \citep{1961PASP...73..206W,2013ApJ...773...41D,2013ApJ...773...42O,2021MNRAS.508.4241K}, and a normal profile of at least one near-IR DIB \citep{2019MNRAS.485.3398R}. We note that most of the extinction towards the Lagoon~Nebula is related to the dust clouds located within 300~pc from the Sun (Fig.~\ref{fig:7}). Star NGC~6530~151 displays a strangely broad and strong profile of the DIB when compared with the rest of the Lagoon~Nebula targets. The spectral type of this star varies between literature sources, with the most likely classification being an earlier type~A or a type~B (based on the presence of a narrow He~\textsc{i} triplet at {5785~\AA}). Unlike Herschel~36, anomalous red wings in the profiles of DIBs are not observed towards NGC~6530~151.

Curiously, the peripheries of the Lagoon~Nebula show a broader and stronger median DIB profile. In the north, the spectra with the increased broadening correspond to targets located beyond 1.4~kpc. Although two stars in the south are located 100--200~pc closer than the nebula, they are associated with H~\textsc{ii} regions IC~1274 and IC~4685. The star probing IC~4685 shows a FWHM value that is still within the range of values observed in the Lagoon~Nebula. In the case of IC~1274, the two stars HD~165999 (170~pc) and HD~314031 (1250~pc) with available spectra suggest that the broadening might be related to the presence of the dusty wall covering the H~\textsc{ii} region (Barnard~91).

Only one of the targets assigned to the group labelled as Gum~76 actually belongs to this H~\textsc{ii} region. The three stars ($\lesssim1.3$~kpc) located to the east and to the north show a {6196~\AA} DIB profile as narrow as in the Lagoon~Nebula. The single star to the west is actually located $\sim3$~kpc away, suggesting that the increased FWHM and EW are related to the presence of multiple clouds along the line of sight. The DIB towards HD~164492, the tracer of Gum~76, is slightly broadened and stronger than the stars to the north/east, very similar to the Herschel~36 line of sight. We note that HD~164492 does not show signs of an extended tail towards red in the profile of any DIB.

Clusters NGC~6405 and NGC~6475 tell a very interesting story about the nearby ISM traced by the DIBs. Located slightly below the Galactic plane at the distance of 275~pc, NGC~6475 displays the narrowest profile of the {6196~\AA} DIB found in our whole data set, although two measurements show a FWHM of about 20~km\,s$^{-1}$. Of these two measurements, one corresponds to a problematic case that should be regarded as an outlier (HD~162724, see Appendix~\ref{section:C}). The presence of the weak narrow DIB ($\lesssim16$~km\,s$^{-1}$) is based on UVES spectra of sufficient SNR obtained for five stars in the cluster. NGC~6405 ($\sim450$~pc) can be found in the plane of our Galaxy, only $3.5^{\circ}$ to the west of the Galactic centre. The DIB seems to be very slightly broader in the area of this cluster when compared with the objects to the east. In terms of the structure of the line-of-sight extinction (Fig.~\ref{fig:7}), NGC~6405 does not seem to differ from NGC~6475, except for the total amount of extinction. The difference in the DIB broadening seen towards these two clusters might be explained by the presence (or absence) of a foreground dust cloud, specifically the cloud seen in Fig.~\ref{fig:7} at the distance of 400~pc.

\begin{figure}
 \includegraphics[width=\columnwidth]{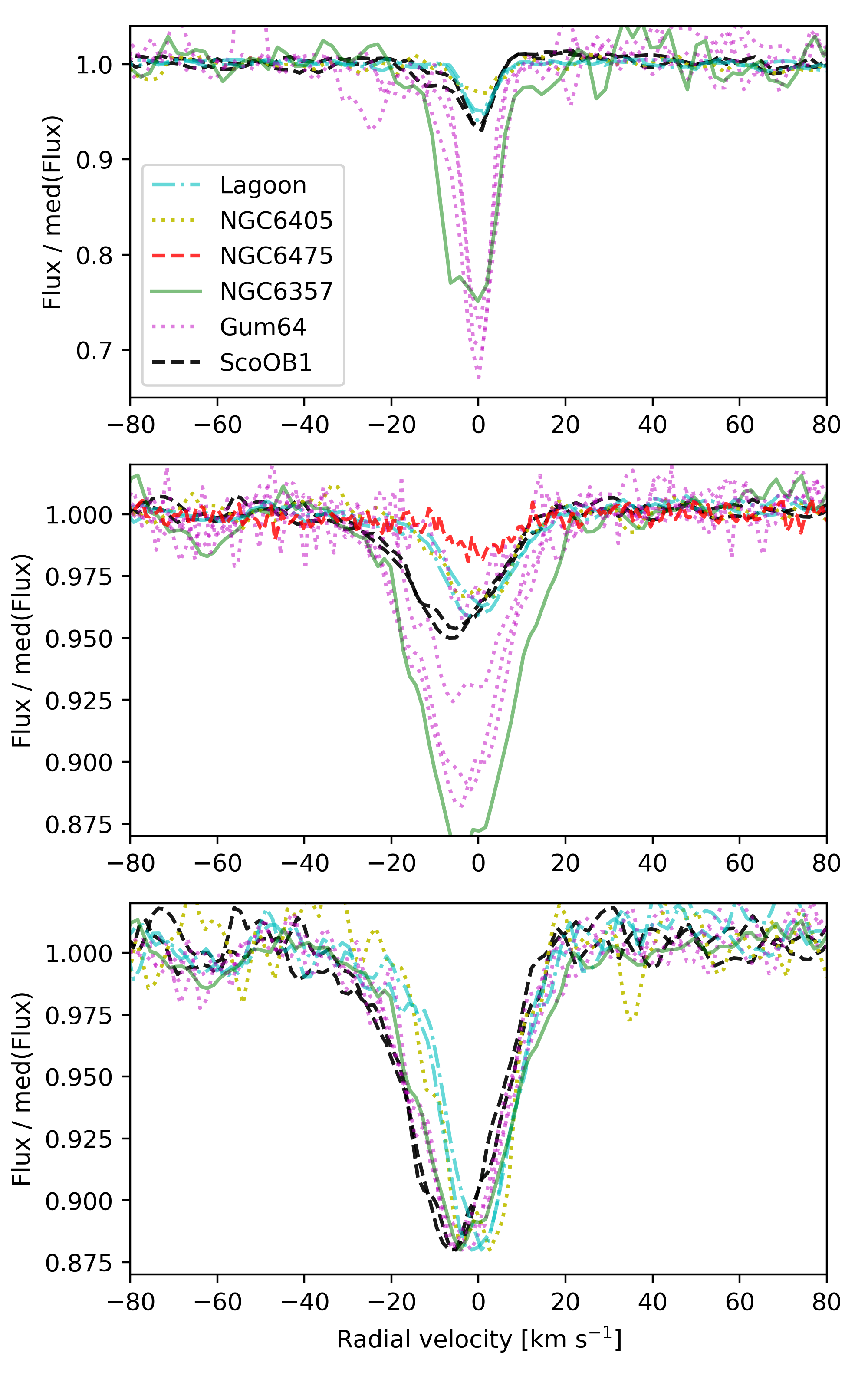}
 \caption{Same as Fig.~\ref{fig:5} but for the Lagoon region. Wavelengths were shifted to the rest position of CH$^*$ (top panel). Middle panel includes a UVES spectrum of the nearby star HD~156369 (low intensity, magenta dotted line) as an example of a very narrow DIB profile. HD~162780 (red dashed line) is also included, but the rest wavelength could only be estimated from the optical sodium doublet line.}
 \label{fig:8}
\end{figure}

Properties of the DIB seen towards the area surrounding Gum~64 depend on the specific line of sight. The three stars located to the north (Gum~63 and northwards) show FWHM of about 22~km\,s$^{-1}$ and a strength of {$\sim38$~m\AA}. A possibly stronger and broader profile (23~km\,s$^{-1}$, {$\sim56$~m\AA}) can be found in the spectra of HD~319703 (Gum~61), HD~319702 (Gum~64b), and HD~156738 (G351.2+0.5). For these six stars, some of the Gaia parallaxes are fairly uncertain, making the determination of the distances difficult. For this reason, we are unable to confirm or rule out small scale variations in the DIB profile around Gum~64. HD~156369, a star to the west of Gum~64, is located approximately 290~pc away and shows a DIB profile as narrow (see Fig.~\ref{fig:8}) as the one observed in NGC~6475.

Spectra of the stars around NGC~6281 support the general findings in the Lagoon region. However, a member of this stellar cluster, CD$-$37$^{\circ}$11229, shows a very narrow DIB that compares with the lines of sight toward the nearby stars (see above). The Gaia parallax measurement is quite precise and very similar to the mean distance based on the parallaxes of the whole cluster. It is worth noting that NGC~6281 is the oldest cluster investigated in this paper \citep[$\approx300$~Myr,][]{2021MNRAS.504..356D}.

Sco~OB1 was included in the Lagoon~region due to its relatively large distance \citep[1.3--1.4~kpc,][]{2020MNRAS.495.1349Y,2021MNRAS.504..356D} and its close proximity to the Galactic disk. We do not detect obvious variations of the FWHM or the EW in the sky. We note the presence of at least 2--3 prominent clouds in the dust maps (Fig.~\ref{fig:7}). Three stars unrelated to the OB association are included in this sample. They are located in the vicinity of the emission nebula IC~4628 at a distance of $\sim1.6$~kpc. We interpret this as a lack of DIB carriers in the ISM very close to Sco~OB1 or IC~4628. The extinction shown in Fig.~\ref{fig:7} agrees with this interpretation -- no detectable amount of dust is located between 1.1~kpc and 1.25~kpc.

Examples of the DIB spectra observed in the Lagoon region are displayed in Fig.~\ref{fig:8}. The Lagoon~Nebula, the nearby star near Gum~64, and NGC~6405 show very similar profile shapes, which suggests similar conditions in the DIB-probed ISM along the lines of sight. The very narrow profiles towards the Lagoon~Nebula and NGC~6475 can serve as useful tools for understanding the intrinsic profile of the DIB. The broad DIB towards Sco~OB1 can be easily explained by Doppler splitting, which is observed in the profiles of CH$^*$ (Fig.~\ref{fig:8}, top panel) and K~\textsc{i}. Gum~64 and NGC~6357 are different from Sco~OB1 when considering the targets near the H~\textsc{ii} regions -- the OB association shows a {6196~\AA} DIB peaking slightly more blue-wards from the rest wavelength. To explain the DIB broadening, we may need more than just the presence of massive stars producing strongly ionised ISM.

\subsection{Sco-Cen}\label{section:4.3}

\begin{figure}
 \includegraphics[width=\columnwidth]{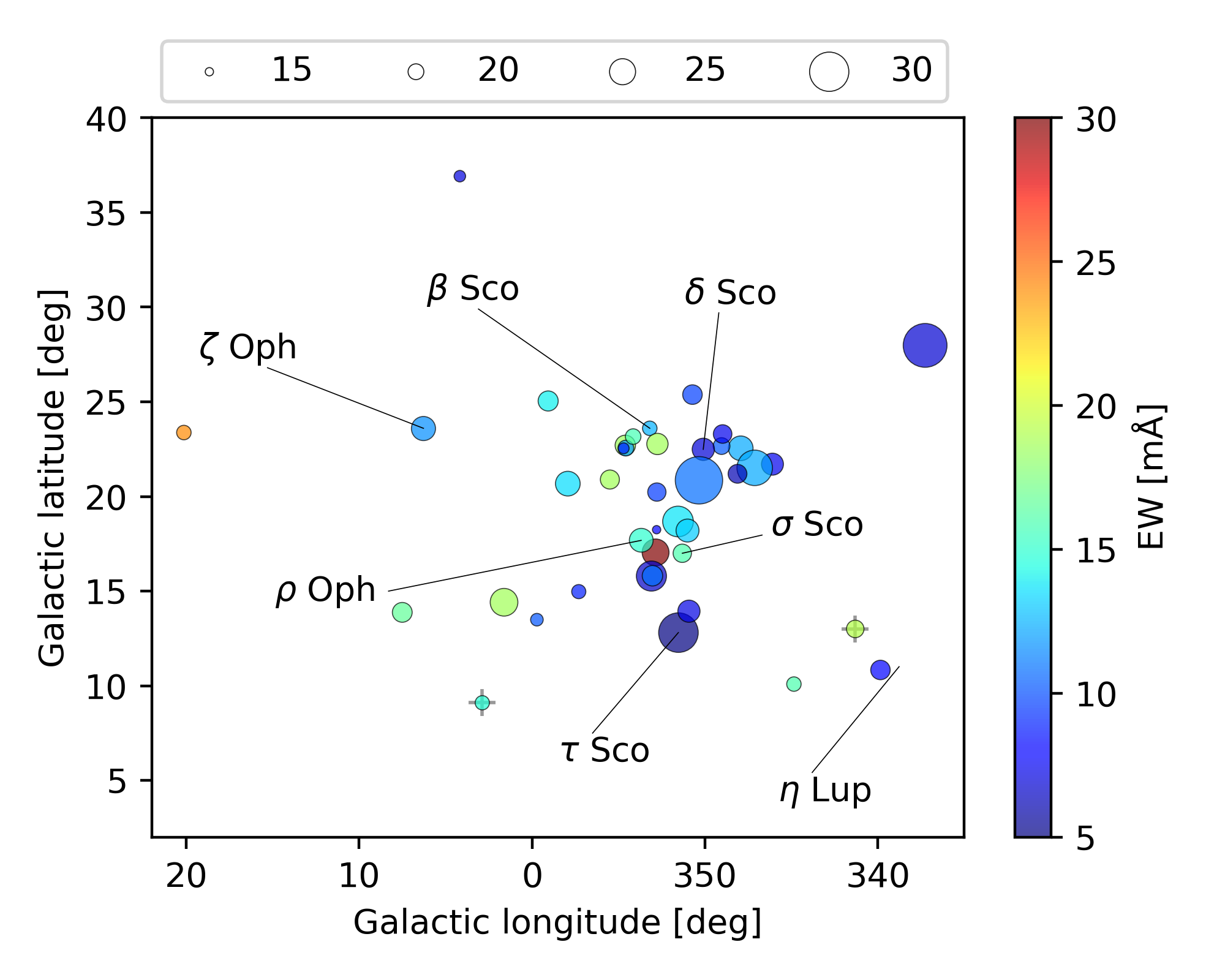}
 \caption{Same as Fig.~\ref{fig:2}, but focusing on USco. The plus symbols indicate stars at distances $>300$~pc.}
 \label{fig:11}
\end{figure}

In our analysis of Sco-Cen, we first focus on the ISM in front of USco. Once again, we note a good agreement with the results obtained by \citet{2023MNRAS.523.4158G}. However, we find the EW and the FWHM to vary between different parts of USco. Looking at Fig.~\ref{fig:11}, we are unable to directly point out any correlations. Overall, we can see that stars surrounding $\beta$~Sco show a somewhat narrower {6196~\AA} DIB when compared with the targets around $\delta$~Sco and $\rho$~Oph. The DIB appears to be stronger towards the $\beta$~Sco and the $\rho$~Oph targets -- this is most likely related to the absence of a prominent dust cloud in the close vicinity of $\delta$~Sco. The high FWHM obtained for HD~149438 ($\tau$~Sco) is likely a problematic measurement resulting from the absence of a strong feature ($\textrm{EW}<4$~{m\AA}, see Appendix~\ref{section:C}). One can explain the broadening and the strength of the DIB in the spectra of $\rho$~Oph and HD~147889 as a nature of stars which are embedded in the Ophiuchus cloud (possibly similar to the case of Herschel~36 in the Lagoon~Nebula). On the other hand, similar arguments cannot be made for the stars with very broad profiles of the DIB, for example HD~148605, HD~144334, or HD~142301.

HD~148605 is located at the distance of about 139~pc, based on its assignment to the $\rho$~Oph population of stars \citep{2023A&A...677A..59R}. The DIB in this line of sight appears fairly weak (only 7~{m\AA}). The presented results are based on the stacking of 17~noisy spectra, although the resulting spectrum together with the two highest quality spectra ($\textrm{SNR}\sim700$) show a clear presence of the DIB. HD~144334 displays the broadest profile identified in USco. It is located about 140~pc away, relatively close to the position of $\delta$~Sco in the sky. Based upon a closer inspection of the archival spectrum (see Appendix~\ref{section:C}), we find it unlikely that the broad profile is the result of our data handling. HD~142301 is a very interesting case. This evolved B-type star shows some changes in the presence of features around {6196~\AA}. Out of the six FEROS spectra we used, the feature end-points were incorrectly identified only in one case prior to stacking. We conclude that the increased DIB width ($\textrm{FWHM}>25$~km\,s$^{-1}$) associated with USco requires further investigation based on additional (very-high quality) spectra.

\begin{figure}
 \includegraphics[width=\columnwidth]{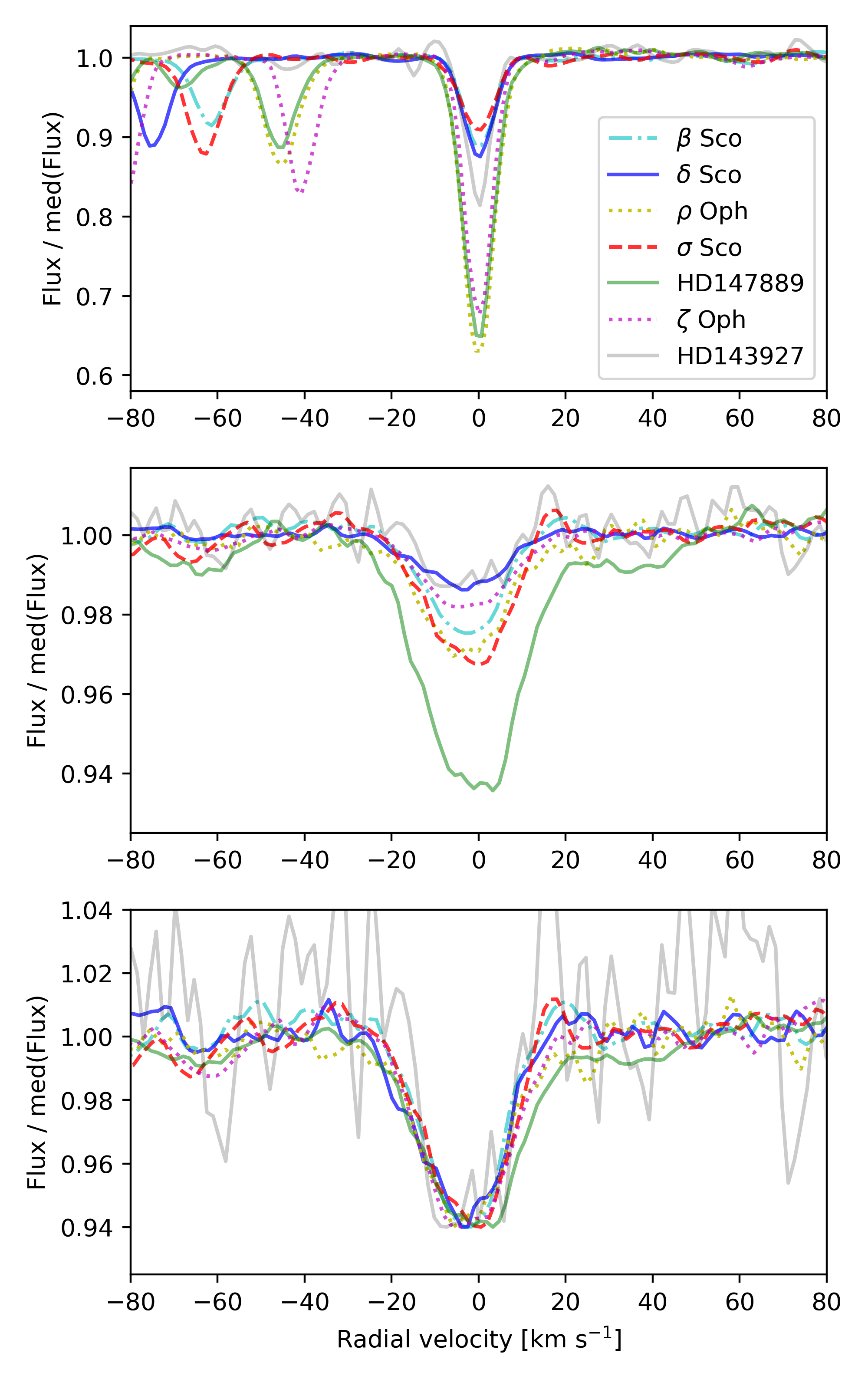}
 \caption{Same as Fig.~\ref{fig:5} but for USco and its surroundings. Wavelengths were shifted to the rest position of the K~\textsc{i} line (top panel). Includes only spectra obtained with FEROS.}
 \label{fig:12}
\end{figure}

Spectra of the most significant stars of USco and its surroundings are shown in Fig.~\ref{fig:12}, including the prototypical stars $\sigma$~Sco and $\zeta$~Oph, which display different properties of the foreground ISM \citep[for example][]{1991PASP..103.1005S} and are often used in the ISM description. The lines of sight towards $\sigma$~Sco and $\rho$~Oph show DIB profiles very similar to each other. The strongest {6196~\AA} DIB in USco is seen towards HD~147889, which shows a relative increase in the absorption within the red wing when compared to the other lines of sight.

\begin{figure}
 \includegraphics[width=\columnwidth]{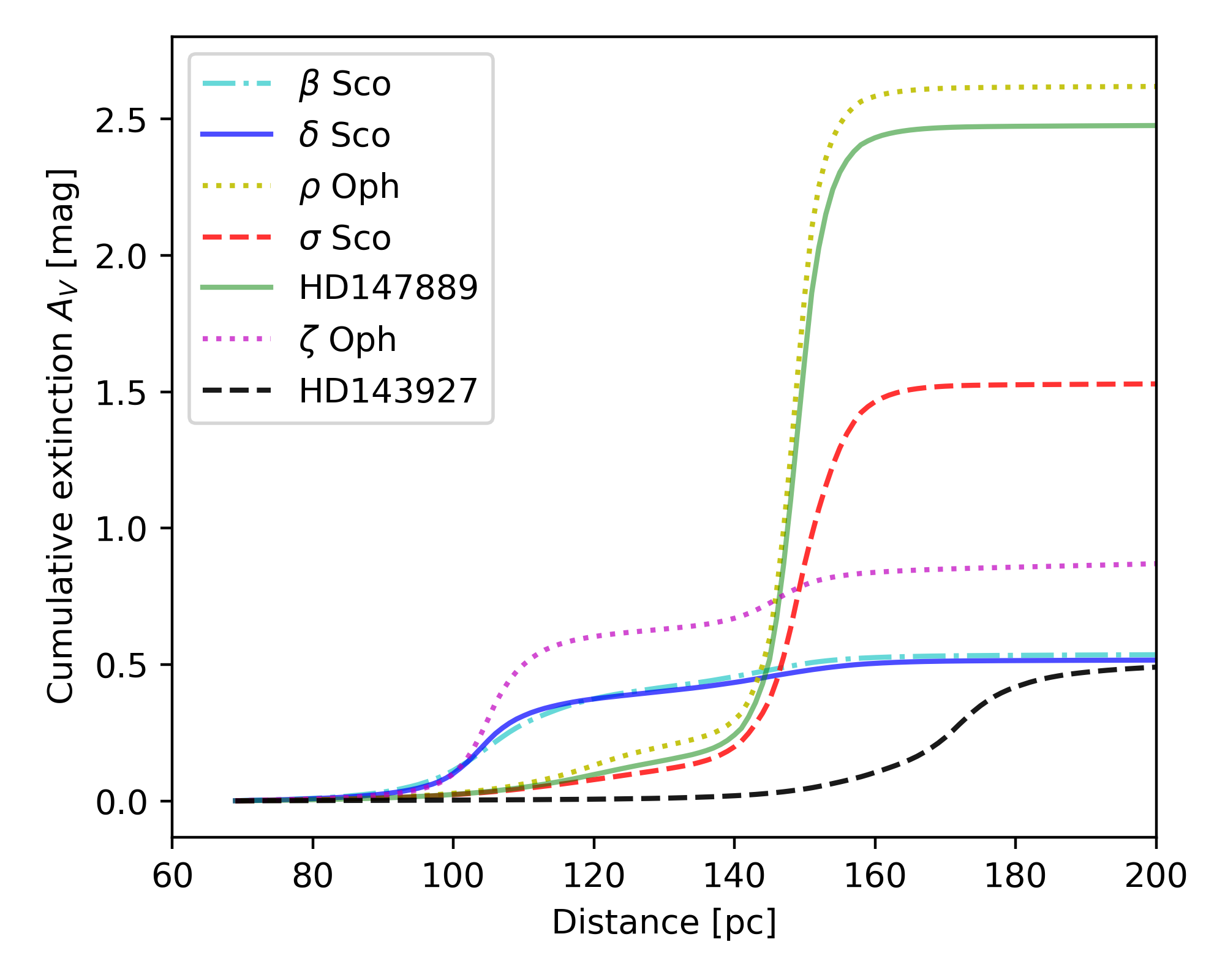}
 \caption{Line-of-sight extinction towards different targets in USco and its surroundings. All of the included stars are located at a distance $<200$~pc.}
 \label{fig:13}
\end{figure}

The $\eta$~Lup group of stars (part of Upper Centaurus-Lupus, UCL), located just to the south-west from the Ophiuchus cloud, shows a statistically narrower DIB ($\sim20$~km\,s$^{-1}$), based on lines of sight with target stars located at 170~pc, 300~pc, and 800~pc. A possible correlation between the EW and the distance can be identified, but we notice a lack of a correlation with the FWHM. On the other hand, the group of stars south-east of the Ophiuchus~cloud shows a little bit of variation in profile parameters. Out of the four south-eastern stars located within 300~pc from the Sun, two (HD~150814, HD~152655) show a very narrow profile when compared with the rest of USco, while the other two (HD~152909, HD~155503) show a broadening similar to that of the $\rho$~Oph group. We note that the angular separation of the latter couple of stars from the Ophiuchus cloud is larger than in the case of the former couple. All of the four stars in the south-east are probing a diffuse wall of dust that can be identified in the dust maps at a distance of about 150~pc. This wall corresponds to the background section of the dust shell surrounding the runaway star $\zeta$~Oph. This O-type star is located much closer to the Sun, $\sim130$~pc, and the only source of absorption at these distance is the foreground wall of the surrounding dust shell. It should be noted that this wall reaches all the way to $\beta$~Sco and $\delta$~Sco (Fig.~\ref{fig:13}).

Focusing on the other parts of Sco-Cen might provide us with important insight into the nearby ISM and its relation to the profile of the DIB. In order to learn something about the Corona~Australis cloud (CrA), the stacking of 12~noisy spectra was required to achieve a good signal around the DIB. While additional measurement are required for a proper verification, this single measurement tells us that within CrA the DIB is narrower than in most lines of sight towards USco. To further analyse the region surrounding CrA, we grouped this measurement with those of two other stars -- a nearby star HD~189103 ($\theta^1$~Sgr, 200~pc, 17~km\,s$^{-1}$) and a more distant star HD~165024 ($\theta$~Ara, 340~pc, 22~km\,s$^{-1}$, no splitting).

The stars included in the UCL-Norma group are not actually connected to the stellar population with the same label mentioned in \citet{2023A&A...677A..59R}, although they do probe the related ISM. HD~150745 has the broadest profile in the region, but its measurement is very uncertain. We find that the estimated median (Table~\ref{table:2}) describes the broadening within UCL-Norma reasonably well, suggesting a FWHM similar to that of UCL~$\eta$~Lup. We find two nearby targets in the area between these two UCL groups. HD~147152 (included under NGC~6188) and HD~145110 (NGC~6067) show a slightly different broadening of 19.5~km\,s$^{-1}$ and 22.1~km\,s$^{-1}$, respectively. However, the measurements for both stars have relatively high uncertainties, and the mean of these values lies within the range of values obtained for the UCL groups. Based on this, we extrapolate that the FWHM of the DIB does not vary much between UCL-Norma and UCL~$\eta$~Lup. Any variation seen towards a more distant target (NGC~6067, NGC~6188) is likely related to the presence of additional clouds along the line of sight.

\begin{figure}
 \includegraphics[width=\columnwidth]{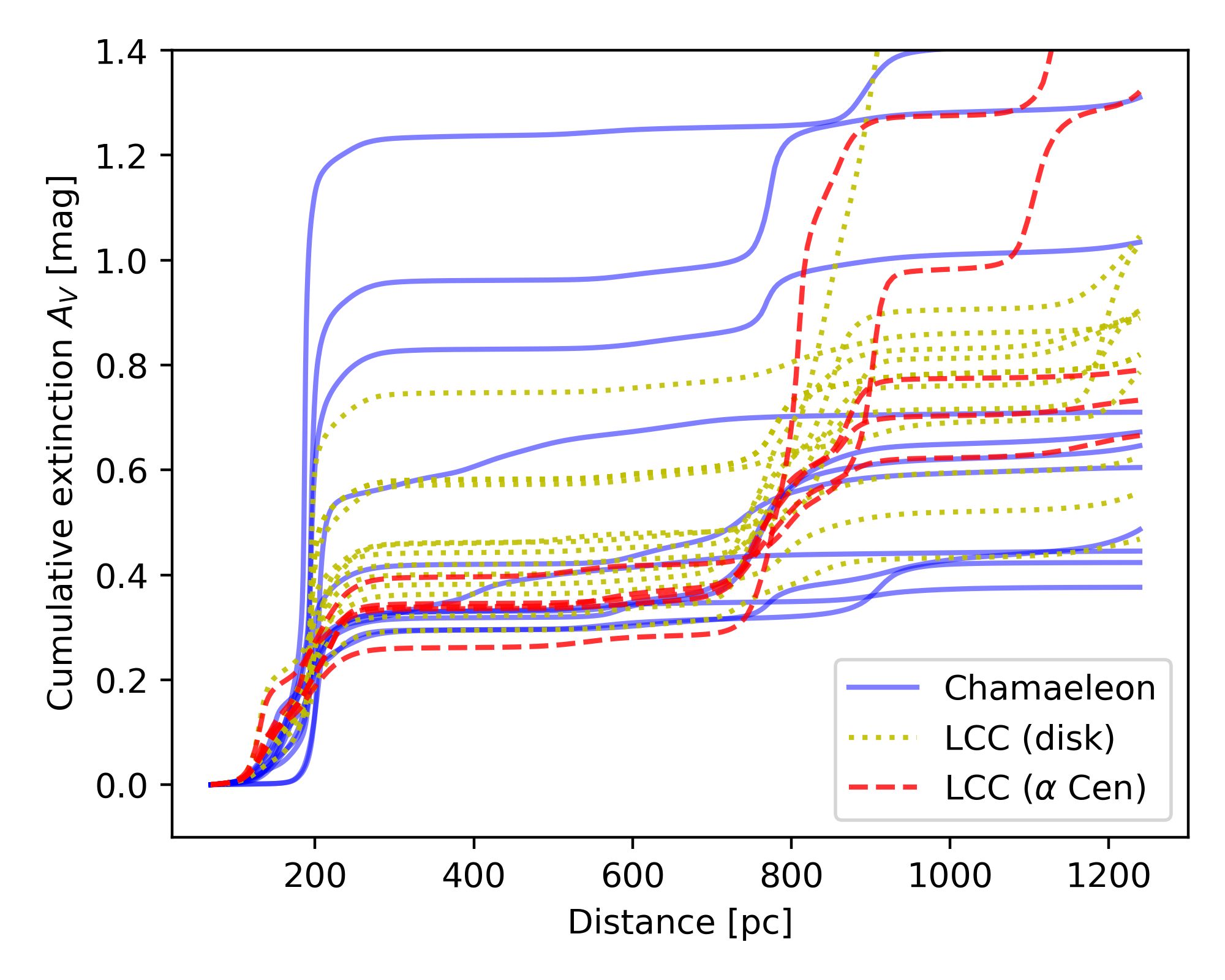}
 \caption{Line-of-sight extinction towards different targets in LCC and Chamaeleon.}
 \label{fig:14}
\end{figure}

\begin{figure}
 \includegraphics[width=\columnwidth]{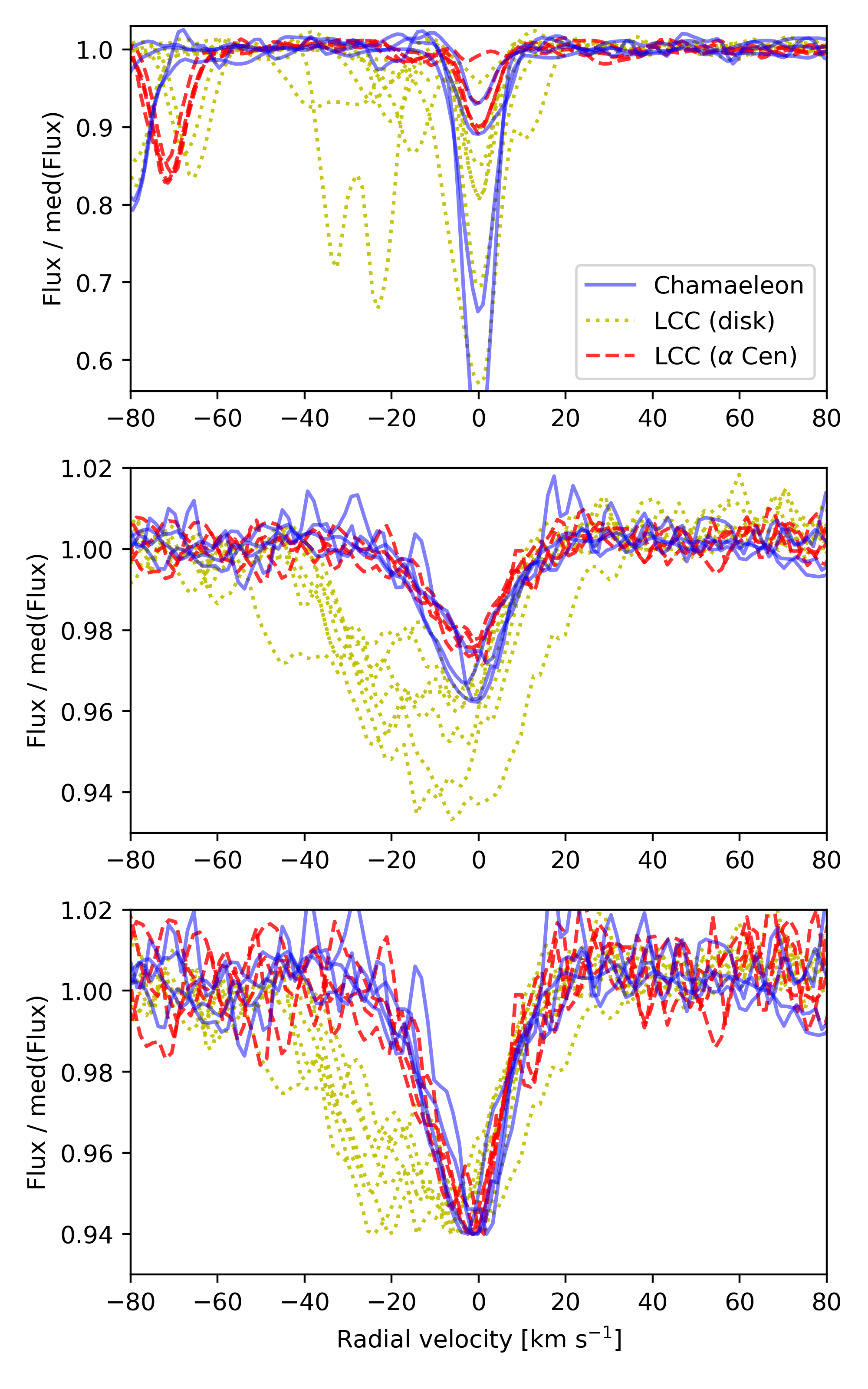}
 \caption{Same as Fig.~\ref{fig:5} but for LCC and Chamaeleon. Wavelengths were shifted to the rest position of the K~\textsc{i} line (top panel). Includes only spectra obtained with FEROS.}
 \label{fig:15}
\end{figure}

Finally, we focus our attention on Lower Centaurus-Crux (LCC) populations of Sco-Cen and the Chamaeleon clouds. We made an arbitrary cut between the Sco-Cen and the Carina regions at the Galactic longitude $l=300^{\circ}$. The rough distinction between LCC-disk and LCC-$\alpha$~Cen lies in the FWHM and in distances -- the former group includes stars (beyond 1.5~kpc) displaying splitting in the profile of the DIB, while the DIB towards the latter (typically within 1~kpc) has a simple and narrow profile. We note an error in the automatic procedure failed to distinguish the splitting towards HD~119646, producing an extremely narrow profile. The line-of-sight extinction for the three analysed groups of targets and their spectra are displayed in Fig.~\ref{fig:14} and Fig.~\ref{fig:15}, respectively.

The amount of splitting in LCC depends heavily on the line of sight, although it correlates with the splitting seen in the profile of K~\textsc{i}. The kinematic components related to ISM further away from the local ISM (beyond 700~pc) must be responsible for the enhanced DIB absorption in the blue wing of the profile. Based on the spectra, we conclude that the ISM probed by LCC-$\alpha$~Cen and the Chamaeleon targets shows little variations. The FWHM and the structure of the profile both suggest the presence of a single strong kinematic component. The correlation between the DIB profile parameters given in Table~\ref{table:2} is biased by the presence of a single target with a high FWHM (HD~111904, closer to the Galactic disk).

The variations in the profile of the DIB seen throughout Sco-Cen can be explained using the arguments from \citet{2016A&A...585A..12B} -- the local conditions within the ISM vary and can be observed as DIB profile changes. Additionally, the data from Table~\ref{table:2} suggest the presence of a FWHM gradient in the sky, with the highest value found in USco, and the lowest in Chamaeleon. We provide additional discussion about this finding in Section~\ref{section:5}.

\subsection{Carina}\label{section:4.4}

We analysed the stars towards IC~2944 and the peripheries of this nebula separately. Out of all targets in the peripheries, only one is located within 500~pc from the Sun. This means that we are unable to extract information about the local clouds towards these lines of sight. However, we can note that there appear to be little-to-no variations in FWHM and EW within IC~2944. The small dense ISM regions surrounding the cluster within the nebula show no signs of an ongoing high-mass star formation process \citep{1997A&A...327.1185R}.

The available spectra prevent us from properly analysing the nearby ISM when looking towards NGC~3572 and NGC~3576 (stars typically beyond 2~kpc), similar to the case of IC~2944. As expected from stellar distances, the profiles of K~\textsc{i} and CH$^*$ lines show Doppler splitting, which can explain the broadened nature of the DIB. The presence of only two distinct kinematic components at these distances suggests that the relatively low EW of the DIB might correlate with the smaller number of clouds along the lines of sight.

Based on the observed properties of the DIB, we distinguish between six sub-regions within the Carina~Nebula. Within the close proximity of $\eta$~Car, there are two groups of stars. The group~A shows a consistently narrow profile of the DIB. It includes $\eta$~Car, the Keyhole~Nebula, and a few stars to its north-east -- \citet{2024AcA....74...69K} investigated the stars that probe the same region of the nebula. In contrast, the profile of the DIB seems to vary a lot towards group~B and the typical broadening is much higher when compared with group~A. The included stars are mostly located within the southern section of the cavity (seen in infrared and radio) surrounding the star $\eta$~Car. The broadening can be easily explained by the observed splitting in the atomic K~\textsc{i} line -- the amount of splitting depends on the line of sight. While HD~93250 shows two distinct kinematic components of different widths, no significant amount of splitting can be identified towards $\eta$~Car.

\begin{figure}
 \includegraphics[width=\columnwidth]{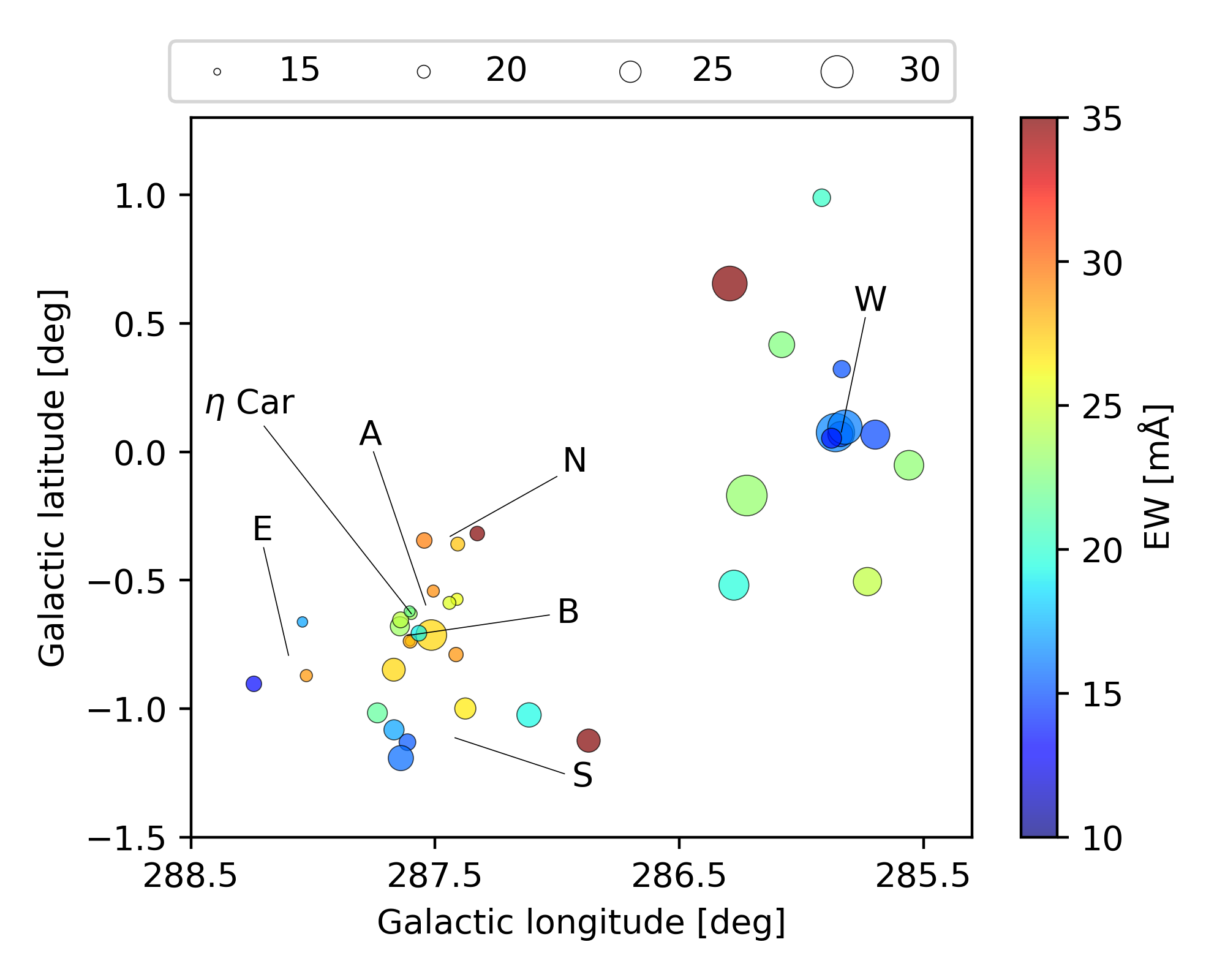}
 \caption{Same as Fig.~\ref{fig:2}, but focusing on the Carina~Nebula. None of the stars is located within 800~pc. Labels N, E, S, and W correspond to north, east, south, and west, respectively. Labels A and B include the core of the nebula and probe slightly different conditions.}
 \label{fig:16}
\end{figure}

We find systematic variations in the EW and the FWHM of the DIB across the Carina~Nebula (Fig.~\ref{fig:16}). Groups A, B, and N (north) show the strongest DIB in the region. The DIB seems to be generally weaker when compared with other regions at similar distances (see the Eagle~Nebula and the Omega~Nebula in Table~\ref{table:2}) -- together with IC~2944, NGC~3572, and NGC~3576, this supports the idea of a lack of a higher number of DIB clouds along the lines of sight towards the Carina region. The southern (Collinder~228) and the western (Gum~31, NGC~3293) groups of the Carina~Nebula show a DIB profile that is both broad and at low peak intensity. Group~E (east, Bochum~11) does not follow this trend. As we can see in Table~\ref{table:2}, there is a strong correlation between the distance of the targets and the measured EWs in the peripheries of the nebula. The lack of a correlation between the EW and the FWHM complicates the explanation of the weaker and broader lines of sight.

\begin{figure}
 \includegraphics[width=\columnwidth]{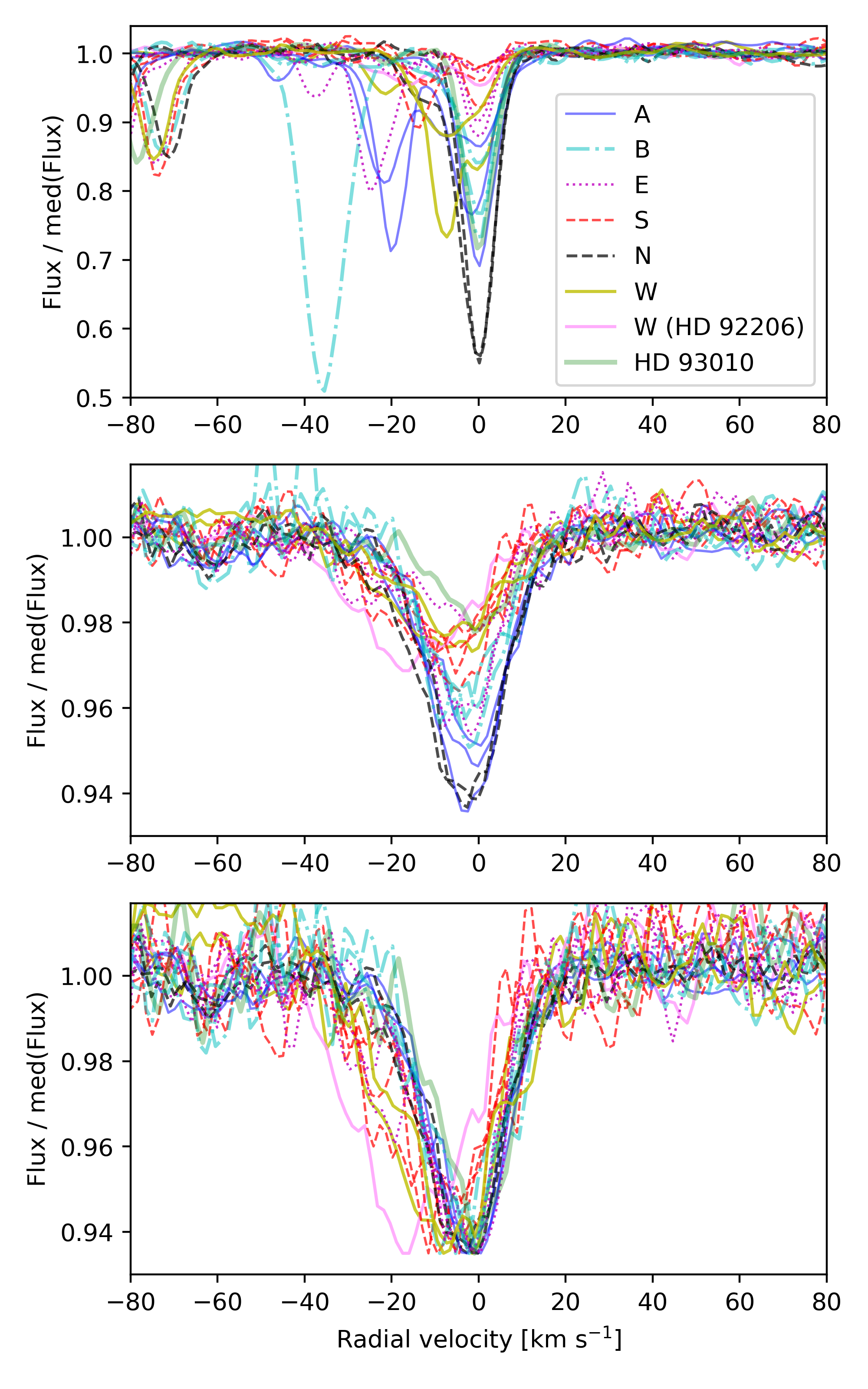}
 \caption{Same as Fig.~\ref{fig:5} but for the Carina~Nebula. Wavelengths were shifted to the rest position of the K~\textsc{i} line (top panel). Includes only spectra obtained with FEROS.}
 \label{fig:17}
\end{figure}

Spectra of the targets from the Carina~Nebula (Fig.~\ref{fig:17}) show that splitting occurs mostly in the blue wing. To be sure about the identification of the local cloud velocity, we made use of HD~93010, a star located at the distance of about 400~pc. In some lines of sight we note the presence of a blended kinematic component in the profiles of K~\textsc{i} and CH$^*$, making a precise identification of the local ISM component difficult. In the case of HD~92206, we see that the DIB is significantly blueshifted compared to the other lines of sight, which might be explained by a relative underdensity in the local ISM when compared with other line of sight in Carina-west. Except for this star, it appears that a large fraction of the DIB absorption in Carina-west occurs in the nearby ISM. This kinematic component also covers a large fraction of the DIB profile in Carina-south. The increased broadening and strength of the DIB towards HD~93683 and HD~91824 tells us that most of the DIB carriers might be located within 2~kpc from the Sun, except for Carina-north, group~A, and group~B. Confirming this result requires an enhanced sample of stars located in this region at distances between 1--2~kpc.

Not much information can be extracted about the line of sight from the dust maps (Fig.~\ref{fig:C1}). In general, all lines of sight show positions of clouds at 140--190~pc and at 720~pc. The total integrated extinction along the line of sight is very low towards the Carina~Nebula \citep{2008hsf2.book..138S} with variations that depend on the line of sight \citep{1995RMxAC...2...51W}. At the distance of 1.25~kpc, we find a median value of $A_V\approx0.21$~mag, much lower than in any other region investigated in this work. No other structures can be identified in the dust maps.

Most of the results extracted for the Carina region can be explained in terms of multiple kinematic components producing the observed broad profiles of the DIB -- this is especially apparent in the EW-FWHM correlations towards IC~2944, its periphery, and the periphery of the Carina~Nebula. The lack of a significant correlation towards the inner parts of the Carina~Nebula is most likely the result of a limited distance sampling (only two stars are located within 2~kpc). Finally, we highlight the contrast between the groups A and B in the Carina~Nebula -- the behaviour of the DIB profile changes on small spatial scales, which seems to indicate changes in the physics conditions within the ISM.

\subsection{Vela}\label{section:4.5}

We identify a very strong correlation between the EW and the distance towards Vela targets (Table~\ref{table:2}, hints can also be seen in Fig.~\ref{fig:C2a}). On the other hand, we find no correlation between the FWHM and the EW. This might point to the broadening not being strongly connected to Doppler splitting within the profile of the DIB. The dust maps show that a relatively large number of clouds is present within 600~pc towards Vela, depending on the line of sight (Fig.~\ref{fig:C2b}). The amount of extinction per cloud does not seem to exceed $A_V\approx0.2$~mag (except for one target, HD~74804).

Since we are unable to learn more about the region by following the group definitions from Table~\ref{table:2}, we decided to divide the region into the Vela-disk ($|b|\leq5^{\circ}$) and the Vela-periphery ($|b|>5^{\circ}$) sub-regions. Vela-periphery includes only stars located between 200~pc and 500~pc away from the Sun. Already within this narrow range of distances we notice a weak correlation between the EW and the distance from the star. There is a relative weakness of the DIB compared to the other targets in the region. Vela~OB2 and NGC~2516 cover most of the Vela-periphery and the associated FWHM of the DIB is relatively high. We note that the DIB is weaker towards Vela~OB2 when compared with NGC~2516. A narrower profile is found towards HD~84201, the only star to the north of Vela-disk. The dust cloud associated with this star is visible also in optical and its distance is approximately 200~pc according to the dust maps. We highlight the mysteriously low EW determined from the profile of the DIB towards this target.

Compared to the peripheries, the Vela-disk sub-region shows much higher EWs, which can be expected based on the larger typical distance to the stars. While HD~80573 ($\sim500$~pc) shows a very weak DIB similar to Vela~OB2, the EW is significantly higher towards HD~73882 ($\sim500$~pc), a star probing Gum~14. The latter target is missing from the list of Gaia astrometric data and the Hipparcos parallax is very uncertain \citep[$\varpi=2.17 \pm 0.91$~mas,][]{2007A&A...474..653V}. Based on the available information, we infer that the distance of this star is likely to be much higher than 500~pc. Besides the distant HD~69464, the stars in Vela-disk show a relatively narrow profile of the DIB with a median value of about 20~km\,s$^{-1}$.

One of the observed stars, HD~77581 (GP~Vel), is a member of the high-mass X-ray binary system Vela~X-1 \citep{1972ApJ...173L.105B}. In contrast with HD~71304, located at a similar distance from the Sun, HD~77581 displays a more narrow DIB profile. We note that a major difference between these two lines of sight is their position with respect to the Gum~Nebula -- while HD~77581 can be found to the north of the nebula, HD~71304 is located beyond it. This suggests that the broadened profile towards HD~71304 might be significantly influenced by the DIB carriers within the Gum~Nebula, although this result remains uncertain. Given the location of Vela in the Galactic disk (opposite of the direction of Galactic rotation), a precise identification of the local ISM motion is difficult, resulting in a lack of kinematic support for the stated hypothesis. Another major influence on the DIB profile observed towards HD~77581 might be coming from the ISM surrounding the star, where an IR bow shock was previously detected \citep{1997ApJ...475L..37K,2022MNRAS.510..515V} -- high-quality spectra of other stars surrounding HD~77581 located beyond \citep[CD$-$39$^{\circ}$5049, CD$-$44$^{\circ}$5201, HD~79186, based on][]{2021MNRAS.504.2968P} and within 1~kpc from the Sun (HD~74803, HD~74824, HD~75272, HD~75500, HD~77302, HD~77957) are need to test this alternate hypothesis. At present, only spectra of HD~79186 are available in the archives, with our measurement of the profile FWHM lying between those of HD~77581 and HD~71304, and the EW being significantly lower.

The young stellar populations in Vela show a very complex star-formation history that is connected to the supernova-driven motion of the ISM in the region \citep{2019A&A...621A.115C}. The effect of these conditions on the presence of DIB carriers in ISM is still poorly understood. Further studies of the DIB profiles (Fig.~\ref{fig:C2c}) in this region are required since our data set does not allow us to provide a more detailed analysis. Any such investigation is going to be hindered by the absence of a strong absorption within the profiles of DIBs.

\subsection{CMa-Mon}\label{section:4.6}

\begin{figure}
 \includegraphics[width=\columnwidth]{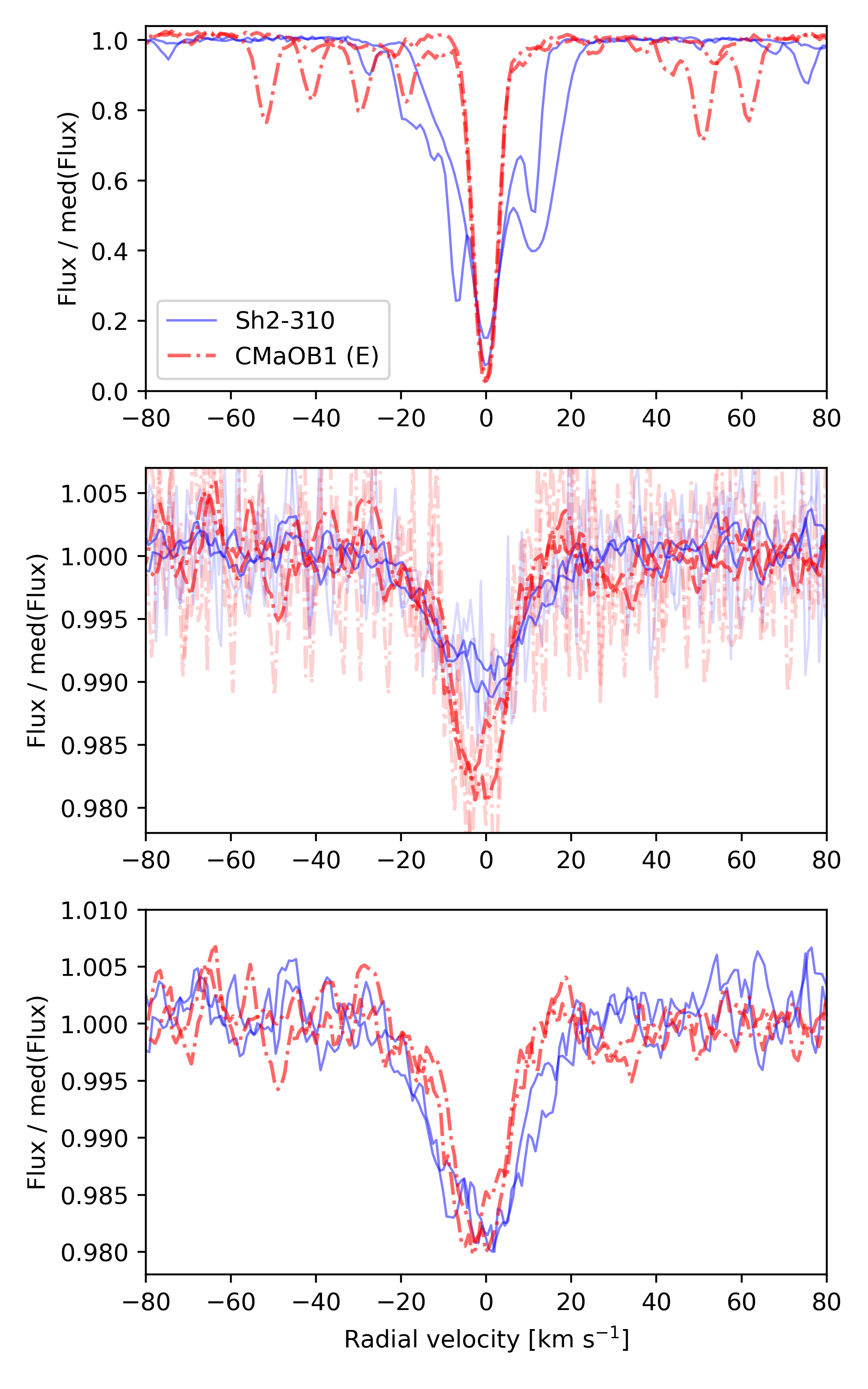}
 \caption{Same as Fig.~\ref{fig:5} but for Sh2-310 and the eastern periphery of CMa~OB1. Wavelengths were shifted to the rest position of the Na~\textsc{i} line (top panel) due to the lack of strong K~\textsc{i} and CH$^*$ lines. The more transparent lines in the middle panel show the raw data, while the less transparent lines represent smoothed data based on the median flux by including the six nearest data points.}
 \label{fig:18}
\end{figure}

The two stars from the Sh2-310 group (at 1.1~kpc and 1.7~kpc) display weak DIBs and K~\textsc{i} lines, hinting at a lack of dense clouds along these lines of sight (up to a distance of at least 1.1~kpc). Four stars from the peripheries of CMa~OB1 are located closer than 500~pc from the Sun and show a DIB profile that is statistically narrower when compared with Sh2-310, but has a similar EW (Fig.~\ref{fig:C3a}). We find at least two important differences between these lines of sight. Firstly, we note the lack of H~\textsc{ii} regions in the peripheries (except for the more distant HD~51452). Secondly, the dust maps (Fig.~\ref{fig:C3b}) suggest that the nearby cloud within 300~pc is more diffuse towards Sh2-310, while the cloud at about 1~kpc has a similar density in all of the lines of sight, highlighting its relatively low influence on the profile of the DIB. The spectra plotted in Fig.~\ref{fig:18} suggest that the broadening of the DIB towards Sh2-310 is symmetrical around the rest velocity. We also notice hints of splitting in K~\textsc{i} lines that are not visible in the CMa~OB1 peripheries. It is worth mentioning that the stars around NGC~2384, located about 2.5~kpc away \citep{2021MNRAS.504..356D}, also show an increase in DIB broadening. However, the DIB is stronger by a factor of at least two when compared with the previously mentioned targets, and its profile can be likely explained by Doppler splitting.

Regardless of the distance, the stars in the close proximity of CMa~OB1 show an increase in the EW when compared with the stars located further away in the sky, such as HD~55879 and HD~57682. The measured FWHM of the DIB is fairly large ($>20$~km\,s$^{-1}$). According to \citet{2019A&A...628A..44F} and references therein, the dust clouds in this region were previously shaped (to a smaller degree) by supernovae. The presence of a cloud in front of the members of the OB association might explain deviations from the EWs seen towards Sh2-310 and the CMa~OB1 peripheries. On the other hand, the $\textrm{EW}\gtrsim15$~{m\AA} for the stars within 500~pc still needs to be explained. Looking at the dust maps, we note the presence of a cloud $\sim180$~pc away from the Sun. This cloud covers not only the eastern peripheries of CMa~OB1 but partially also some of the targets assigned to the CMa~OB1 group. Another cloud is located at 900~pc and it obscures only the OB association and regions to its south. The dust cloud related to the association is found at the distance edge of the dust map. Overall, we find that the amount of extinction in these lines of sight can easily explain the high EWs.

The final group of stars in this region that we aim to explore traces the ISM in front of the Rosette~Nebula (distance of $\approx1.5$~kpc). Six of the selected stars probe the interior of the nebula and show one of the most consistent FWHM measurements presented in this work (Table~\ref{table:2}, note the small value of standard deviation). The behaviour of the DIB in the peripheries of the nebula depends on the line of sight. Moving away from the centre in the sky, we see a small decrease in the FWHM towards HD~46485 and HD~259440, but the value increases as we move towards targets significantly further away from the nebula.

\begin{figure}
 \includegraphics[width=\columnwidth]{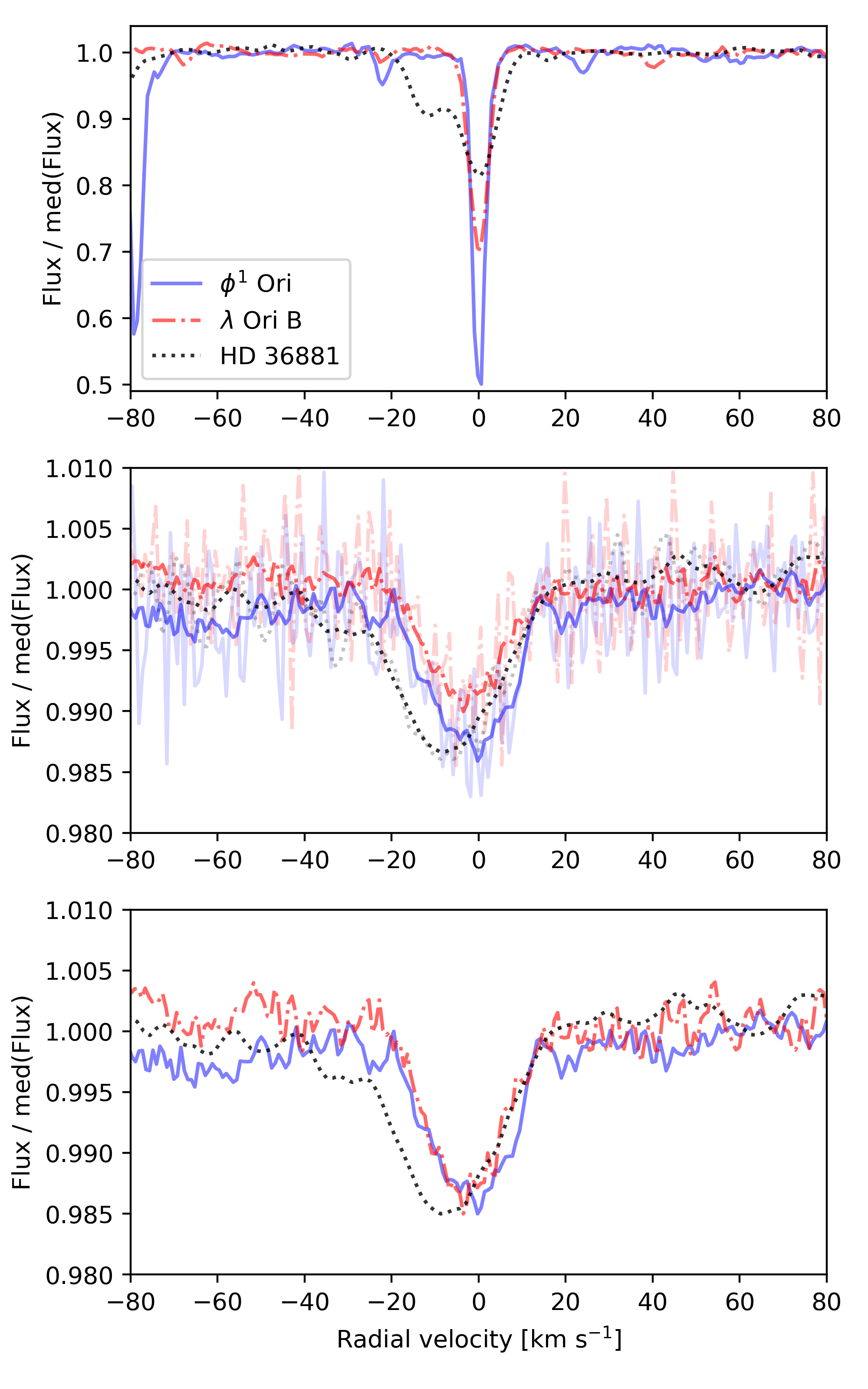}
 \caption{Same as Fig.~\ref{fig:18} but for Orion ($\lambda$~Ori). Wavelengths were shifted to the rest position of the K~\textsc{i} line (top panel).}
 \label{fig:19}
\end{figure}

\subsection{Orion}\label{section:4.7}

The targets in Orion were separated into groups based on their proximity to Orion~A (Nebula), Orion~B (Belt), and $\lambda$~Ori (Fig.~\ref{fig:C4a}). For the peripheries, we considered the angular separation from the prominent H~\textsc{ii} regions and the distance of the target. For example, while HD~35149 lies outside of a dusty region, it is located at $\sim500$~pc and one can identify the presence of ionised medium in its proximity. On the other hand, HBC~197 is associated with the Mon~R2 stellar population found at the distance of 800--900~pc, as well as with the dust structure found at about the same distance in the dust maps.

The Orion~Nebula (or Orion~A) shows strong variations in the EW and the FWHM, depending on the line of sight. The broadest and strongest DIB can be found in the spectra of HD~37061, a star located farthest away from the dense dusty core of the nebula (OMC~1) -- this is in agreement with the previous discovery of weaker DIBs in denser environments \citep[see][and references therein]{1994A&A...281..517J}. HD~37017 and HD~36629 are located on the edge of the dust cloud related to the nebula. We find no distinction between these lines of sight when looking at the foreground extinction in dust maps (Fig.~\ref{fig:C4b}). It should be noted that the profile measurements are quite uncertain due to the low strength of the DIB in the Orion region (within 500~pc from the Sun).

\begin{figure}
 \includegraphics[width=\columnwidth]{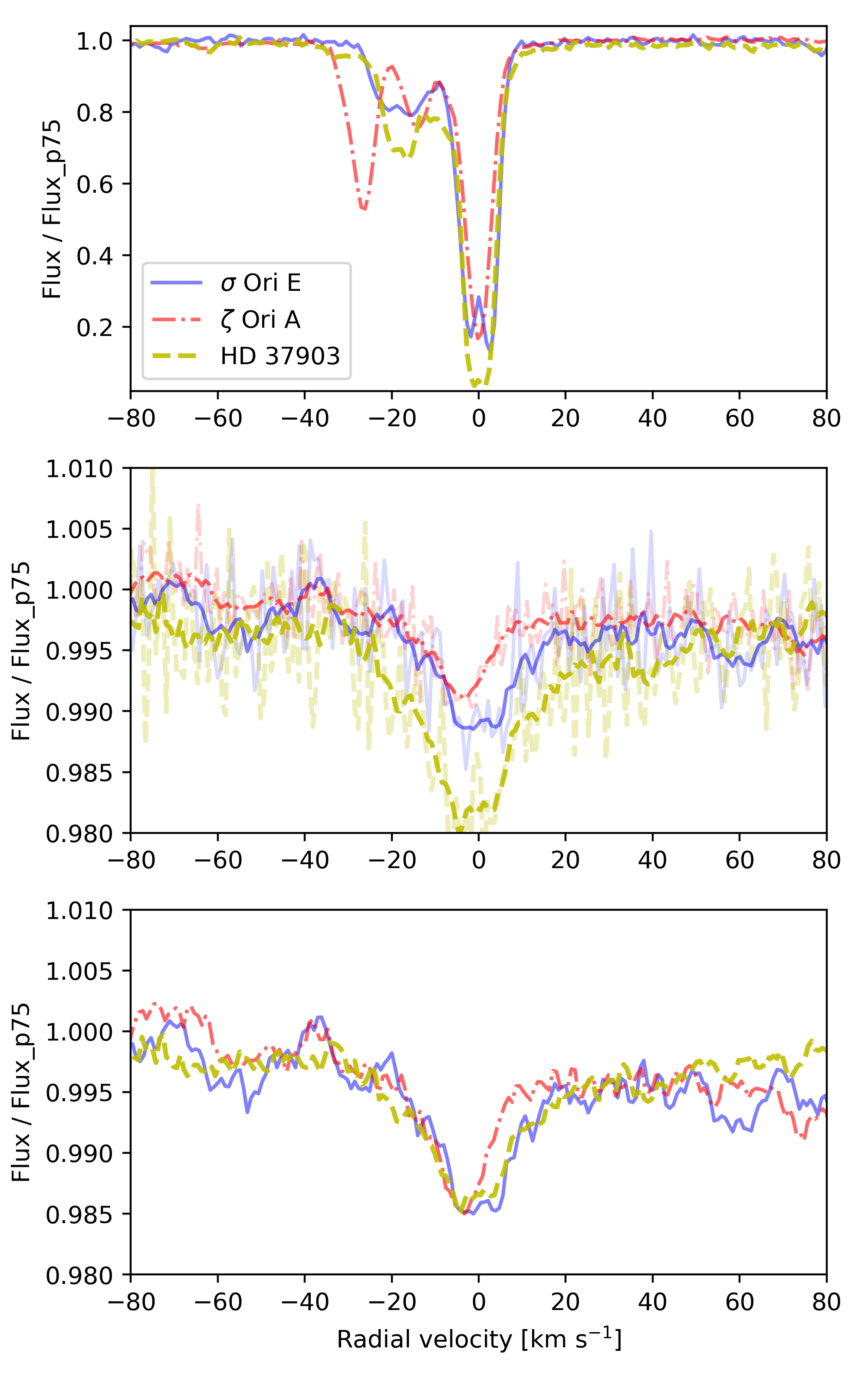}
 \caption{Same as Fig.~\ref{fig:18} but for Orion (Belt). Wavelengths were shifted to the rest position of the Na~\textsc{i} line (top panel) due to the lack of strong K~\textsc{i} and CH$^*$ lines. The 75th percentile of the flux was used for the scalar scaling of the spectrum -- median provided worse scaling for the region around the DIB.}
 \label{fig:20}
\end{figure}

The three stars surrounding $\zeta$~Ori (Orion~B section of the Orion~Belt) show consistent values of the FWHM and the EW. All of these stars can be found at distances $\gtrsim390$~pc. On the other hand, $\zeta$~Ori shows a much weaker and a much narrower profile of the DIB. It is possible that the inconsistency in the properties of the DIB in Orion~B could be explained by $\zeta$~Ori being located closer than 390~pc. This would be consistent with the Hipparcos parallax, the distance obtained using interstellar calcium lines, and the distance obtained by analysing the orbit of this binary system \citep{2013A&A...554A..52H}. The other two targets from the Orion~Belt group show a broader profile ($\textrm{FWHM}>25$~km\,s$^{-1}$). The DIB is broadest and strongest towards HD~37140, with the profile being similar to one of the nearby stars assigned to the Orion's interior peripheries, HD~290862. Curiously, both targets are located close to the reflection nebulae IC~426 and NGC~2068 (M78), with HD~290862 being a prominent star inside NGC~2068.

Similarly to the Orion~B group, the $\lambda$~Ori group of stars presents a small deviation in the DIB profile parameters. In this case, HD~36881 is located in the background at a distance of $\sim500$~pc. The increase in the EW can be explained by HD~36881 probing through the whole dust cloud associated with $\lambda$~Ori. We note that the $\lambda$~Ori group suggests a narrower DIB than the Orion~Nebula and the Orion~Belt targets -- 20~km\,s$^{-1}$ after excluding HD~36881, compared to 23~km\,s$^{-1}$. However, making a proper comparison is difficult due to the low number statistics.

Fig.~\ref{fig:19} and Fig.~\ref{fig:20} show the spectra from the $\lambda$~Ori and the Orion~Belt sub-regions, respectively. We notice that the star HD~36881 located at a larger distance shows an increase in absorption in the blue wing of the profile. In contrast, $\zeta$~Ori shows a lack of absorption in the red wing of the profile. This shows how switching from the closer to the more distant targets in the Orion region influences the observed profile of the {6196~\AA} DIB, probing clouds with different velocities at different distances. We note that these results should not be extrapolated to the other parts of the Orion region.

Focusing on the interior peripheries, we first decided to exclude HD~290862 and HD~35715 from our analysis. While the assignment of the former would better fit the Orion~Belt group, the measurements of the DIB for the latter result in highly uncertain profile parameters. We notice that the FWHM towards HD~34880, HD~34827, and $\eta$~Ori (HD~35411) is extremely large ($>30$~km\,s$^{-1}$), although all three stars are located within $\sim350$~pc from the Sun. The presence of a dust cloud (at 260~pc, according to the dust maps, not to be confused with LDN~1634) can be noted in front of HD~34880 and HD~34827, providing a possible explanation for the relatively large EWs. However, this cloud reaches almost towards $\eta$~Ori, which makes it difficult to explain the relative weakness of the DIB seen towards this star. One possibility is a lower distance compared to the other two stars, as suggested by the weaker optical ISM lines -- this would make $\eta$~Ori the closest star to the Sun with FWHM larger than 30~km\,s$^{-1}$. For the remaining targets, we notice a possible anti-correlation between the EW and the distance of the star.

Regardless of the distance, we find that the FWHM in outer peripheries is generally lower ($\sim18$~km\,s$^{-1}$) than in the previously discussed sections of Orion. The only exception to this rule is HD~30677, which shows a broadened profile of the DIB. This is difficult to explain in terms of Doppler splitting due to the lack of Doppler splitting in K~\textsc{i} and CH$^*$. In the case of the nearby star HD~30492, the increased EW can be explained by the presence of the foreground segment of the Orion-Eridanus superbubble \citep{1979ApJ...229..942R,2019A&A...631A..52J}. Finally, we would like to point out a decrease of the FWHM as a function of the angular separation from the Orion star-forming regions -- this effect is especially obvious in the south-west, where the nearby ($<250$~pc) stars HD~26912 ($\mu$~Tau) and HD~28114 show very narrow profiles (17--18~km\,s$^{-1}$).

\subsection{Per-Tau}\label{section:4.8}

Seven targets located towards the Perseus~cloud with available archival spectra are included in our investigation (Fig.~\ref{fig:C5a}). We find a very good agreement between our measurements of the FWHM and the values obtained by \citet{2023MNRAS.523.4158G}. Two lines of sight seem to show somewhat broader ($\sim 20$~km\,s$^{-1}$) DIB. HD~281159 is a young stellar object surrounded by a nebulosity in optical -- we presume that an additional kinematic component influences the profile, with the ISM in the close proximity of the star being responsible for almost half of the strength of the DIB (26~{m\AA} compared to 15~{m\AA} for the nearby stars). In the case of HD~23478, we find that the automatic procedure slightly overestimates the size of the region around the line in the spectrum of this star, which influences the FWHM measurement. However, the estimated FWHM error of 1.25~km\,s$^{-1}$ means that the typical FWHM in Perseus~cloud lies within the error range of the measurement. The dust maps suggest that the lines of sight towards the Perseus cloud probe ISM with significantly different foreground densities (Fig.~\ref{fig:C5b}).

\begin{figure}
 \includegraphics[width=\columnwidth]{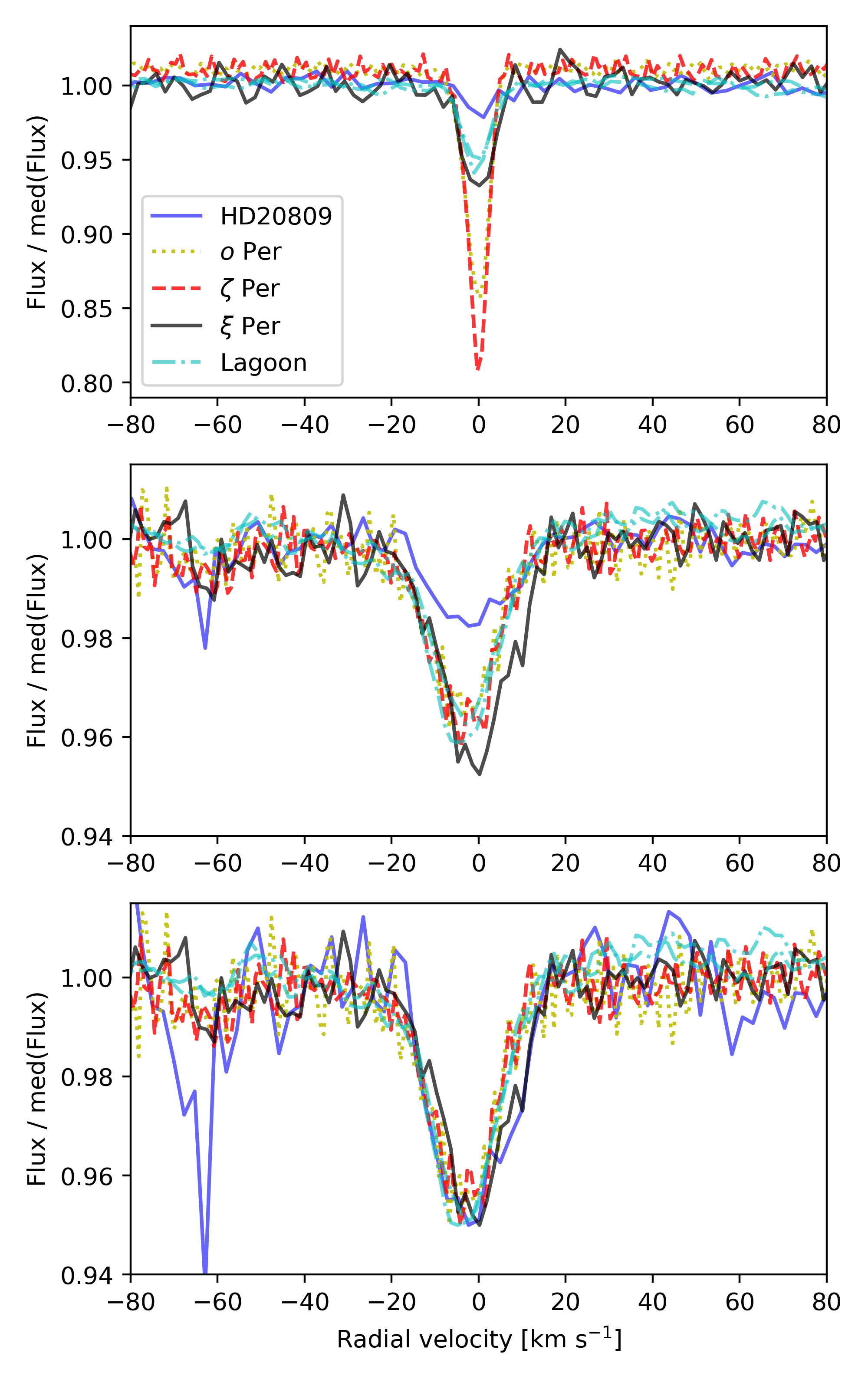}
 \caption{Same as Fig.~\ref{fig:5} but for the Per-Tau region and the Lagoon~Nebula. Wavelengths were shifted to the rest position of CH$^*$ (top panel). Includes spectra obtained with FEROS (Lagoon~Nebula), UVES (Perseus cloud), and ELODIE (HD~24912).}
 \label{fig:9}
\end{figure}

\begin{figure}
 \includegraphics[width=\columnwidth]{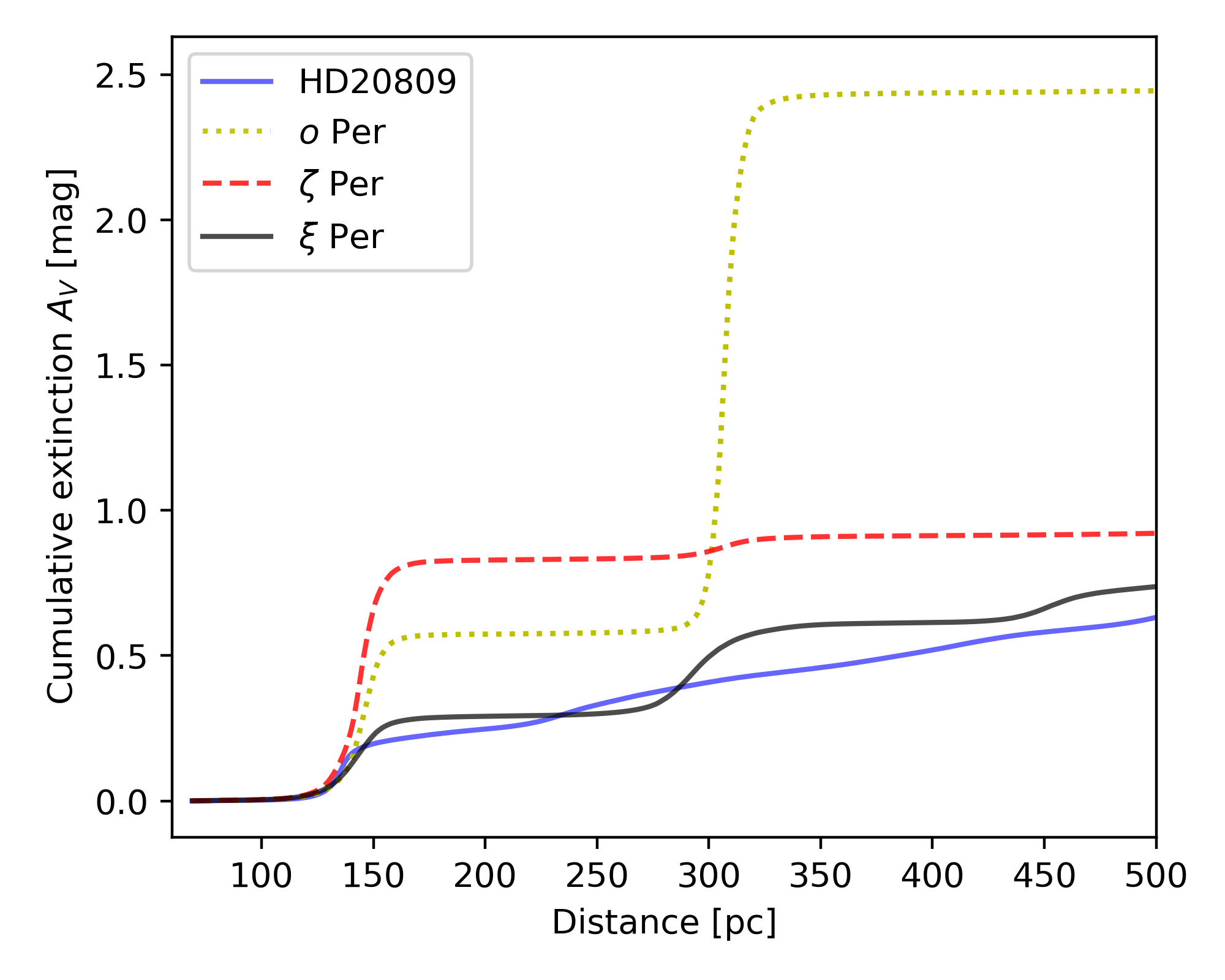}
 \caption{Line-of-sight extinction towards different targets in the Perseus region. We note that the spectra of all targets are being influenced by the presence of the Taurus~cloud ($\sim150$~pc).}
 \label{fig:10}
\end{figure}

Making use of ELODIE was necessary to obtain the spectra of the members of the $\alpha$~Per open cluster (Melotte~20). Two lines of sight were included and both of them show a similar increase in the FWHM of the DIB in comparison with the Perseus~cloud. While the stars probing the ISM near the extinction clouds are located at the distance of 250--400~pc, the 80--90~Myr old $\alpha$~Per cluster is only $\sim175$~pc away from the Sun \citep{2019A&A...628A..66L,2021MNRAS.504..356D}. We would like to point out the similarity between the typical FWHM and EW of this region when compared with the nearby stars from USco (see Tables~\ref{table:2} and \ref{table:3}).

In Fig.~\ref{fig:9}, we can compare the spectra of the Perseus cloud targets with those obtained for the $\alpha$~Per cluster members. We can clearly see the broadening in the $\alpha$~Per cluster (HD~20809) when compared with the other stars. Interestingly, $\xi$~Per shows an offset in the position of the DIB, although this line of sight differs from the others due to the star's connection to the nearby the California~Nebula \citep[for example, see][]{1978ApJ...219..105E}. We note that all lines of sight related to the Perseus~cloud are heavily influenced by the outer parts of the Taurus~cloud ($\sim150$~pc, see Fig.~\ref{fig:10}), both belonging to a structure known as the Per-Tau shell \citep{2021ApJ...919L...5B}.

\subsection{Gem-Aur, Heart, and Sadr}\label{section:4.9}

The two stars around NGC~2168 (not members, see Fig.~\ref{fig:C5a}) and the star associated with Sh2-247 show the same value of the FWHM. The measured EWs seem to correlate with distance, showing that at about 1~kpc the strength of the DIB is $\sim15$~{m\AA}, while at 2~kpc it increases to $\sim80$~{m\AA}. This strong increase for the most distant target is most likely related to the presence of a dust cloud surrounding Sh2-247, within which the target is located. There is a connection between the star-forming clouds Sh2-247 and Sh2-252, the latter being associated with the cluster NGC~2175 \citep{1989A&A...221..295K,2013ApJ...768...72S}.

IC~405 (Flaming Star Nebula) is directly associated with HD~34078 (AE~Aur). Located at the distance of 350--400~pc, this runaway star is moving through the surrounding ISM at a very large velocity, resulting in bow shocks \citep[for example,][]{2007ApJ...655..920F}. The available UVES spectrum does not show clear signs of splitting in the DIB or other interstellar lines, except for Ca~\textsc{ii}. The dust maps (Fig.~\ref{fig:C5b}) show a presence of up to three clouds -- a diffuse feature at a distance of 170~pc (associated either with Per-Tau shell or the Orion-Eridanus superbubble), a cloud at 220~pc, and another (denser) cloud at 370~pc that is associated with IC~405. We report that HD~34364 (at 140~pc, no measurement provided in this work), a nearby star located only a small angular distance from HD~34078 (at 390~pc), shows no signs of the {6196~\AA} DIB. We suggest a negligible influence of diffuse cloud on the DIB towards IC~405. Due to low sampling and distances of IC~410 and IC~417, we are unable to extract much information besides what we already noticed towards the much more nearby IC~405. Spectra of HD~35239, a B-type star located $\approx 250$~pc away from the Sun, might provide an additional insight on the importance of the cloud at 220~pc for studies of DIB profiles.

No ESO spectra are available for regions between IC~405 and Sco-Cen. To obtain any information about the DIB in these lines of sight (Fig.~\ref{fig:C6a}), we need to rely on the limited coverage of the spectra obtained from ELODIE archive. For IC~1848 (800--900~pc away), we find that the DIB is relatively strong. On the other hand, dust maps (Fig.~\ref{fig:C6b}) reveal a very complex line of sight affected by multiple dust clouds within 800~pc from the Sun. The same results are obtained for the more distant Heart~Nebula. However, we note that the increase of the EW from $\sim800$~pc (IC~1848) to $\sim2$~kpc (Heart~Nebula) is minimal.

In the Sadr region, we can notice a dichotomy in terms of the EW. Three of the targets from IC~1318 group show a generally large EW (40~{m\AA}), probing most likely the dusty clouds in the region at a distance of about 1.7~kpc \citep{2020ApJ...896...29K}. Two other stars from this group show EWs that are twice smaller, although the broadening is the same. Parallaxes and interstellar line EWs (DIBs, CH$^*$) suggest a distance that is possibly smaller than that of IC~1318. Only a single measurement is available for Sh2-119 and North America Nebula ($\lesssim800$~pc). We find that the two lines of sight coincide in the EW (19 and 14~{m\AA}, respectively) but differ in the broadening of the profile (19 and 23~km\,s$^{-1}$, respectively, see also Fig.~\ref{fig:C6c}). We estimate that the DIB profile towards IC~1318 is mainly affected by at least two DIB clouds -- one at a distance $<800$~pc and another somewhere near 1.7~kpc.

%-------------------------------------------------------------------
\section{Discussion}\label{section:5}

\begin{figure*}
 \centering
 \includegraphics[width=2\columnwidth]{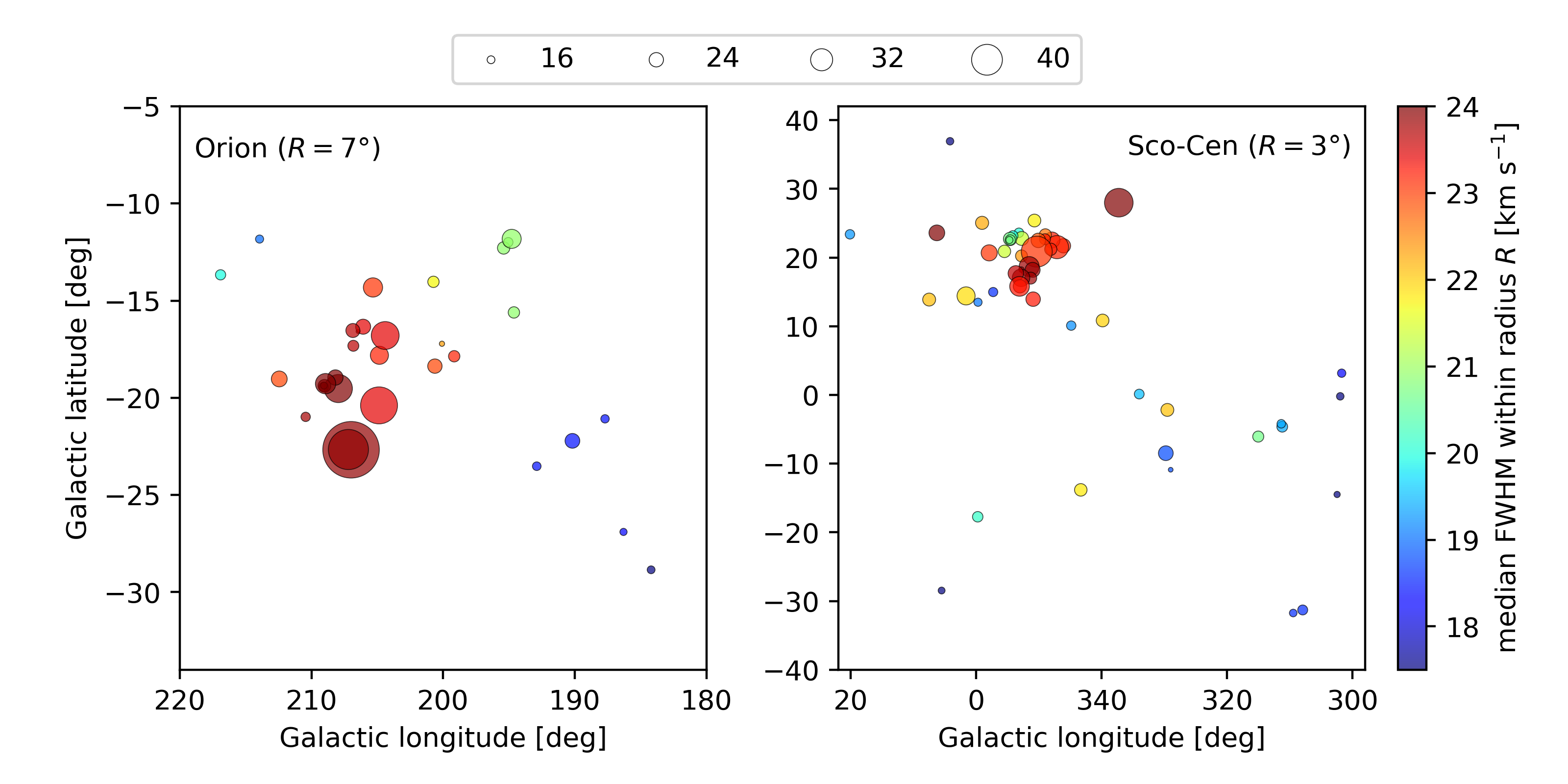}
 \caption{Sky map of the targets within the Orion and the Sco-Cen regions. The size of a data point indicates the measured FWHM (values in~km\,s$^{-1}$), while the colour indicates the median FWHM calculated within a radius $R$ from a data point. For Sco-Cen, we filtered out stars located beyond 500~pc and excluded the extreme cases of $\tau$~Sco and HD~114886.}
 \label{fig:21}
\end{figure*}

\begin{figure}
 \includegraphics[width=\columnwidth]{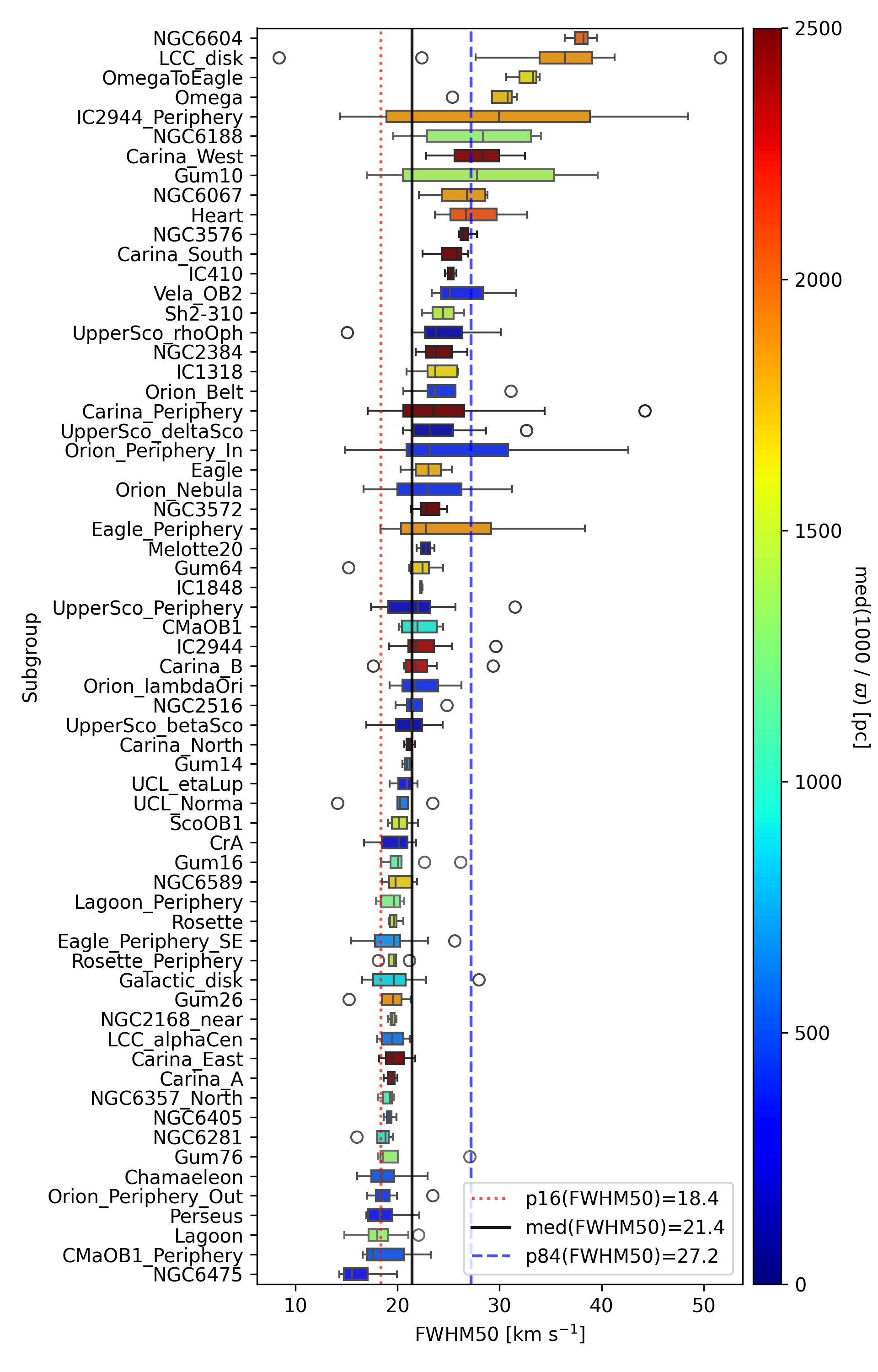}
 \caption{Box plot of the FWHM extracted in this work. The targets not assigned to an object (or "Subgroup") are excluded, as well as objects with only a single measurement. The horizontal lines highlight the 16th percentile, 50th percentile, and the 84th percentile based on all targets.}
 \label{fig:C7}
\end{figure}

Let us discuss the results of our analysis presented in Section~\ref{section:4}, Table~\ref{table:2}, and Table~\ref{table:3}. The key points can be found summarised in Table~\ref{table:4} and Fig.~\ref{fig:C7}. We find at least five objects towards which the profile of the {6196~\AA} DIB is broadened in a similar way as towards USco: Melotte~20, Orion (Nebula and Belt), Vela~OB2, HD~61899 (Vela region, NGC~2451B, assigned to our Gum~10 group), and possibly NGC~2516. We label this behaviour of the DIB as a USco-type \citep[based on the original highlight of this property by][]{2023MNRAS.523.4158G}. These lines of sight are characterised by a typical FWHM of 21--25~km\,s$^{-1}$, the presence of a single strong distinguished dust cloud, and the lack of Doppler splitting in K~\textsc{i} or CH$^*$ lines. Stars towards the Orion~Nebula and Orion~Belt show the same characteristics as those towards USco -- a high median FWHM and a significant scatter around the mean. However, some differences can be noted already in the profile of the DIB. For example, the DIB is significantly weaker towards Orion and Orion lines of sight are likely affected by the presence of another slightly weaker cloud. Moreover, the most prominent broadening of $\approx34$~km\,s$^{-1}$ is detected towards HD~61899, making it one of the more curious targets. While NGC~2516 seems to display the presence of another cloud along the line of sight, we note the lack of Doppler splitting in other interstellar lines. Other stellar populations might also be displaying the USco-type of broadening, but a reliable identification is impossible due to the presence of overlapping Doppler split components originating from different clouds. We note that a splitting in Na~\textsc{i} and Ca~\textsc{ii} lines can be observed towards all USco targets, with the secondary (blueshifted) component being related to an outflow originating from Sco-Cen \citep{2024A&A...689A..84P}. More complex profiles can be seen towards more distant targets (e.g. Orion). The use of these atomic lines might result in a poor identification of the kinematic component of the nearby (denser) clouds, which has been used as the reference frame throughout this paper.

Although we characterise Vela~OB2 as USco-type, it should be pointed out that the lines of sight display relatively strong K~\textsc{i} lines compared to the strength of the DIB, which is in a stark contrast to most of the other regions, where a measurable K~\textsc{i} line usually means the presence of a strong DIB. The most extreme is the line of sight towards HD~62542 (370~pc, not measured in our work), which shows a very weak {6196~\AA} DIB when compared with the strong lines of K~\textsc{i} and CH$^*$. This suggests that the carrier of the DIB is fairly underabundant within the dust wall in front of Vela~OB2 (260--290~pc, $A_V \sim 0.2$~mag).

The stellar populations used to identify the USco-type of broadening have very different characteristics. While USco and Orion and known to be active sites of star formation with massive stars, strong H~\textsc{ii} regions, and are surrounded by dense dusty clouds, star clusters NGC~2516 and Melotte~20 are much older and nowhere close to their sites of origin \citep[for example, see cluster ages and discussions from][]{2021A&A...645A..84M,2024Natur.631...49S}. Vela~OB2 is surrounded by a more evolved ISM than in the case of USco, although both clusters include multiple stellar populations \citep{2022MNRAS.517.5704A,2023A&A...677A..59R}. Together with the data obtained in this work, all the available evidence suggests that a difference in stellar population cannot fully explain the USco-type of broadening. The distances between stellar populations and the intervening clouds might also not be the answer to this puzzle. This argument is based on comparing the values of 30--40~pc for Melotte~20 and USco with the distances of 100--150~pc for the other mentioned lines of sight. However, conclusive results cannot be reached due to possible systematic errors in the cloud distances. For example, we find that the extinction profile as a function of the distance looks quite different when comparing the dust maps by \citet{2020A&A...639A.138L} and \citet{2024A&A...685A..82E} in front of Vela~OB2. The differences are much smaller for the stars in USco and negligible towards Melotte~20.

Curiously, we can note from Table~\ref{table:2} that there appears to be a decrease in the FWHM from the star-forming centres towards the outskirts of Sco-Cen and Orion. This effect is especially evident towards Sco-Cen, where the broadening is highest towards USco, decreases towards UCL and LCC, and even more towards CrA and Chamaeleon. Similarly for the Orion, we see a decrease from the constellation of Orion towards Taurus and to the south. Both cases are displayed in Fig.~\ref{fig:21}. The common property for USco and the Orion~Nebula is that both sites include recently born OB stars -- it is possible that past supernovae and stellar winds that shaped the ISM surrounding these objects also created conditions that are reflected in the observed profile of the DIB. Given the lack of a correlation between the stellar population ages and the DIB broadening in the local ($<500$~pc) ISM, we propose that stellar radiation from current OB stars is likely not the primary mechanism responsible for the observed FWHM gradients.

In contrast to the USco-type lines of sight, we identify several lines of sight displaying narrow DIB profiles ($<19$~km\,s$^{-1}$) that can in future serve to extract empirical profiles of this DIB. We label such lines of sight as Perseus-type, since multiple targets (HD~22951, HD~24398, HD~24534) observed towards Perseus show this property. The same is also true for the Orion stars HD~30492 and HD~30836 ($\pi^4$~Ori), and the Taurus stars HD~26912 ($\mu$~Tau) and HD~28114. Some of the targets probing the ISM in Sco-Cen also show a narrow DIB, for example HD~110336 and HD~116852 (Chamaeleon, significantly different distances), or HD~189103 ($\theta^1$~Sgr, south of CrA).

Excluding the abnormal lines of sight towards Herschel~36 and NGC~6530~151, we find a generally narrow profile also towards the more distant Lagoon~Nebula. The Lagoon~Nebula lines of sight show profiles very similar to Perseus -- this can be seen in Fig~\ref{fig:9}. A comparison with two nearby objects in the periphery of the nebula (HD~162978 and HD~165999, at about 900~pc and 200~pc, respectively) shows that the ISM at smaller distances from the Sun dominates in the observed profile of the DIB. Since measurements of the EW are very similar for these stars and the targets towards the 
Lagoon~Nebula, it follows that the foreground ISM close to the nebula has either an extremely low density or there is a significant lack in the number density of the carrier of this DIB. Unfortunately, the nebula is located at a distance at which an analysis of the intervening cloud positions becomes uncertain and possible only with the lower-resolution 2-kpc map from \citet{2024A&A...685A..82E}. We note that two stars located south-east of the Ophiuchus cloud (HD~150814, HD~152655) show a profile similar to the one observed in the Lagoon~Nebula (more narrow than in the case of most USco targets). As all targets from our USco sample, these are located very close to the Sun ($\lesssim200$~pc). We also highlight the discovery of the narrowest profiles towards NGC~6475.

The profile broadening extracted from the lines of sight towards NGC~6589 (Eagle~region) is comparable with what we found towards the Lagoon~Nebula, the latter being slightly weaker but also slightly narrower (see Table~\ref{table:2} and Fig.~\ref{fig:5}). This is the result of a lack of notable changes in the conditions within the nearby clouds probed by both lines of sight. This result is further supported by the similarities in the line of sight extinction profiles shown in Fig.~\ref{fig:4} and Fig.~\ref{fig:7}. In contrast, the Eagle~Nebula, the Omega~Nebula, and NGC~6604 all show extremely broad profiles that are the result of clear kinematic splitting being present in the profile of the DIB. Similar behaviour can also be observed in the LCC-disk sample of lines of sight. We label objects displaying such a behaviour in the spectra as Eagle-like. Sh2-310 and CMa~OB1 show DIB profiles as broad as in the case of USco. However, the observed spectra for these target groups can be distinguished from a USco-type line of sight by the presence of Doppler split components in the profiles of atomic lines, highlighting the fact that very large FWHM is not always observed in Eagle-like lines of sight.

Focusing on the stars in the Carina~Nebula, we confirm the previously identified \citep{2024AcA....74...69K} lack of significant profile variations in what we label as group~A stars of the Carina~Nebula. On the other hand, group~B stars display similar variations as observed towards USco and Orion. The typical angular separation between two stars in groups~A and B is between $0.09^{\circ}$ and $0.13^{\circ}$. Assuming that the distance from the nebula is $\sim2.5$~kpc, a typical projected distance of $\sim5$~pc can be calculated between the stars. Given the location of the DIB cloud between us and the stars, the scales at which we see variations in group~B are $\lesssim5$~pc. On the other hand, the lack of variations is seen at the same scale towards group~A. Since the two groups are bordering each other, a sudden change in the properties of either the DIB cloud or the dust cloud is required to explain the observations. A good model of these small-scale changes is required to reconcile observations around $\eta$~Car with the proposition from \citet{2024AcA....74...69K} that the DIB carriers occupy a different physical space than dust grains (towards these lines of sight). Reliable and precise measurements of the extinction must be provided before any conclusions can be reached regarding this interpretation of the observations. We note that the standard deviations in the obtained profile parameters (together with the median FWHM) towards the Rosette~Nebula, NGC~6405, and NGC~6357 (north) are very similar to what we see in group~A of the Carina~Nebula. Furthermore, we highlight the similarities between the FWHM of the profiles towards the Carina group~A and the Lagoon~Nebula.

\begin{figure}
 \includegraphics[width=\columnwidth]{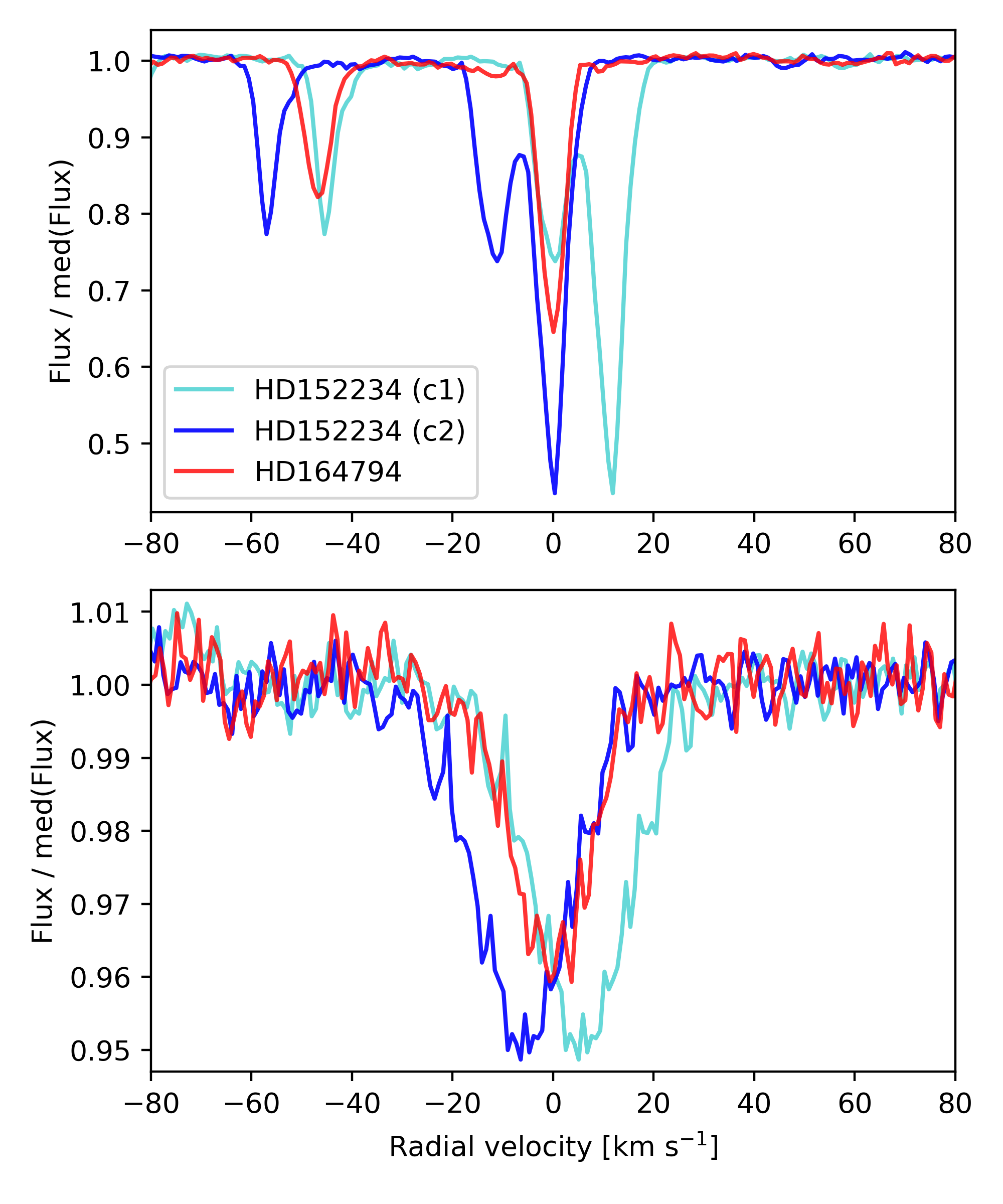}
 \caption{Same as Fig.~\ref{fig:5} but for Sco~OB1 and the Lagoon~Nebula. Scaling to the same line depth is ignored. Wavelengths were shifted to the rest position of the K~\textsc{i} line (top panel). For the Sco~OB1 target, both distinguished kinematic components of K~\textsc{i} were used.}
 \label{fig:22}
\end{figure}

It is worth pointing out that the correlation between the number of clouds along the lines of sight (often itself correlated with splitting in atomic lines) and the DIB profiles cannot be used to rule out the hypothesis explored above. These only highlight that the presence of dust clouds often (if not always) suggest the existence of DIB carriers around the same clouds, very similar to the general correlation between the colour excess and EWs of DIBs, or the correlation between the motion of CO gas and DIB carriers \citep{2023ApJ...954..141S}. The results obtained by \citep{2024AcA....74...69K} hint at physics that is happening on relatively small scales compared to the dimensions of the dust clouds seen in 3D dust maps.

Finally, we aim to explore the reported wavelength shifts of the investigated DIB. Based on the available literature \citep{2006MNRAS.366.1075G,2008PASP..120..178G,2015PASP..127..356G,2015MNRAS.451.3210K}, the DIBs are often seen shifted when compared with the atomic or simple molecular lines. While previous studies revealed a redshift towards Orion (not seen in {6196~\AA}), a blueshift was registered towards Sco~OB1. The presence of Doppler split (approximately 5--10~km\,s$^{-1}$) components in the profiles of K~\textsc{i} and CH$^*$ lines might explain the mentioned blueshift, similar to what we see in Fig.~\ref{fig:5}. For a better illustration, we compare the profiles of the DIBs observed towards HD~152234 (Sco~OB1) and HD~164794 (Lagoon~Nebula) in Fig.~\ref{fig:22}. We can see that regardless the choice of the kinematic components, the position of the maximum absorption of the DIB shifts by $\approx6$~km\,s$^{-1}$. Assuming that the chosen kinematic component of the atomic line was the stronger component, we can see that the reported blueshift is a result of combining two kinematic components in the profile. This explanation for the offset in the position of the DIB was previously suggested by \citet{2015PASP..127..356G}. The evidence for this effect can be further highlighted by the red and the blue wings looking similar to what we observe towards the more simple line of sight of HD~164794, depending on the choice of the rest wavelength component.

The shifts seen towards Orion still need to be investigated further. Since no offset was previously reported for the {6196~\AA} DIB, the data presented in this work are not very helpful for this region. From the perspective of our analysis, the lack of an offset is reasonably correlated with the non-detection of splitting in atomic lines (besides Na~\textsc{i} and Ca~\textsc{ii}, which can trace more diffuse ISM than the DIB).

%-------------------------------------------------------------------
\section{Conclusions}\label{section:6}

We confirm the previous findings from \citet{2023MNRAS.523.4158G} that point at different ISM conditions being probed by the {6196~\AA} DIB towards USco and Perseus. The USco region includes lines of sight showing both narrow and broad profiles of the DIB, with the differences being most prominent in the red wing of the profile. Based on our analysis of the other regions, this should hint at the presence of unresolved kinematic components. However, no redshifted components can be identified in any of the typical tracers of the interstellar gas: K~\textsc{i}, Na~\textsc{i}, Ca~\textsc{ii}, or CH$^*$. In addition to USco, we find a few more regions in the sky that show a similar behaviour (increased median and dispersion of the FWHM) of the DIB, including Orion, Vela~OB2, and Melotte~20. Most interesting is also the existence of star-forming regions towards which this behaviour is not observed, such as Perseus, the Lagoon~Nebula, and the Rosette~Nebula. Presently, we cannot determine whether the USco-type of line of sight is more or less typical than the Perseus-type.

No correlation can be found between the broadening and the ages of the observed stars or their distances from the clouds. On the other hand, we find hints of a decrease in the FWHM as a function of the distance from the centre of star formation in Sco-Cen and Orion -- to our knowledge, no such gradient in the FWHM across the sky was previously reported. While a detailed analysis of the ISM in these regions of space is required to fully understand this DIB behaviour, this is beyond the original aim and scope of this study. We suggest that future investigations should account for the influence of supernovae and stellar winds on the conditions within the ISM probed by the DIB.

The observed lack of DIB profile variations towards the core of the Carina~Nebula \citep{2024AcA....74...69K} still requires further investigation. We verify this lack of variations but also identify a neighbouring region (south from $\eta$~Car) within which the profile of the DIB changes appreciably. Additionally, we find that only a few other regions show the same consistency in the profile of the DIB, most notably the Rosette~Nebula and NGC~6405. As was highlighted by \citet{2024AcA....74...69K}, such regions are of a great interest when searching for a possible disconnection between the interstellar dust and the carriers of the DIBs. It should be mentioned that a similar small-scale structure within Local Bubble was previously noted by \citet{2016A&A...585A..12B}. The available data suggest that the proximity of the foreground cloud to the stellar population and the age of the stellar population do not show a strong correlation with the broadening of the DIB and its variations.

The presented evidence is strong enough to confirm Doppler splitting as the main mechanism responsible for the observed shifts in the central wavelength of the {6196~\AA} DIB \citep[e.g.][]{2008PASP..120..178G}. For the most part, this splitting can be hinted at by the presence of Doppler split components within the profile of K~\textsc{i}. However, as can be seen in USco, the presence of a detectable K~\textsc{i} line is not necessary to observe DIBs. On the other hand, we find that the presence of a K~\textsc{i} kinematic component often correlates with the presence of a DIB kinematic component. An interesting exception is the dust wall in front of Vela~OB2.

A continuous development in our understanding of the physics hidden within the profiles of DIBs (and other interstellar features) is key to understanding the physics of the ISM. The presence or absence of certain features may hint at densities of the foreground clouds. The broadening of a single kinematic component tells us something about the gas temperatures or possibly the isotopic content in molecules \citep[see][and references therein]{2009A&A...498..785K}, although as we see for DIBs the story might be a little more complicated. Future studies might want to try and correlate the broadening and the intensity of the DIB with variations in the extinction law. Assuming that we have a complete understanding of the processes affecting the profile of a DIB, a reconstruction of the properties of the DIB clouds extracted from the profiles can be done even for complicated lines of sight (such as the Omega~Nebula) by making use of targets located at various distances from the Sun. Such reconstructions have not only been successful in creating 3D dust maps and providing information about interstellar magnetic fields \citep{2024A&A...685A..82E,2024A&A...690A.314H}, but also produced information about interstellar velocity fields based on an infrared DIB \citep{2018AJ....156..248T}.

\section*{Data availability}

Information provided in Tables~\ref{table:2}, \ref{table:3}, and \ref{table:4} is based on individual target measurements, which are only available in electronic form at the CDS via anonymous ftp to \href{https://cdsarc.u-strasbg.fr/}{cdsarc.u-strasbg.fr} (\href{ftp://130.79.128.5}{130.79.128.5}) or via \href{http://cdsweb.u-strasbg.fr/cgi-bin/qcat?J/A+A/}{http://cdsweb.u-strasbg.fr/cgi-bin/qcat?J/A+A/}.

%-------------------------------------------------------------------
%-------------------------------------------------------------------
%-------------------------------------------------------------------
%-------------------------------------------------------------------
%-------------------------------------------------------------------
\begin{acknowledgements}
This paper made use of data obtained from the ESO Science Archive Facility, specifically from the following ESO programmes: 0100.D-0621(A), 0100.D-0621(B), 0101.C-0180(A), 0102.A-9010(A), 0102.C-0040(B), 0102.C-0547(A), 0102.D-0885(A), 0104.A-9001(A), 0106.A-9006(A), 0106.A-9009(A), 0108.A-9004(A), 0110.A-9029(A), 0111.A-9002(A), 072.D-0235(B), 072.D-0410(A), 073.C-0337(A), 073.C-0733(C), 073.D-0072(A), 073.D-0126(A), 073.D-0234(A), 073.D-0291(A), 073.D-0609(A), 074.B-0455(A), 074.D-0021(A), 074.D-0240(A), 074.D-0300(A), 075.D-0021(A), 075.D-0061(A), 075.D-0103(A), 075.D-0177(A), 075.D-0342(A), 075.D-0369(A), 075.D-0532(A), 076.B-0055(A), 076.C-0164(A), 076.C-0431(A), 076.D-0172(A), 076.D-0294(A), 077.B-0348(A), 077.C-0295(A), 077.C-0575(A), 077.D-0009(A), 077.D-0146(A), 077.D-0164(A), 077.D-0384(A), 077.D-0390(A), 077.D-0478(A), 077.D-0605(C), 077.D-0635(A), 077.D-0712(A), 078.C-0403(A), 078.D-0080(A), 078.D-0669(A), 079.A-9008(A), 079.B-0856(A), 079.C-0597(A), 079.C-0789(A), 079.D-0031(A), 079.D-0564(A), 079.D-0564(B), 079.D-0564(C), 079.D-0567(A), 079.D-0718(B), 080.D-2006(A), 081.A-9006(A), 081.C-0475(A), 081.C-2003(A), 081.D-0656(D), 081.D-2002(A), 081.D-2005(C), 081.D-2008(A), 082.C-0390(A), 082.C-0427(C), 082.C-0566(A), 082.C-0831(A), 082.D-0061(A), 082.D-0136(A), 082.D-0933(A), 083.A-9003(A), 083.A-9011(A), 083.A-9013(A), 083.A-9014(A), 083.C-0503(A), 083.C-0676(A), 083.C-0794(D), 083.D-0034(A), 083.D-0040(A), 083.D-0475(A), 083.D-0589(A), 083.D-0589(B), 084.A-9016(A), 084.B-0029(A), 084.C-1002(A), 084.D-0067(A), 084.D-0481(C), 084.D-0611(A), 085.A-9027(B), 085.C-0614(A), 085.C-0614(B), 085.D-0093(A), 085.D-0185(A), 086.A-9019(A), 086.D-0078(A), 086.D-0236(A), 086.D-0449(A), 086.D-0997(A), 086.D-0997(B), 087.A-9005(A), 087.A-9029(A), 087.D-0010(A), 087.D-0099(A), 087.D-0264(F), 087.D-0946(A), 088.A-9003(A), 088.D-0064(A), 088.D-0424(D), 089.C-0006(A), 089.D-0153(A), 089.D-0730(A), 089.D-0975(A), 090.D-0358(A), 090.D-0600(A), 091.C-0713(A), 091.D-0122(A), 091.D-0221(A), 091.D-0622(A), 092.A-9018(A), 092.A-9020(A), 092.C-0019(A), 092.C-0173(A), 092.D-0363(A), 094.A-9012(A), 094.C-0946(A), 094.D-0355(A), 095.A-9029(D), 095.A-9032(A), 095.D-0234(A), 096.A-9018(A), 096.A-9024(A), 096.A-9027(A), 096.A-9030(A), 096.A-9039(A), 096.D-0008(A), 097.A-9013(A), 097.A-9024(A), 097.A-9039(C), 097.C-0409(A), 097.C-0741(A), 097.C-0979(A), 098.A-9007(A), 098.A-9039(C), 098.C-0463(A), 099.A-9029(A), 099.A-9032(A), 099.C-0637(A), 099.D-0057(A), 105.207T.001, 106.211J.002, 109.22V6.001, 109.23K8.001, 112.262P.001, 167.D-0173(A), 178.D-0361(B), 178.D-0361(D), 178.D-0361(F), 178.D-0361(H), 179.C-0197(A), 179.C-0197(B), 179.C-0197(C), 179.C-0197(D), 182.D-0356(A), 182.D-0356(C), 182.D-0356(D), 182.D-0356(E), 182.D-0356(F), 185.D-0056(A), 185.D-0056(B), 185.D-0056(C), 185.D-0056(D), 185.D-0056(E), 185.D-0056(F), 185.D-0056(I), 185.D-0056(J), 185.D-0056(K), 185.D-0056(L), 188.B-3002(M), 188.B-3002(Q), 188.B-3002(R), 188.B-3002(S), 190.D-0237(A), 190.D-0237(E), 193.B-0936(U), 193.B-0936(W), 194.C-0833(A), 194.C-0833(B), 194.C-0833(C), 194.C-0833(D), 194.C-0833(E), 194.C-0833(F), 194.C-0833(G), 194.C-0833(H), 266.D-5655(A), 278.D-5041(A), 285.D-5042(A), 295.D-5025(A), 382.D-0705(A), 384.B-0066(A), 60.A-9036(A), 60.A-9122(B), 60.A-9700(A), 60.A-9700(G), 60.A-9709(G), 65.I-0498(A), 66.B-0320(A), 67.D-0047(A), 69.B-0277(A), and 71.C-0367(A).

This paper also makes use of data obtained at Observatoire de Haute-Provence that are available in the ELODIE archive (\url{http://atlas.obs-hp.fr/elodie/}).

This work made use of data from the European Space Agency (ESA) mission {\it Gaia} (\url{https://www.cosmos.esa.int/gaia}), 
processed by the {\it Gaia} Data Processing and Analysis Consortium (DPAC, \url{https://www.cosmos.esa.int/web/gaia/dpac/consortium}). 
Funding for the DPAC has been provided by national institutions, in particular the institutions participating in the {\it Gaia} 
Multilateral Agreement.

The following Python libraries were used in this work: \texttt{numpy} \citep{numpy}, \texttt{scipy} \citep{scipy}, \texttt{astropy} \citep{astropy1,astropy2,astropy3}, \texttt{matplotlib} \citep{matplotlib}, and \texttt{seaborn} \citep{Waskom2021}.

This project has received funding from the European Research Council (ERC) under the European Union’s Horizon 2020 research and innovation programme (Grant agreement No. 101055318).

\end{acknowledgements}

%-------------------------------------------------------------------
\bibliographystyle{aa} 
\bibliography{aanda}

%-------------------------------------------------------------------
%-------------------------------------------------------------------
%-------------------------------------------------------------------
%-------------------------------------------------------------------
%-------------------------------------------------------------------
\begin{appendix}

\section{Identifying end-points of the DIB profile}\label{section:A}

\begin{figure}
 \includegraphics[width=\columnwidth]{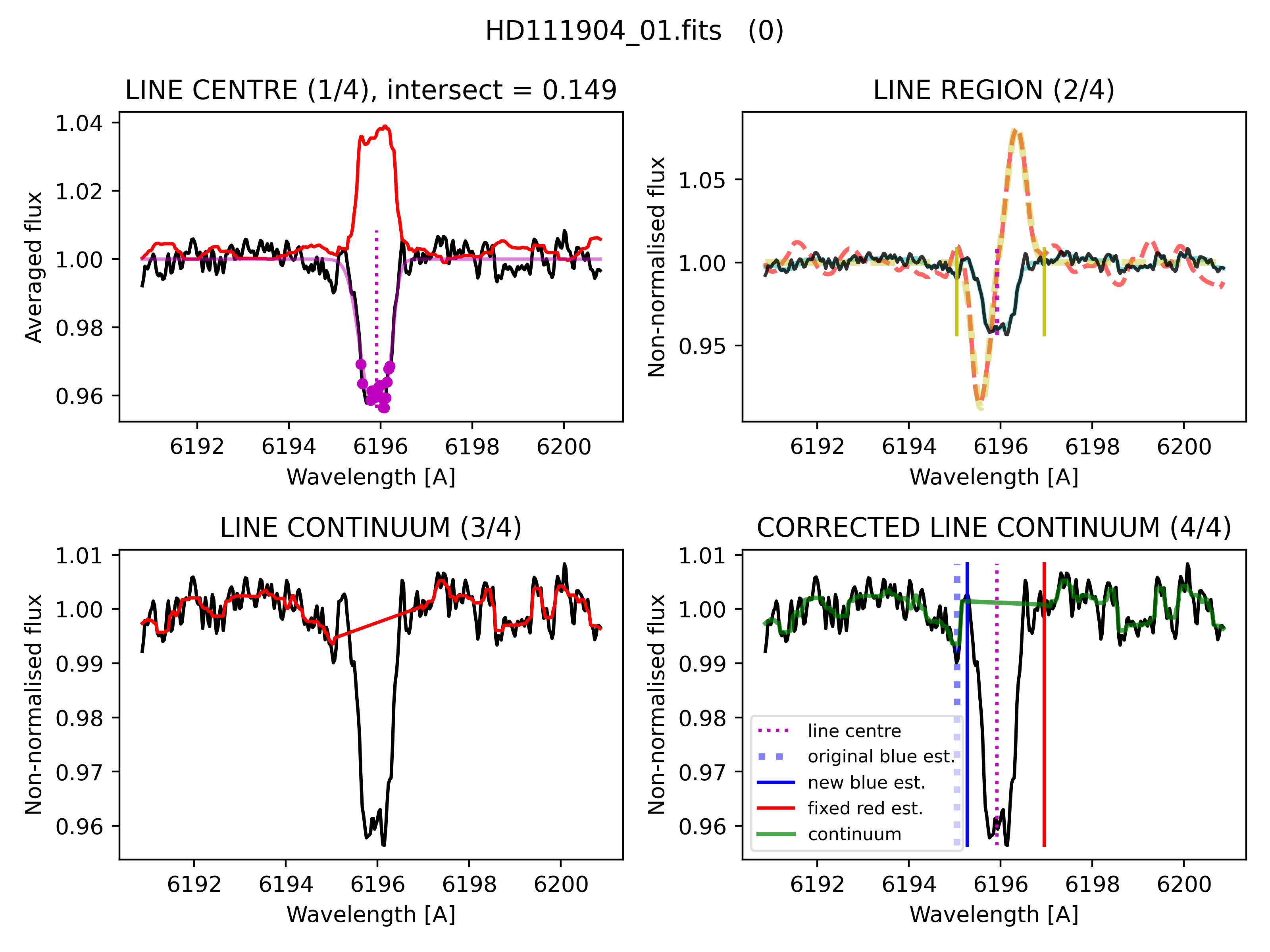}
 \caption{Graphical representation of the intermediate steps of our continuum normalisation procedure. Top-left: roughly stacked spectrum is used to identify the position of the DIB. Top-right: the gradient of the spectrum and the initial estimates of profile end-points. Bottom-left: initial estimate of the continuum. Bottom-right: the final continuum model based on an improved estimate of the blue end-point.}
 \label{fig:A}
\end{figure}

Our end-point detection algorithm is based on a reasonable initial estimate of the central wavelength of the DIB. The spectrum was first smoothed by using the information obtained at a given wavelength together with the 10~nearest data points. We use this set of 11~points to calculate the gradient in the curve of the spectrum as a function of wavelength. Assuming the presence of a single symmetric absorption feature and the absence of noise, the feature can be detected by searching for a decrease followed up by an increase in the gradient of the spectrum.

The gradient curve is more complex in the case of real spectra, but we can still make use of the idealised case. The initial estimate of the position of the DIB was used to identify the position at which the idealised gradient structure is expected to be present. Next, we made use of the fact that this structure can be fitted with a combination of two Gaussians -- a blueshifted component with a negative amplitude and a redshifted component with a positive amplitude. The end-points of the absorption feature were then estimated by calculating the 3-sigma distance from the centre of the Gaussian (redwards from the red component, bluewards for the blue component). This estimate is typically good enough for the red wing of the DIB.

In the case of the blue wing, we need to pay attention to the presence of another (much weaker) DIB at around {6195~\AA}. This additional feature influences the estimate of the blue end-point of the DIB in a significantly large number of spectra and cannot be ignored. Let us take $M$-nearest wavelengths to the original estimate of the blue wing end-point. For each instance, we can continue with the continuum normalisation procedure and fit a simplified model of the DIB (two Gaussians). If the weaker DIB (ignored during fitting) remains present in the spectrum, there should be a larger difference between the fit and the reduced spectrum when compared with the case of the weaker feature being absent. However, if we move the end-point too close to the centre of the DIB, the EW noticeably decreases.

We made use of all of this information to extract the best estimate of the wavelength of the blue wing end-point. For each of the $M$-estimates, we calculated the sum of squared residuals $\Sigma_m$ (differences between the fit and the reduced spectrum). These sums were normalised by $\sigma_1 A_1 + \sigma_2 A_2$ ($\sigma_i$ and $A_i$ are the standard deviation and the amplitude of the Gaussian, respectively) to account for the possible decrease of the EW towards the centre of the DIB. The best estimate should have a very small normalised value of $\Sigma_m$, although minimasing this quantity is not optimal. Instead, we calculated the median wavelength of the end-points for which $\Sigma_m$ is smaller or equal to the 16th percentile of all $\Sigma$. This is motivated by the fact that multiple neighbouring data points can ambiguously serve as the end-point if the weak and the strong DIBs are not blended.

An example of the whole continuum normalisation procedure is displayed in Fig.~\ref{fig:A}. All individual spectra of HD~111904 presented the same problem with identifying the blue end-point of the DIB. The roughly stacked spectrum in the top-left panel was produced by stacking only spectra rescaled using the median of the flux -- it is only used for estimating the central wavelength of the DIB.

\section{Summary tables}\label{section:B}

The summary tables mentioned in Sections~\ref{section:4} and \ref{section:5} are provided below. Tables~\ref{table:2} and \ref{table:3} are focused on the individuals objects (clouds or stellar clusters) and capture the statistical properties of the investigated DIB profile with the following measures: the median of the~FWHM, the standard deviation of the~FWHM, the median of the~EW, and the range of EW values. Correlations between the~FWHM, the~EW, and the distance were calculated for objects containing at least eight targets (except for the Perseus cloud, where $N=7$).

% ------------------- START TABLE
\begin{table*}
\caption{Measured {6196~\AA} DIB properties in the well-sampled regions}\label{table:2}
\small
\centering
\begin{tabular}{lll|rrrrrr}
\hline\hline
Region ID & Object & Targets & med(FWHM) & std(FWHM) & med(EW) & range(EW) & corr(EW, FWHM) & corr(EW, $d$)  \\
 &  &  & [km\,s$^{-1}$] & [km\,s$^{-1}$] & [m\AA] & [m\AA] & &   \\
  \hline
  
1 & Omega~Nebula & 4 &  30.8 &  2.5 &  72 &  56--91 & ... & ...  \\
1 & Omega-Eagle & 3 &  33.3 &  1.4 &  56 &  48--62 & ... & ...  \\
1 & Eagle~Nebula & 12 &  23.0 &  1.6 &  50 &  41--62 & 0.356 & 0.229   \\
1 & NGC~6604 & 6 &  38.2 &  1.0 &  63 &  55--75 & ... & ...   \\
1 & \textit{all above} & 25 &  25.4 &  6.4 &  56 &  41--91 & 0.488 & -0.011   \\
1 & \textit{periphery} & 14 &  22.8 &  5.7 &  53 &  19--87 & 0.788 & 0.598   \\
1 & \textit{periphery SE} & 10 &  19.6 &  2.8 &  28 &  6--48 & 0.845 & 0.602   \\
1 & NGC~6589 & 7 &  20.2 &  2.0 &  24 &  23--64 & ... & ...   \\
\hline
2 & Lagoon~Nebula & 20 & 18.0 & 1.7 & 20 & 14--35 & 0.622 & -0.128  \\
2 & \textit{periphery} & 6 & 19.7 & 1.1 & 27 & 17--44 & ... & ...  \\
2 & Gum~76 & 5 & 18.6 & 3.4 & 20 & 14--70 & ... & ...  \\
2 & NGC~6475 & 7 & 15.6 & 1.9 & 6 & 5--7 & ... & ...  \\
2 & NGC~6405 & 4 & 19.2 & 0.5 & 15 & 14--19 & ... & ...  \\
2 & \textit{NGC~6357 and N} & 6 & 19.4 & 1.9 & 16 & 12--75 & ... & ...  \\
2 & Gum~64 & 7 & 22.4 & 2.8 & 41 & 12--62 & ... & ...  \\
2 & NGC~6281 & 4 & 18.8 & 1.4 & 28 & 11--30 & ... & ...  \\
2 & Sco~OB1 & 13 & 20.2 & 0.9 & 26 & 16--35 & 0.297 & 0.135  \\
2 & \textit{all above} & 72 & 19.2 & 2.4 & 20 & 5--75 & 0.780 & 0.579  \\
2 & \textit{Galactic disk} & 10 & 19.4 & 3.3 & 20 & 7--88 & 0.167 & 0.979  \\
\hline
3 & USco ($\rho$~Oph) & 10 & 23.8 & 3.8 & 12 & 3--33 & -0.011 & 0.014  \\
3 & USco ($\beta$~Sco) & 9 & 21.4 & 2.1 & 14 & 9--19 & 0.556 & -0.318  \\
3 & USco ($\delta$~Sco) & 8 & 23.2 & 3.9 & 9 & 9--12 & 0.559 & -0.423  \\
3 & \textit{periphery} & 11 & 22.0 & 3.9 & 12 & 7--24 & -0.119 & 0.285  \\
3 & UCL ($\eta$~Lup) & 3 & 20.9 & 1.1 & 16 & 8--19 & ... & ...  \\
3 & UCL (Norma) & 5 & 20.3 & 3.1 & 13 & 8--14 & ... & ...  \\
3 & CrA & 3 & 20.2 & 2.1 & 6 & 4--14 & ... & ...  \\
3 & NGC~6067/6188 & 8 & 26.8 & 4.7 & 28 & 8--45 & 0.935 & 0.723  \\
3 & LCC ($\alpha$~Cen) & 6 & 19.5 & 2.7 & 12 & 7--21 & ... & ...  \\
3 & LCC (disk) & 15 & 36.4 & 9.4 & 31 & 18--66 & 0.493 & 0.552  \\
3 & Chamaeleon & 10 & 18.5 & 2.0 & 14 & 9--18 & -0.028 & 0.591  \\
\hline
4 & Carina~Nebula (A) & 5 & 19.4 & 0.4 & 26 & 22--29 & ... & ...  \\
4 & Carina~Nebula (B) & 7 & 21.8 & 3.4 & 26 & 19--29 & ... & ...  \\
4 & Carina~Nebula (N) & 3 & 21.1 & 0.4 & 30 & 28--37 & ... & ...  \\
4 & Carina~Nebula (S) & 8 & 25.4 & 1.4 & 20 & 15--45 & 0.283 & 0.786  \\
4 & Carina~Nebula (E) & 3 & 19.5 & 1.5 & 17 & 13--29 & ... & ...  \\
4 & Carina~Nebula (W) & 13 & 28.7 & 3.2 & 20 & 14--62 & 0.324 & 0.400  \\
4 & \textit{periphery} & 27 & 23.3 & 5.7 & 25 & 3--96 & 0.418 & 0.744  \\
4 & IC~2944 & 11 & 21.8 & 2.7 & 20 & 15--33 & 0.825 & 0.462  \\
4 & \textit{periphery} & 11 & 30.0 & 11.2 & 15 & 9--29 & 0.740 & 0.786  \\
4 & NGC~3572 & 7 & 22.8 & 1.2 & 29 & 28--37 & ... & ...  \\
4 & NGC~3576 & 4 & 26.4 & 0.7 & 31 & 25--93 & ... & ...  \\
\hline
5 & NGC~2516 & 4 & 21.4 & 1.8 & 13 & 11--17 & ... & ...  \\
5 & Vela~OB2 & 3 & 25.1 & 3.6 & 6 & 6--10 & ... & ...  \\
5 & Gum~14 & 2 & 21.0 & 0.5 & 15 & 14--17 & ... & ...  \\
5 & Gum~(10, 16, 26) & 18 & 20.1 & 5.8 & 18 & 3--75 & 0.083 & 0.827  \\
\hline
6 & Rosette~Nebula & 6 & 19.7 & 0.5 & 35 & 32--39 & ... & ...  \\
6 & \textit{periphery} & 5 & 19.6 & 1.0 & 36 & 21--48 & ... & ...  \\
6 & CMa~OB1 & 9 & 22.0 & 1.8 & 18 & 9--22 & 0.203 & 0.224  \\
6 & \textit{periphery} & 7 & 17.5 & 2.4 & 8 & 3--36 & ... & ...  \\
6 & Sh2-310 & 2 & 24.5 & 2.1 & 8 & 7--10 & ... & ...  \\
6 & NGC~2384 & 4 & 23.9 & 1.8 & 30 & 20--35 & ... & ...  \\
\hline
7 & Orion (Nebula) & 6 & 22.9 & 4.8 & 6 & 2--13 & ... & ...  \\
7 & Orion (Belt) & 6 & 23.3 & 5.4 & 6 & 3--15 & ... & ...  \\
7 & Orion ($\lambda$~Ori) & 3 & 21.7 & 2.9 & 7 & 6--14 & ... & ...  \\
7 & \textit{periphery, inside} & 11 & 23.2 & 8.2 & 10 & 2--18 & 0.238 & -0.056  \\
7 & \textit{periphery, outside} & 7 & 18.3 & 2.0 & 17 & 5--30 & ... & ...  \\
\hline
8 & Pleiades & 1 & 20.6 & ... & 6 & ... & ... & ...  \\
8 & Perseus~cloud & 7 & 18.3 & 1.8 & 16 & 11--26 & 0.569 & -0.198  \\
8 & Melotte~20 & 2 & 22.8 & 0.9 & 9 & 8--9 & ... & ...  \\
\hline

\end{tabular}
\tablefoot{
Distances $d$ were calculated by inverting the parallaxes obtained by Gaia. A label \textit{periphery} refers to a group of nearby peripheral targets that cannot be associated with the investigated stellar clusters in a given region, while \textit{all above} refers to a set of targets belonging to all of the above mentioned objects in the region. Directions in the sky are occasionally included in the label of the object -- the Carina~Nebula~(N) indicates the location of the stars in the northern section of the nebula. Carina~A and B represent regions very close to $\eta$~Car. HD~119646 was excluded from the analysis of LCC~(disk), see Section~\ref{section:4.3}. Correlations were calculated only of objects with a number of targets $>7$ (except for the Perseus~cloud).
}
\end{table*}
% ------------------- END TABLE
% ------------------- START TABLE
\begin{table*}
\caption{Measured {6196~\AA} DIB properties in the poorly sampled regions}\label{table:3}
\small
\centering
\begin{tabular}{lll|rrrrrr}
\hline\hline
Region ID & Object & Targets & med(FWHM) & std(FWHM) & med(EW) & range(EW) & corr(EW, FWHM) & corr(EW, $d$)  \\
 &  &  & [km\,s$^{-1}$] & [km\,s$^{-1}$] & [m\AA] & [m\AA] & &   \\
  \hline
  
9 & Sh2-247 & 1 &  19.3 & ... &  83 & ... & ... & ...  \\
9 & \textit{near NGC~2168} & 2 &  19.5 & 0.4 & 32 & 17--47 & ... & ...  \\
9 & IC~405 & 1 & 23.1 & ... & 20 & ... & ... & ...  \\
9 & IC~410 & 3 &  25.3 & 0.5 & 41 & 29--49 & ... & ...  \\
9 & IC~417 & 1 & 21.4 & ... & 34 & ... & ... & ...  \\
\hline
10 & IC~1848 & 2 &  22.3 & 0.1 & 44 & 43--45 & ... & ...  \\
10 & Heart~Nebula & 3 &  26.7 & 3.8 & 51 & 50--68 & ... & ...  \\
\hline
11 & Sh2-119 & 1 & 19.2 & ... & 19 & ... & ... & ...  \\
11 & North~America & 1 & 23.3 & ... & 14 & ... & ... & ...  \\
11 & IC~1318 & 5 &  23.7 & 1.9 & 41 & 23--48 & ... & ...  \\
\hline

\end{tabular}
\end{table*}
% ------------------- END TABLE
% ------------------- START TABLE
\begin{table*}
\caption{Summary of the key results presented in Sections~\ref{section:4} and \ref{section:5}}\label{table:4}
\small
\centering
\begin{tabular}{l|rrrrrr|rrrr|l}
\hline\hline
Object & med($l$) & med($b$) & med($d$) & range($d$) & $A^{\dagger}_V$ & Foregr. & FWHM & Line & FWHM & FWHM & LoS \\
 & [$^{\circ}$] & [$^{\circ}$] & [pc] & [pc] & [mag] & clouds & value & split & variation & gradient & type  \\
  \hline

Omega Nebula & 15.1 & $-0.8$ & 1741 & 1529--1856 & 0.83* & 3 & 2 & yes & yes & yes & Eagle \\
Eagle Nebula & 17.0 & $+0.8$ & 1780 & 1664--2078 & 1.14* & 3 & 1 & yes & yes & yes & Eagle \\
NGC 6604 & 18.4 & $+1.8$ & 2008 & 1375--2105 & 1.21* & 2--3 & 2 & yes & yes & yes & Eagle \\
NGC 6589 & 11.5 & $-1.6$ & 1383 & 1140--1900 & 0.70* & 1--2 & 1 & yes & yes & yes & Eagle \\
Lagoon Nebula & 6.1 & $-1.3$ & 1312 & 1221--1526 & 0.77* & 2 & 0 & no & yes & no & Perseus \\
NGC 6475 & 355.9 & $-4.5$ & 277 & 245--315 & 0.24 & 2 & 0 & no & small/yes & no & Perseus \\
NGC 6405 & 356.5 & $-0.7$ & 502 & 467--543 & 0.52 & 3 & 0 & no & small & no & Perseus \\
Gum 64 & 351.2 & $+0.8$ & 1666 & 291--1867 & 2.10* & 1 & 1 & yes & yes & no & Eagle \\
Sco OB1 & 343.5 & $+1.2$ & 1534 & 1392--1789 & 1.42* & 2 & 1 & yes & minimal & no & Eagle \\
USco (all) & 352.8 & $+20.9$ & 143 & 124--214 & 0.36 & 1 & 1 & no & yes & yes & USco \\
UCL (all) & 330.3 & $-7.1$ & 518 & 175--1135 & 0.39 & 1 & 1 & no & yes & yes & USco \\
LCC ($\alpha$ Cen) & 311.4 & $-4.2$ & 606 & 411--1202 & 0.34 & 2 & 1 & no & yes & yes & USco \\
LCC (disk) & 309.2 & $-0.7$ & 1929 & 1652--3814 & 0.82* & 1--2 & 2 & yes & yes & no & Eagle \\
Chamaeleon & 304.2 & $-8.3$ & 516 & 259--3542 & 0.59 & 1--2 & 0 & no & yes & yes & Perseus \\
Carina Nebula (A) & 287.5 & $-0.6$ & 2430 & 2258--2601 & 0.21* & 3 & 0 & yes & minimal & no & Per/Eag \\
Carina Nebula (B) & 287.6 & $-0.7$ & 2321 & 2246--2555 & 0.21* & 3 & 1 & yes & yes & no & Eagle \\
Carina Nebula (N) & 287.4 & $-0.3$ & 2468 & 2415--2521 & 0.21* & 3 & 1 & no & small & no & USco \\
Carina Nebula (S) & 287.6 & $-1.1$ & 2731 & 2438--4141 & 0.22* & 3 & 1 & no & yes & no & USco \\
Carina Nebula (E) & 288.0 & $-0.9$ & 2470 & 2406--2762 & 0.23* & 3 & 0 & yes & yes & no & Per/Eag \\
Carina Nebula (W) & 285.9 & $+0.1$ & 2447 & 1859--2687 & 0.21* & 3 & 2 & yes & yes & no & Eagle \\
IC 2944 & 294.8 & $-1.6$ & 2361 & 954--2810 & 0.70* & 1 & 1 & yes & yes & no & Eagle \\
NGC 2516 & 273.8 & $-15.9$ & 411 & 380--445 & 0.29 & 2 & 1 & no & yes & no & USco \\
Vela OB2 & 260.2 & $-10.3$ & 368 & 285--459 & 0.22 & 1 & 1 & no & yes & no & USco \\
Rosette Nebula & 206.3 & $-2.1$ & 1514 & 1411--1566 & 0.68* & 3 & 0/1 & yes & minimal & no & Per/Eag \\
CMa OB1 & 224.4 & $-1.9$ & 1068 & 129--2476 & 0.22 & 2 & 1 & yes & yes & no & Eagle \\
Orion (no periphery) & 206.8 & $-17.3$ & 407 & 336--511 & 0.19 & 2 & 1 & no & yes & yes & USco \\
Orion (outside) & 190.2 & $-22.2$ & 437 & 155--1618 & 0.45 & 1--2 & 0 & no & yes & yes & Perseus \\
Perseus cloud & 160.5 & $-17.1$ & 350 & 252--614 & 0.83 & 2 & 0 & no & yes & no & Perseus \\
Melotte 20 & 147.0 & $-6.4$ & 170 & 164--176 & 0.22 & 1 & 1 & no & small & no & USco \\

\hline

\end{tabular}
\tablefoot{
The provided interstellar extinction $A^{\dagger}_V$ (integrated up to the median distance) is obtained from the 1.25-kpc extinction map from \citet{2024A&A...685A..82E}, and thus provides only limited information about the extinction along the line of sight towards more distant stars -- such cases are marked with an asterisk. The same extinction map was also used to provide a lower estimate for the number of prominent foreground clouds. We define small (0), intermediate (1), and large (2) FWHM values using the ranges of $\textrm{med(FWHM)} \leq 19.5$~km\,s$^{-1}$, $19.5~\textrm{km\,s}^{-1}<\textrm{med(FWHM)}<26.0$~km\,s$^{-1}$, and $\textrm{med(FWHM)} \geq 26.0$~km\,s$^{-1}$, respectively. We consider the variation of the FWHM within an area in the sky to be small if $\textrm{std(FWHM)}<1.0$~km\,s$^{-1}$, and minimal if also the number of observed targets is $\geq 5$. Atomic (K~\textsc{i}, Na~\textsc{i}) or molecular (CH$^*$) line is considered to be splitting if the number of distinguished (no significant blending) kinematic components is two, or higher. An apparent gradient in the FWHM across the sky is noted when the object, either alone or in combination with other objects in the region, exhibits this characteristic. The line of sight classification is based on the prototype cases: USco (intermediate FWHM, no clear line splitting), Perseus (small FWHM, typically no splitting), and Eagle (typically intermediate-to-large FWHM, distinguished splitting). In case of NGC~6475, the variation decreases if two outlying measurements are excluded from the calculation. HD~110432 and stars in the direction of Chamaeleon at distances $>1$~kpc show signs of splitting, although the measured DIB profile widths are as narrow as in the case of Perseus (except for HD~109867). Three of the targets (HD~92607, HD~93222, and HD~303135) in the Carina Nebula~(S) show signs of splitting, making the classification of this object uncertain (USco-type or Eagle-type). Possible splitting can also be noted in Orion, specifically towards HD~37903 (K~\textsc{i} line) .

}
\end{table*}
\FloatBarrier
% ------------------- END TABLE

\section{Problematic measurements}\label{section:C}

\begin{figure}
 \includegraphics[width=\columnwidth]{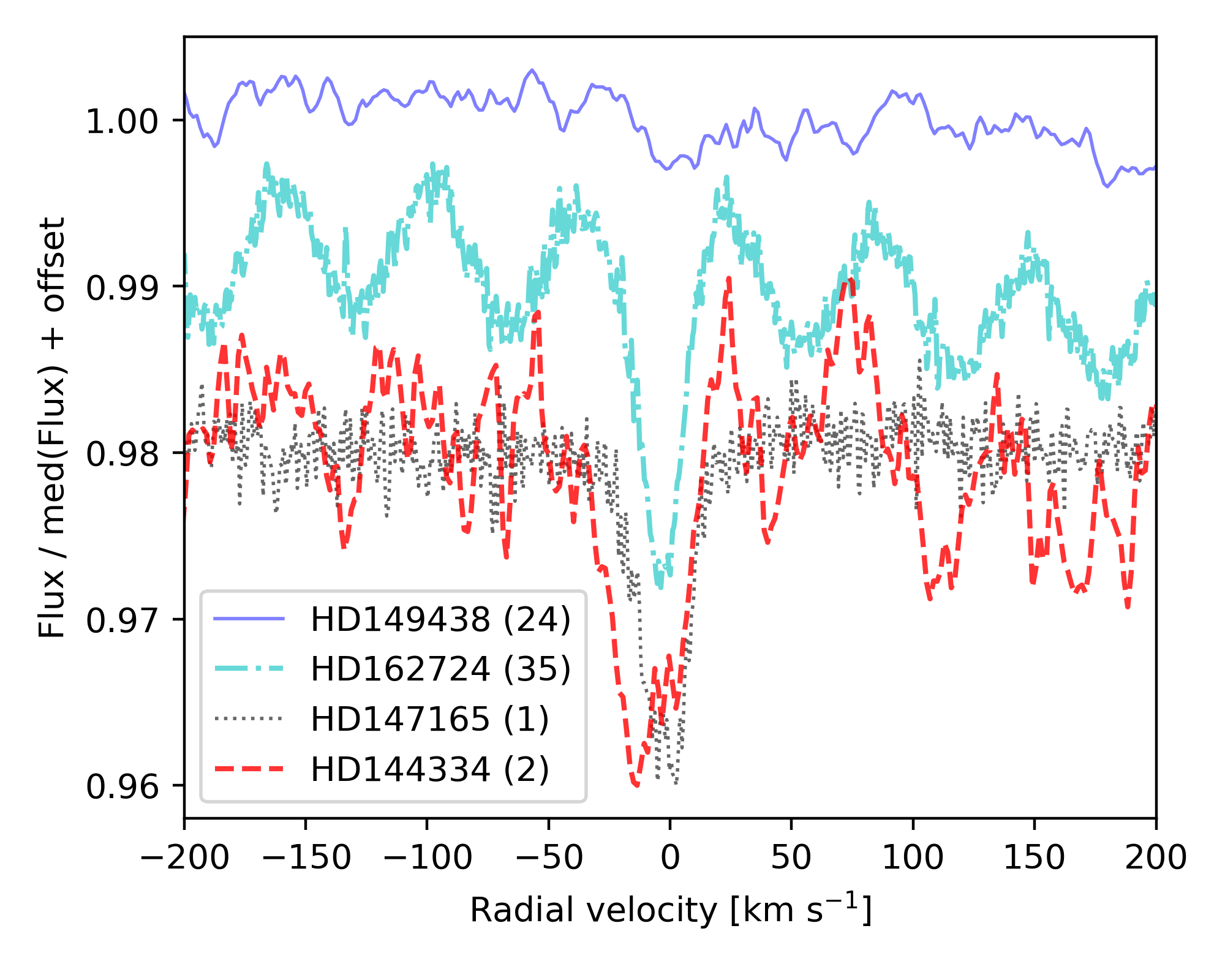}
 \caption{Similar to Fig.~\ref{fig:5} but only showing the profile of the DIB. Wavelengths were shifted to the rest position of the Na~\textsc{i} line, except for HD~147165 where K~\textsc{i} was used. The profiles of the DIB towards HD~147165 and HD~144334 are compared by scaling the profiles to the same line depth. The numbers in brackets displayed within the legend correspond to the number of stacked spectra.}
 \label{fig:B}
\end{figure}

The spectra of two targets, HD~144334 and HD~149438 ($\tau$~Sco), that are very important for the analysis of USco represented a problem for our fitting procedure. In case of HD~149438, the DIB appears to be extremely weak and difficult to identify even after stacking multiple spectra. HD~144334 shows a more easily distinguished profile, but the obtained FWHM of the profile was unexpectedly large even when compared with $\sigma$~Sco (HD~147165). The stacked spectra of HD~144334, HD~149438, and $\sigma$~Sco (for comparison) are displayed in Fig.~\ref{fig:B}. We can clearly see that the increased broadening of the DIB towards HD~144334 is a valid piece of information extracted from the spectra.

With the exception of $\sigma$~Sco, we were forced to use Na~\textsc{i} to identify the rest wavelength due to the absence of a strong K~\textsc{i} line. Since the optical Na~\textsc{i} lines are splitting similar to what has been observed for Ca~\textsc{ii} \citep[e.g.][]{2024A&A...689A..84P}, we pick the red component to represent the velocity of the local ISM. The possibly problematic use of Na~\textsc{i} for the rest wavelength identification does not have an influence on the main message behind Fig.~\ref{fig:B}, where we focus on the broadening and the strength of the DIB.

As was mentioned in Section~\ref{section:4.2}, our procedure failed to obtain the correct profile parameters of the DIB towards HD~162724. The main problem lies in the "wavy" nature of the HARPS pipeline reduced spectrum (see Fig.~\ref{fig:B}). Our procedure assumes that the structure of the continuum within the DIB profile is close to linear and the lack of a more complicated continuum modelling results in an incorrect identification of the profile end-points. To deal with this problem, we included a simple harmonic model of the fringe pattern and removed it prior to fitting. This was necessary for three targets: HD~51193, HD~113904, and HD~162724.

\section{List of additional figures}\label{section:D}

Here we list the remaining figures mentioned throughout Section~\ref{section:4}. These are similar to Fig.~\ref{fig:2}, Fig.~\ref{fig:4}, and Fig.~\ref{fig:5}, and are based on various regions in the sky.

\begin{figure}
 \includegraphics[width=\columnwidth]{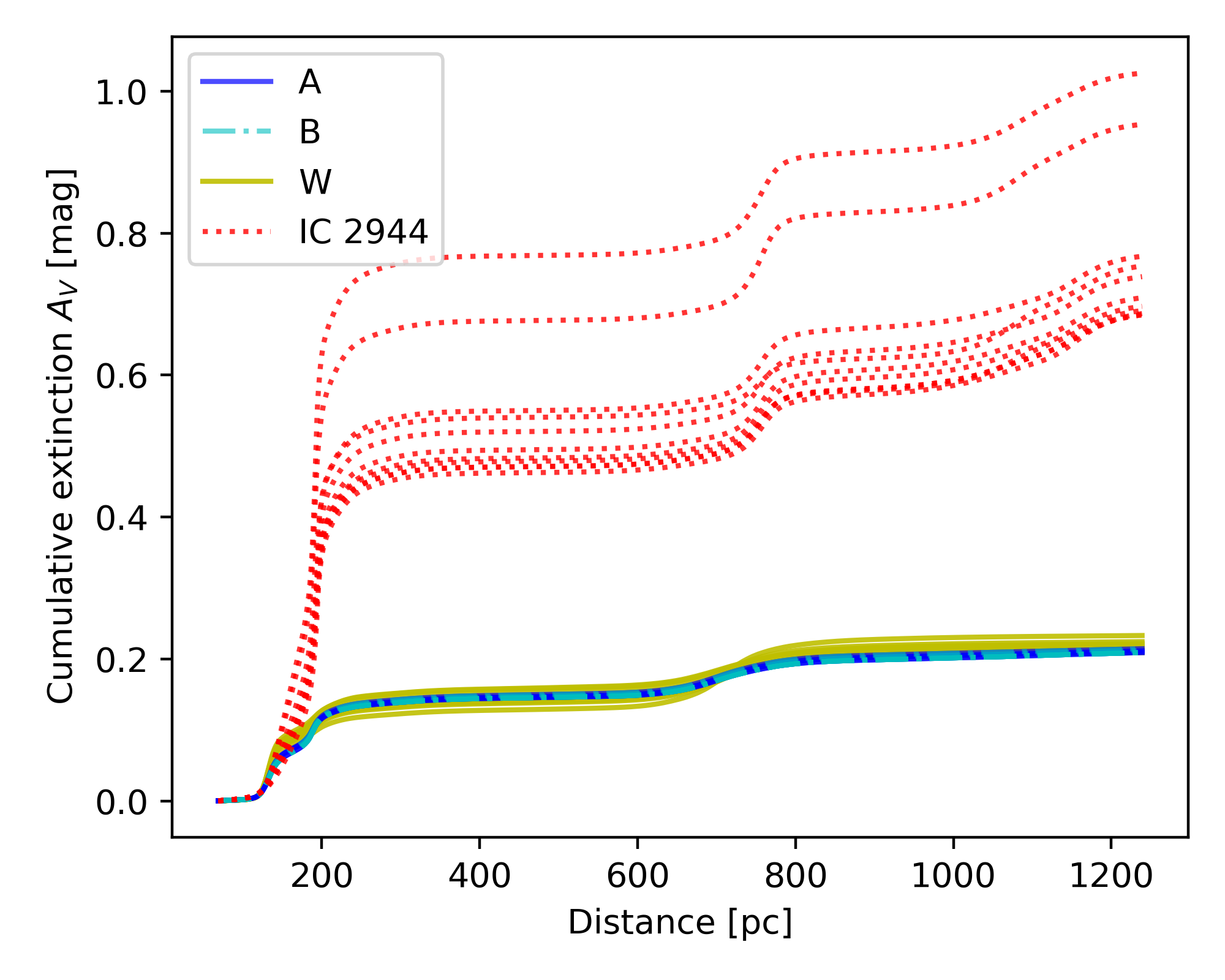}
 \caption{Line-of-sight extinction towards different targets in the Carina region. Together with the western portion of the nebula, Carina groups A and B probe a low-density ISM within 1.25~kpc from the Sun.}
 \label{fig:C1}
\end{figure}

\begin{figure}
 \includegraphics[width=\columnwidth]{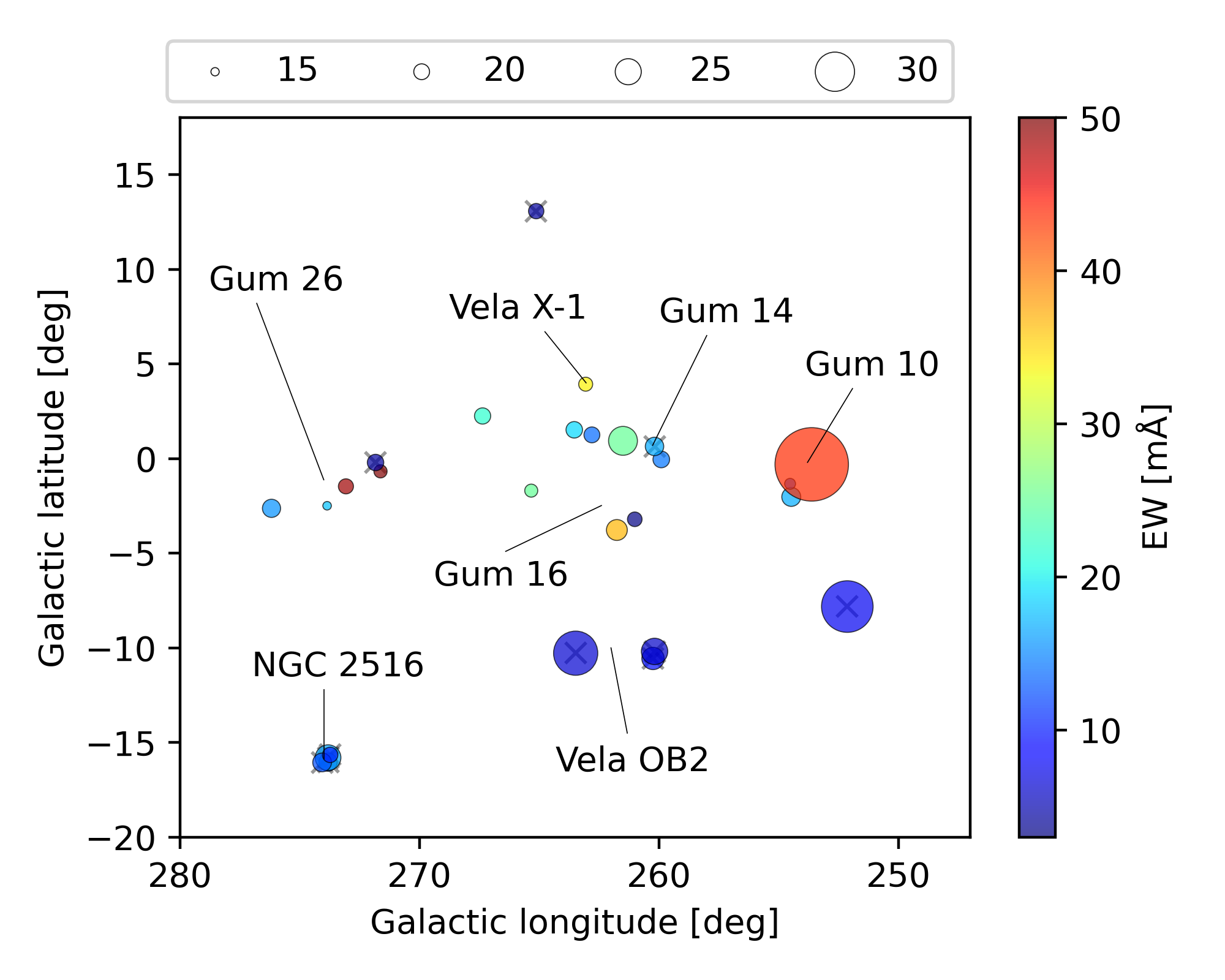}
 \caption{Sky map of the targets within the Vela region. The size of a data point indicates the measured FWHM (values in~km\,s$^{-1}$), while the colour indicates the strength of the DIB. The most prominent groups of stars are highlighted. Grey crosses indicate stars located within 500~pc from the Sun.}
 \label{fig:C2a}
\end{figure}

\begin{figure}
 \includegraphics[width=\columnwidth]{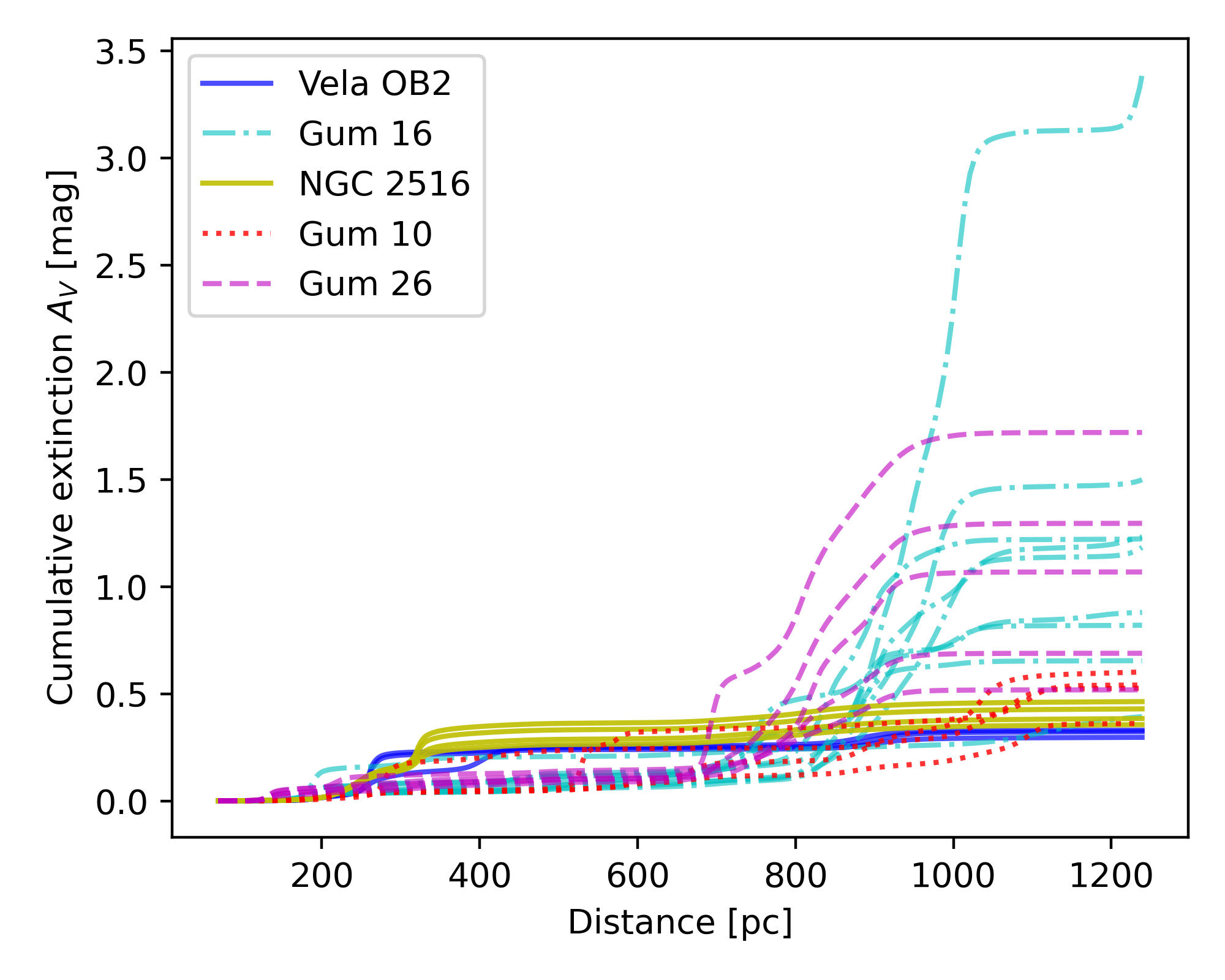}
 \caption{Line-of-sight extinction towards different targets in the Vela region.}
 \label{fig:C2b}
\end{figure}

\begin{figure}
 \includegraphics[width=\columnwidth]{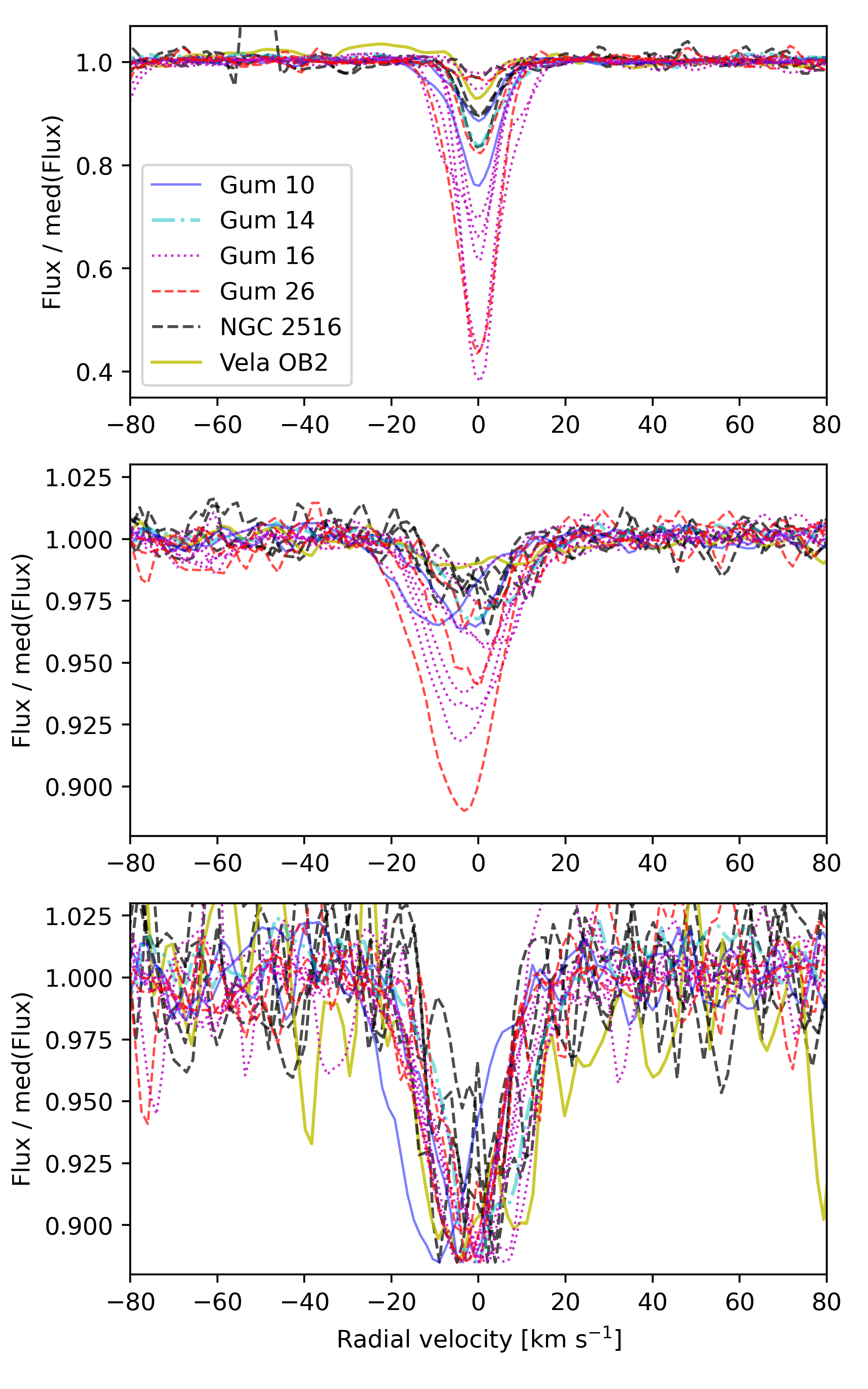}
 \caption{Similar to Fig.~\ref{fig:5} but for the Vela region. Wavelengths were shifted to the rest position of the K~\textsc{i} line. The difficulty in distinguishing between the local ISM and the more distant ISM should be noted in this region (opposite direction of Galactic rotation).}
 \label{fig:C2c}
\end{figure}

\begin{figure}
 \includegraphics[width=\columnwidth]{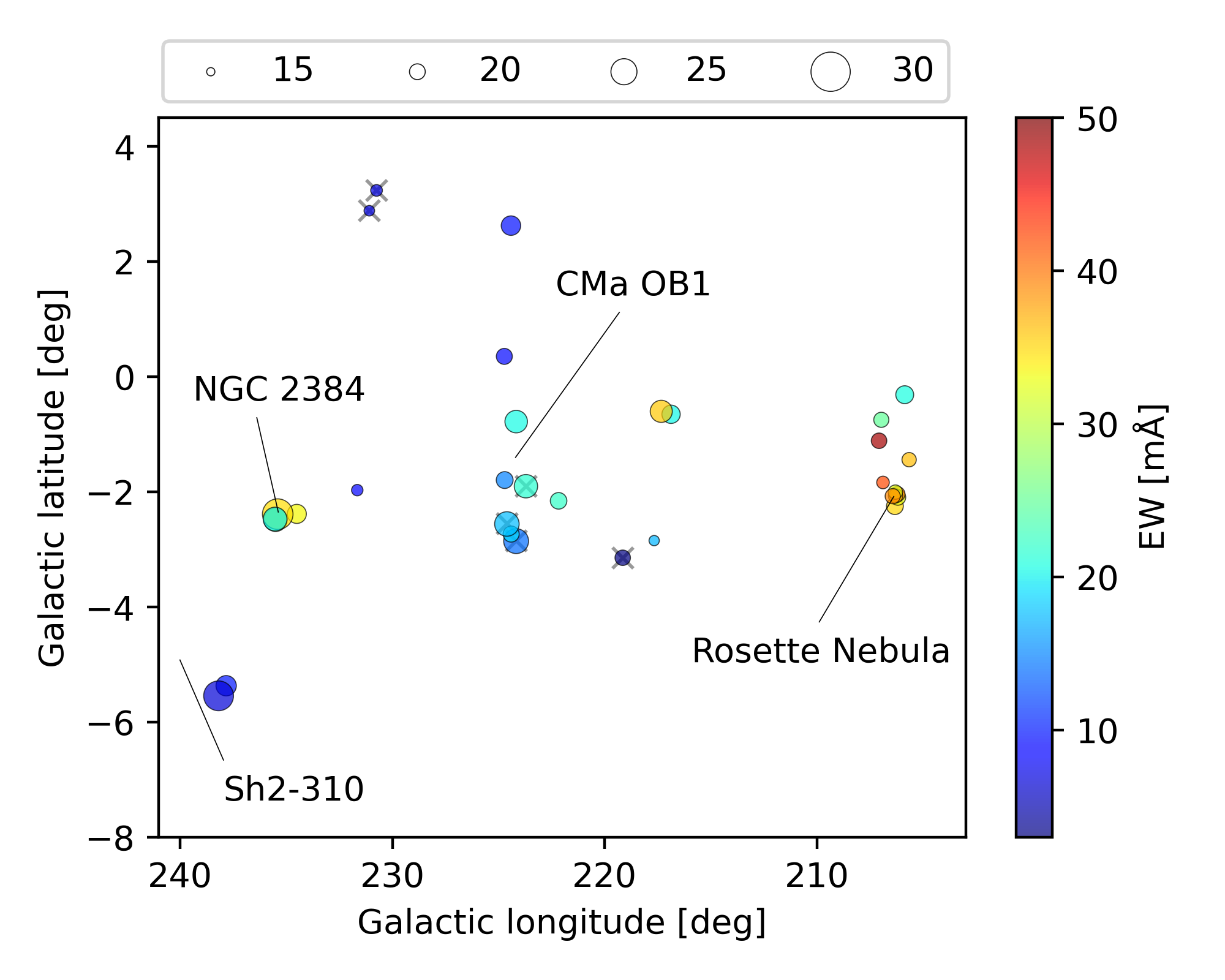}
 \caption{Sky map of the targets within the CMa-Mon region. The size of a data point indicates the measured FWHM (values in~km\,s$^{-1}$), while the colour indicates the strength of the DIB. The most prominent groups of stars are highlighted. Grey crosses indicate stars located within 500~pc from the Sun.}
 \label{fig:C3a}
\end{figure}

\begin{figure}
 \includegraphics[width=\columnwidth]{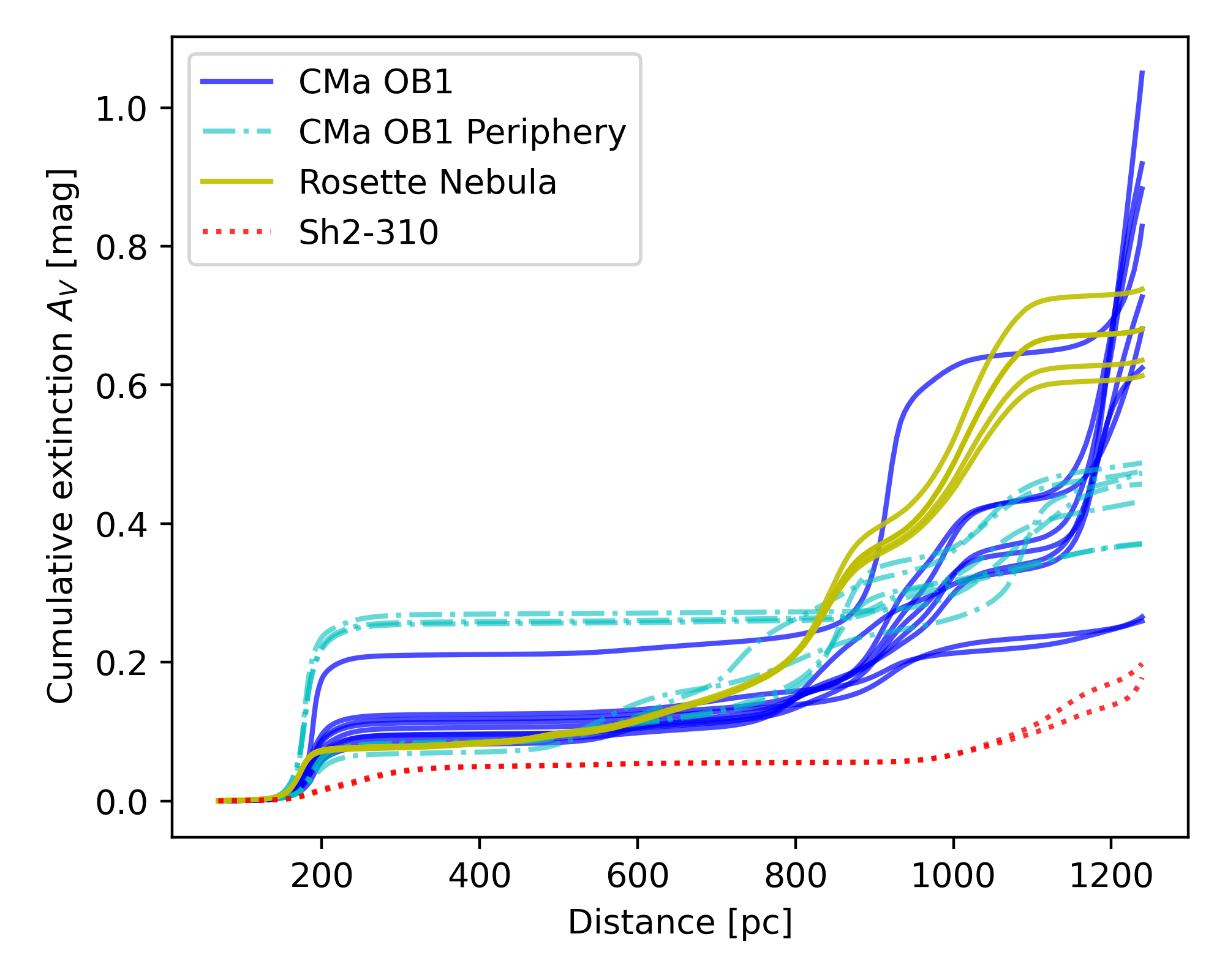}
 \caption{Line-of-sight extinction towards different targets in the CMa-Mon region.}
 \label{fig:C3b}
\end{figure}

\begin{figure}
 \includegraphics[width=\columnwidth]{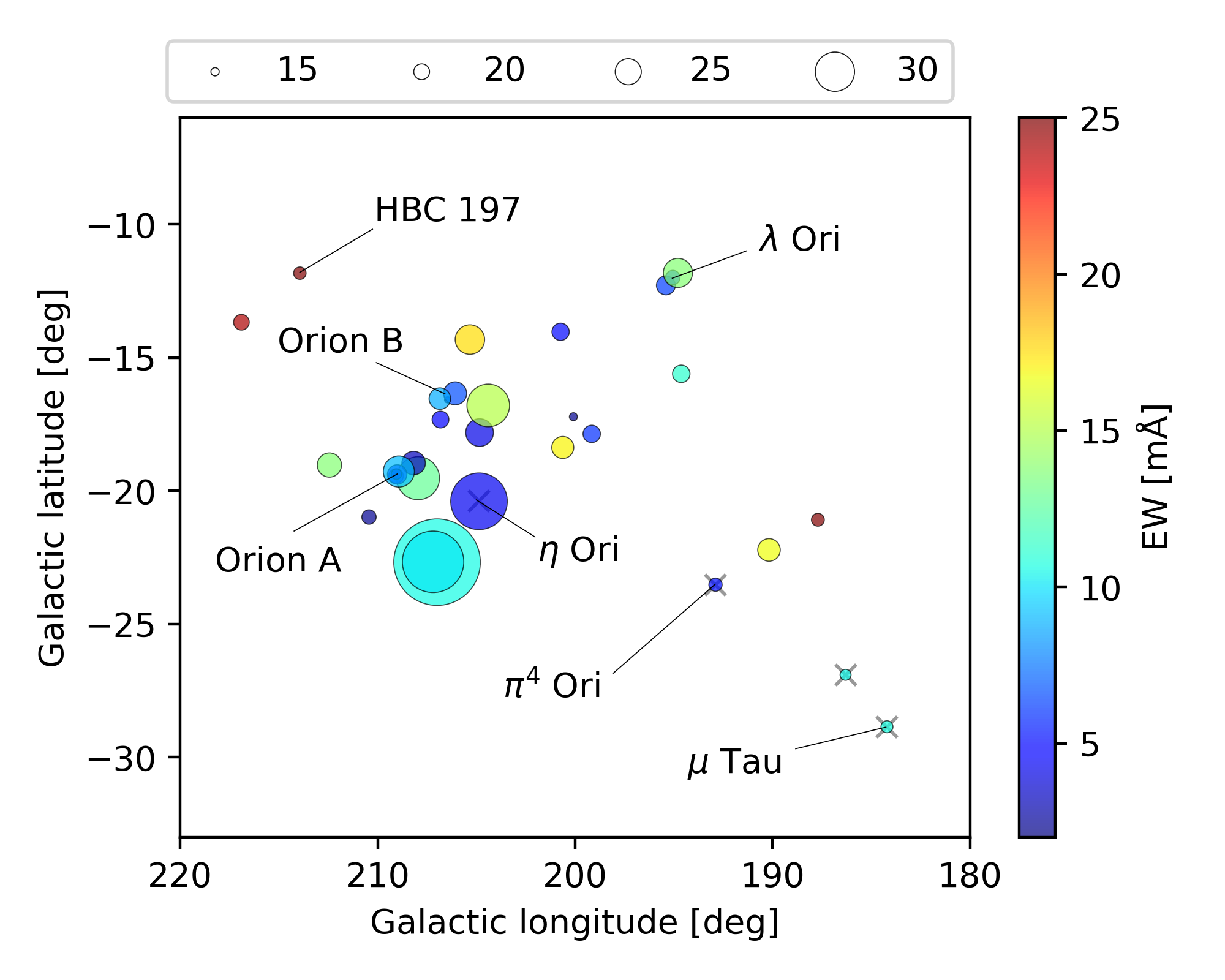}
 \caption{Sky map of the targets within the Orion region. The size of a data point indicates the measured FWHM (values in~km\,s$^{-1}$), while the colour indicates the strength of the DIB. The most prominent groups of stars are highlighted. Grey crosses indicate stars located within 300~pc from the Sun.}
 \label{fig:C4a}
\end{figure}

\begin{figure}
 \includegraphics[width=\columnwidth]{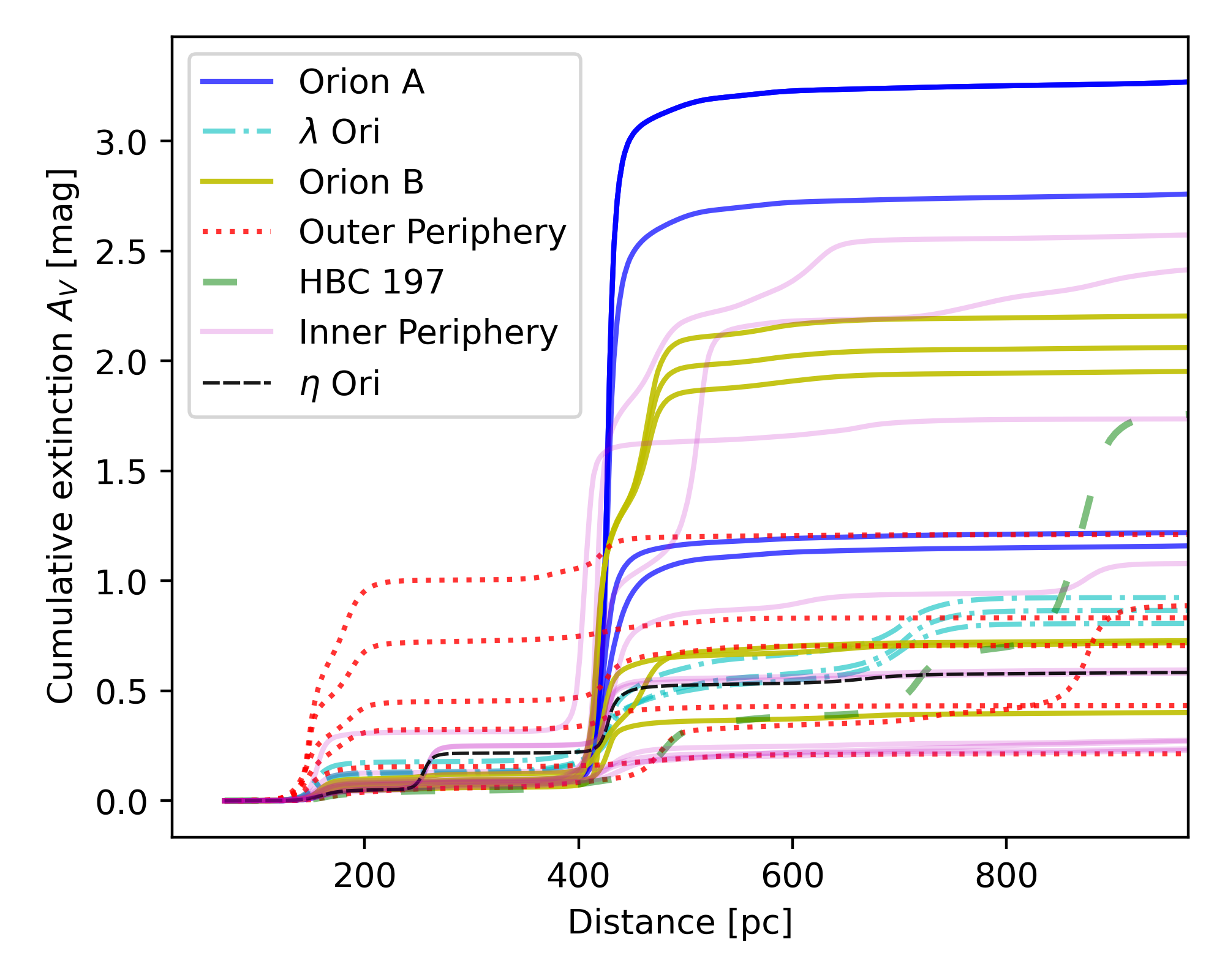}
 \caption{Line-of-sight extinction towards different targets in the Orion region.}
 \label{fig:C4b}
\end{figure}

\begin{figure}
 \includegraphics[width=\columnwidth]{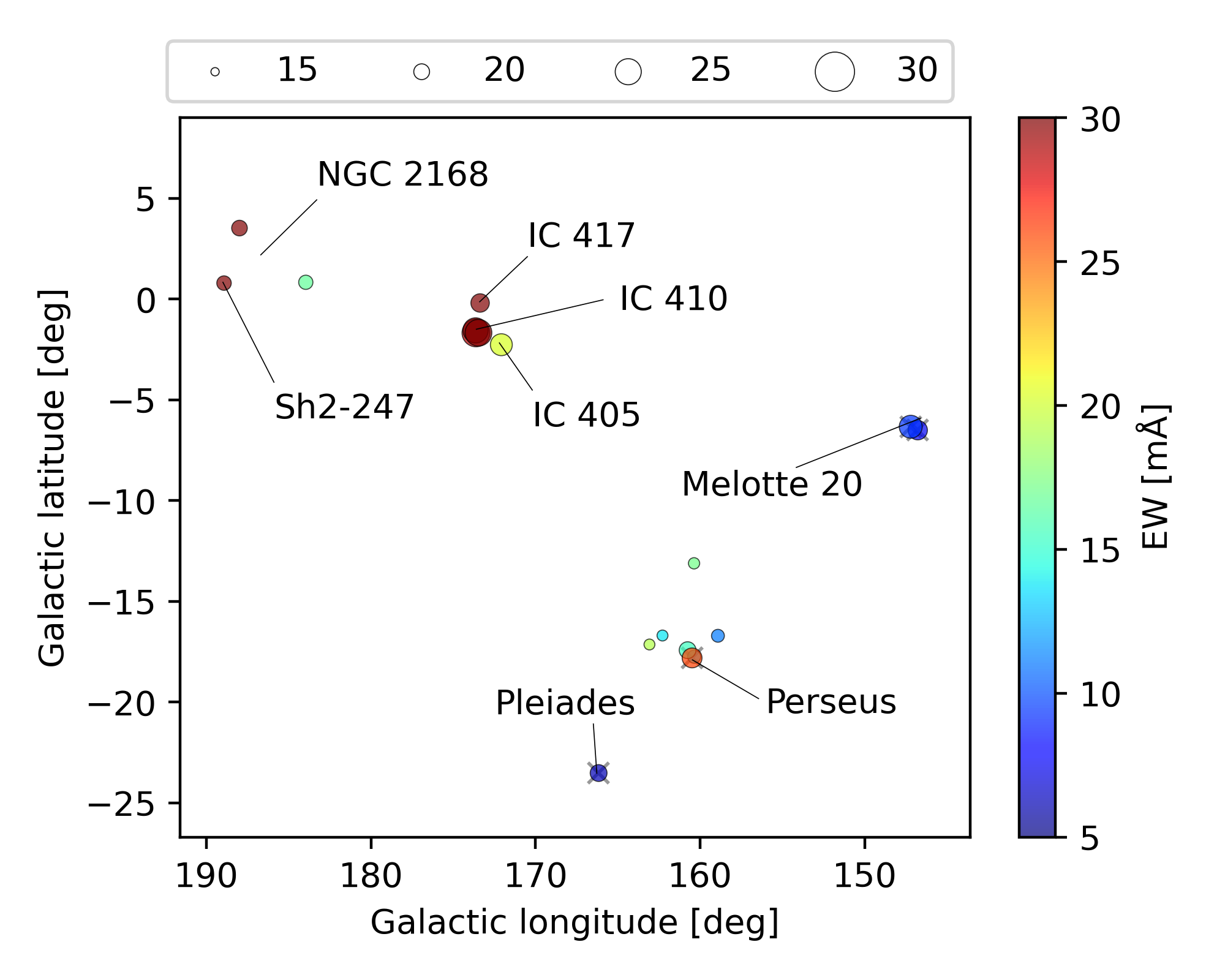}
 \caption{Sky map of the targets within the Per-Tau and Gem-Aur regions. The size of a data point indicates the measured FWHM (values in~km\,s$^{-1}$), while the colour indicates the strength of the DIB. The most prominent groups of stars are highlighted. Grey crosses indicate stars located within 200~pc from the Sun.}
 \label{fig:C5a}
\end{figure}

\begin{figure}
 \includegraphics[width=\columnwidth]{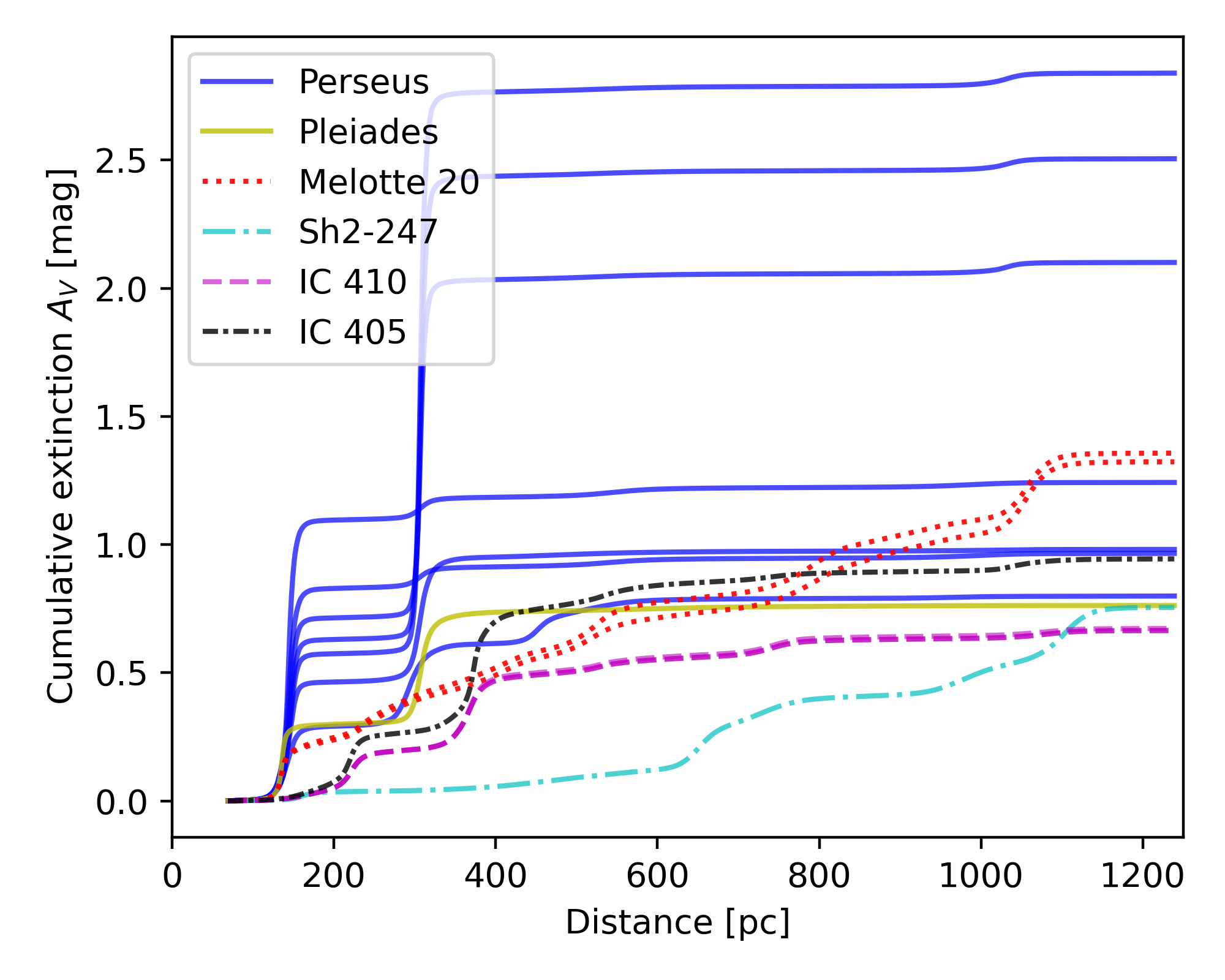}
 \caption{Line-of-sight extinction towards different targets in the Per-Tau and Gem-Aur regions.}
 \label{fig:C5b}
\end{figure}

\begin{figure}
 \includegraphics[width=\columnwidth]{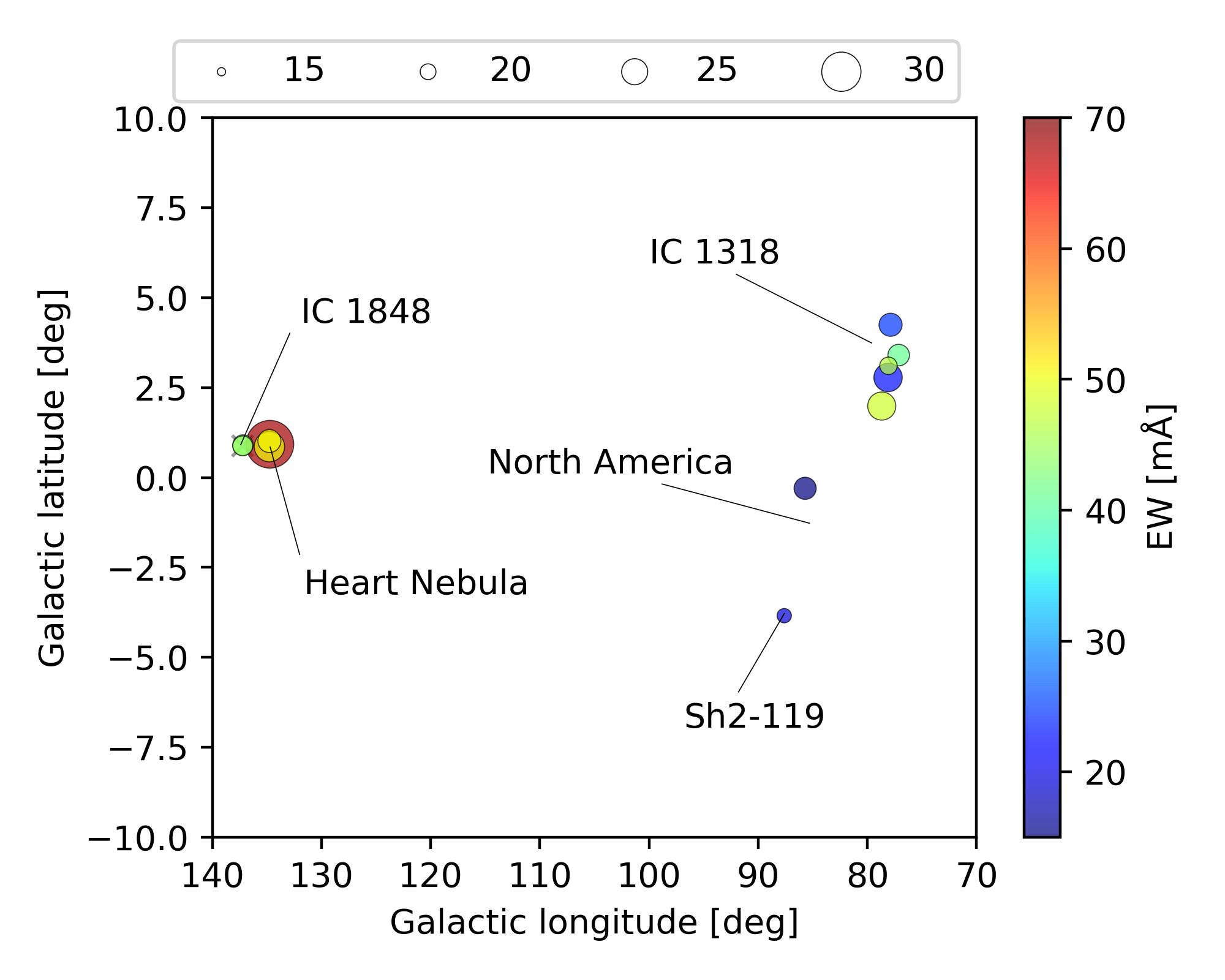}
 \caption{Sky map of the targets within the Heart and Sadr regions. The size of a data point indicates the measured FWHM (values in~km\,s$^{-1}$), while the colour indicates the strength of the DIB. The most prominent groups of stars are highlighted. Grey crosses indicate stars located within 300~pc from the Sun.}
 \label{fig:C6a}
\end{figure}

\begin{figure}
 \includegraphics[width=\columnwidth]{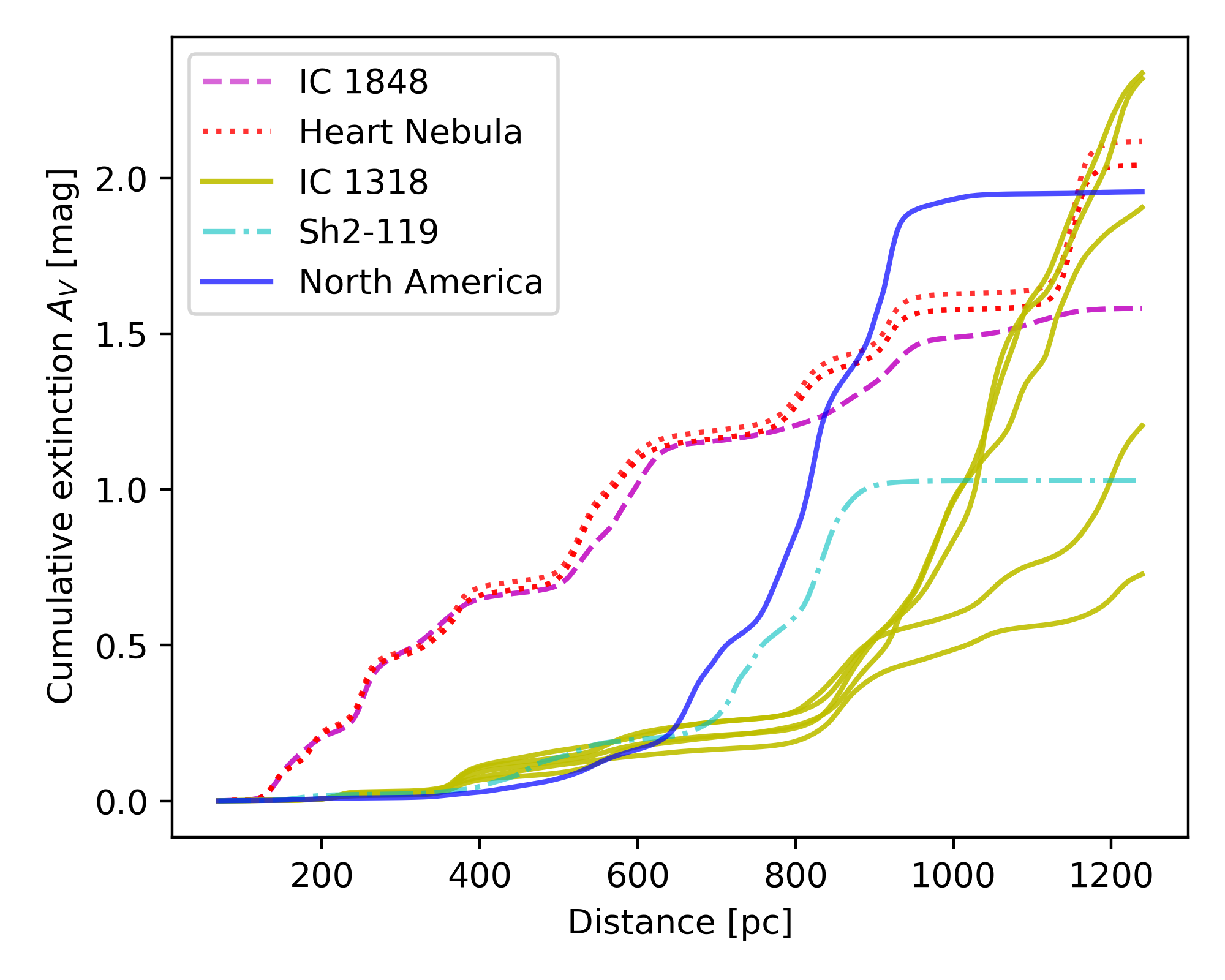}
 \caption{Line-of-sight extinction towards different targets in the Heart and Sadr regions.}
 \label{fig:C6b}
\end{figure}

\begin{figure}
 \includegraphics[width=\columnwidth]{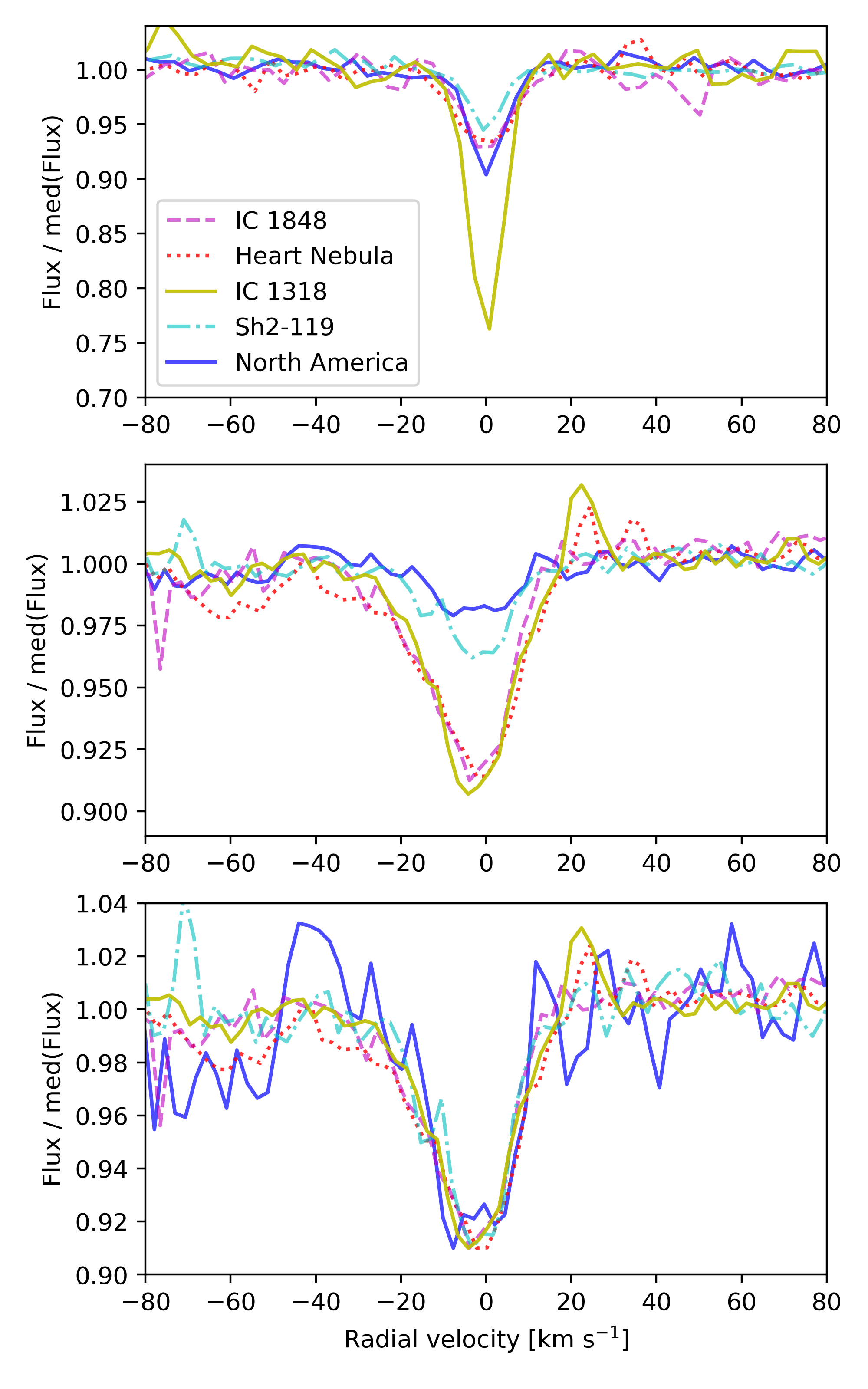}
 \caption{Similar to Fig.~\ref{fig:5} but for the Heart and Sadr regions. Wavelengths were shifted to the rest position of the CH$^*$ line. The difficulty in distinguishing between the local ISM and the more distant ISM should be noted in this region (opposite direction of Galactic rotation).}
 \label{fig:C6c}
\end{figure}

\end{appendix}

\end{document}